\documentclass[aps,prb,twocolumn,amsmath,amssymb,nofootinbib,eqsecnum,superscriptaddress,floatfix]{revtex4}
\usepackage{graphicx}
\usepackage{dcolumn} 
\usepackage{bm}      
\usepackage{longtable}
\usepackage{color}

\begin{document}


\title{
Conductance fluctuations 
in disordered superconductors with broken time-reversal symmetry
near two dimensions 
      }

\author{S.\ Ryu}
\affiliation{Kavli Institute for Theoretical Physics,
	     University of California, 
	     Santa Barbara, 
	     CA 93106, 
	     USA}

\author{A.\ Furusaki}
\affiliation{Condensed Matter Theory Laboratory, 
             RIKEN, 
             Wako, 
             Saitama 351-0198, 
             Japan}

\author{A.\ W.\ W.\ Ludwig}
\affiliation{Department of Physics,
             University of California, 
             Santa Barbara, 
             CA 93106
	     USA}
\affiliation{Kavli Institute for Theoretical Physics,
	     University of California, 
	     Santa Barbara, 
	     CA 93106, 
	     USA}

\author{C.\ Mudry}
\affiliation{Paul Scherrer Institute,
             CH-5232 Villigen PSI, 
             Switzerland}

\date{\today}

\begin{abstract}
We extend the analysis of the conductance fluctuations in 
disordered metals by Altshuler, Kravtsov, and Lerner (AKL)
to  disordered superconductors with broken time-reversal symmetry 
in $d=(2+\epsilon)$ dimensions
(symmetry classes C and D of Altland and Zirnbauer).
Using a perturbative renormalization group
analysis of the corresponding non-linear sigma model (NL$\sigma$M)
we compute the anomalous scaling dimensions of the
dominant scalar operators with $2s$ gradients
to one-loop order. We show that,
in analogy with
the result of AKL for ordinary,  
metallic systems (Wigner-Dyson classes),
an infinite number of high-gradient operators 
would become relevant (in the renormalization group sense)
near two dimensions 
if contributions beyond one-loop order are ignored.
We explore the possibility to compare, in symmetry class D,
the $\epsilon=(2-d)$ expansion in $d<2$ with exact results
in one dimension.
The method we use to perform the one-loop renormalization analysis
is valid for general symmetric spaces of K\"ahler type, 
and suggests that this is a generic property of
the perturbative treatment of NL$\sigma$Ms
defined on Riemannian symmetric target spaces.
\end{abstract}

\maketitle

\section{
Introduction
        }

\subsection{Conductance fluctuations and the non-linear sigma model}
\label{subsec: Conductance fluctuations and the non-linear sigma model}

The non-linear sigma model (NL$\sigma$M)
provides, as first proposed by F.\ Wegner,\cite{WegnerNonLinSigma1979}
an efficient framework within which to formulate (Anderson-type) localization
transitions of noninteracting quantum mechanical particles
subject to (static) random potentials.
This provides some of the simplest
models of electrons in disordered solids.
In a seminal piece of work, Altshuler, Kravtsov
and Lerner (AKL)
\cite{%
Altshuler91,
Altshuler86to89}
showed in the 1980's how to incorporate
the important phenomenon of conductance fluctuations%
\cite{Altshuler85,Lee85,Stone85}
into this framework.
Within a perturbative renormalization group (RG)
analysis of the weakly coupled NL$\sigma$M
in $d=2$ (or $ d=2+\epsilon$) spatial dimensions,
i.e., in the regime 
of large (dimensionless) conductance $g$,
these authors  stressed the importance
of terms in the low-energy effective action
of the NL$\sigma$M
which  possess an arbitrary large number ($=2s$) of spatial
derivatives but do not violate any symmetries.
These terms are called ``high-gradient operators''
and are generalizations of the kinetic 
term in the NL$\sigma$M which has two derivatives ($2s=2$).
High-gradient operators are highly irrelevant in the RG sense
by ``power counting'' 
(i.e., without incorporation of fluctuation corrections).
Specifically, AKL and Yudson studied the anomalous (RG) dimensions
of these operators to one-loop order
in the fluctuation corrections
in $d=2$ and in the
$\epsilon=(d-2)$-expansion
for $d>2$.
\cite{Altshuler91,Altshuler86to89,Kravtsov88,Kravtsov89}
It was 
found that
(i):
 these operators
would  become strongly relevant (in the RG sense)
for a  large enough number $2s$ of derivatives, 
in spite of their strong irrelevance by power counting, 
if contributions beyond one-loop order are ignored,
and  
that (ii): they would
(consequently) dominate the cumulants
of the conductance of sufficiently large order $2s$,
in a sample of (large) linear size $L$.
Based on these observations AKL
were led to argue that for large
mean (dimensionless) conductance $\bar g$,
the main weight of the probability distribution
of the conductance arises from
a narrow peak at $g=\bar{g}$,
while there exist anomalously
long tails (close to being log-normal).
The latter arise from the 
high moments of the conductance,
which are dominated by the
above-mentioned high-gradient operators appearing
in the low-energy effective action.
Upon approaching the regime of stronger disorder 
(smaller mean conductance $\bar g$), 
and in particular upon approaching a metal-insulator transition,
the probability distribution becomes increasingly broad.
AKL find similar behavior for
the probability distributions
of other observables,
including, e.g.,~the local density of states (LDOS),
current densities, and long-time tails of relaxation currents.
Very recently,\cite{Motrunich02,Mudry03} 
the basic observations of
AKL for the LDOS were confirmed in a chiral symmetry class
(nomenclature of Refs.~\onlinecite{Zirnbauer96} and 
\onlinecite{Altland97}),
and were analyzed more completely
in this case.

Ten
years ago, in another seminal piece of work,
Zirnbauer,\cite{Zirnbauer96}
and Altland and Zirnbauer \cite{Altland97} (AZ)
provided a general classification  scheme for  the behavior
of non-interacting quantum mechanical particles
subject to random potentials,
on length scales much larger than the mean
free path in terms of a total of ten symmetry classes.\cite{CartanFootnote}
In every realization of disorder,
the Hamiltonian has the same symmetries
as a corresponding ensemble of random matrices. 
This encompasses the three previously known so-called
``Wigner-Dyson symmetry classes''
relevant for the physics of Anderson localization of electrons in disordered solids
(corresponding to
``orthogonal, unitary, and symplectic'' random matrix ensembles).%
\cite{Efetov83}
Altland and Zirnbauer's extension includes four novel symmetry classes
describing the (Anderson-like) localization physics of  non-interacting
Bogoliubov-De Gennes (BdG) quasiparticles
within a mean-field treatment of pairing
in disordered superconductors.
(Four symmetry classes arise since  SU$(2)$ spin-rotational, or
time-reversal invariance may be present or absent.)
As parameters are varied,
the wave functions of the (BdG) quasiparticles
may be extended or localized due to disorder,
and (Anderson localization type) 
transitions between such phases may occur. 
Since BdG quasiparticles
do not carry a conserved electric charge, 
the difference between phases of localized and extended 
quasiparticle wave functions manifests itself
not in the electrical conductivity,
but instead in the thermal conductivity $\kappa$,
and if spin is conserved, also in the spin  
conductivity $\sigma^{\ }_{\mathrm{spin}}$.\cite{FootnoteSpinConductivity}
In complete analogy with the
fluctuations of the electrical
conductance in a finite sample of disordered metals 
(described by
the Wigner-Dyson symmetry classes),
the system of BdG quasiparticles
also exhibits fluctuations of the appropriate
conductance in a finite-size sample.
These are the fluctuations of the thermal conductance
(divided by temperature $T$),
or of the spin conductance.
Henceforth, for notational convenience,
we will refer to these quantities simply as ``the conductance'',
and denote the corresponding ``dimensionless conductance''
again by the symbol $g$.

Just as for
disordered metals
(i.e., for the Wigner-Dyson symmetry classes),
the NL$\sigma$M also provides an efficient
framework to formulate (Anderson-type)
localization physics 
of BdG quasiparticles in superconductors.
In this article we focus on
systems where time-reversal symmetry
is broken in every realization of
disorder:
the corresponding AZ symmetry classes are conventionally
called ``class D'' if SU$(2)$ spin-rotational
invariance is broken and ``class C'' if it is conserved, 
in every disorder realization.%
\cite{footnoteDIIIandCI}
In parallel, and for comparison,
we will also discuss here localization physics
of ordinary electrons in the absence
of time-reversal symmetry,
which is described by the Wigner-Dyson symmetry class 
conventionally called ``unitary class'' or ``class A''.
(See also Table \ref{tab: RMT and NLSM}.)
When disorder averaging is implemented
using fermionic replicas, 
the target manifolds of the resulting NL$\sigma$Ms
are the symmetric spaces
$\mathrm{O}(2N)/\mathrm{U}(N)$
for symmetry class D,\cite{Senthil00}
and
$\mathrm{Sp}(N)/\mathrm{U}(N)$
for symmetry class C.\cite{Senthil98}
We remind the reader that for the familiar 
unitary symmetry class A,
the corresponding target space\cite{Efetov80} is
$\mathrm{U}(2N)/[\mathrm{U}(N)\times \mathrm{U}(N)]$.
(In all three cases the number $N$ of replicas is taken to zero
at the end of calculations.) 

The aim of this paper is to investigate
perturbatively, in the spirit of the original work of AKL
for ordinary metals,
the conductance fluctuations for the 
novel AZ symmetry classes D and C,
i.e., for BdG quasiparticles in superconductors
lacking time-reversal symmetry,
in the absence (class D)
or presence (class C)
of spin-rotational symmetry.
Our main results are briefly summarized
in Sec.~\ref{SummaryResults} below. 
We also point out that it might be possible
in the case of symmetry class D
to make contact with exact nonperturbative results
obtained in $d=1$ dimensions
from the so-called Dorokhov-Mello-Pereyra-Kumar 
(Fokker-Planck) analysis\cite{DorokhovMPK}  
of the probability distribution of the conductance. 

We end this section by recalling, as an aside, that
the NL$\sigma$Ms in all three symmetry classes, D, C, and A, 
share certain essential features associated with
time-reversal symmetry breaking.
First, in $d=2$ dimensions, 
all three target manifolds allow for a topological term 
(a term that does not modify the equations of motion),
the so-called ``theta- or Pruisken term''. 
Its presence is responsible for the appearance of 
thermal, 
spin, 
or charge Hall insulating phases, 
and corresponding plateau transitions in $d=2$ dimensions.
\cite{Senthil98,Kagalovsky99,Marston99,Chalker01,Pruisken84}
Second, the target manifold for symmetry class D stands out 
in that it is not simply connected.\cite{Zirnbauer96,Read00,Bocquet00}
Consequently, there are certain discrete $\mathbb{Z}^{\ }_2$
domain wall excitations
that may give rise to a rich phase diagram
for certain types of disorder.
\cite{Gruzberg01,Gruzberg05,Chalker01} 
In this paper we shall ignore the physics associated 
with such $\mathbb{Z}^{\ }_2$
domain walls, which
requires an extension
of the NL$\sigma$M
description
in symmetry class D.
These features are not present in certain
realizations of this symmetry class,
on which we focus in this paper.
Finally, all three target spaces have in common that
they are K\"ahler manifolds from a geometrical point of view. 
This means that each target space defines a complex manifold
endowed with a Hermitian metric derived from a K\"ahler potential.%
\cite{Helgason78,Nakahara03} 
This property will allow us to 
perform the perturbative RG analysis in a unified way 
for symmetry classes D, C, and A.
(See Sec.~\ref{SummaryMethodCalculation} for a brief summary.)

\begin{table}
\begin{center}
\begin{tabular}{cccccc}\hline\hline
RMT 
& 
$\mathcal{T}$ 
& 
$\mathrm{SU}(2)$   
&  
NL$\sigma$M target space      
& 
$\vartheta$ &$\mathcal{N}$
\\ \hline 
D    
& 
no         
& 
no    
& 
$\mathrm{O}(2N)/\mathrm{U}(N)$ 
& 
$+1$ 
& 
$N$ 
\\ 
C    
& 
no         
& 
yes   
& 
$\mathrm{Sp}(N)/\mathrm{U}(N)$ 
& 
$-1$ 
& 
$N$ 
\\ \hline 
A   
&   
no    
&  
---   
& 
$\mathrm{U}(p+q)/[\mathrm{U}(p)\times \mathrm{U}(q)]$ 
& 
0 
& $p+q$ 
\\ \hline \hline
\end{tabular}
\end{center}
\caption{
\label{tab: RMT and NLSM}
Fermionic replica target spaces of the NL$\sigma$M describing
weakly disordered superconductors and metals
when time-reversal symmetry ($\mathcal{T}$) is broken.
Symmetry classes D, C, and A refer to the symmetric spaces defined by
the RMT limits of the NL$\sigma$M.
The number of replicas $\mathcal{N}$ is $N$ 
for the two BdG symmetry classes 
whereas it is $p=q=N$ for the unitary symmetry class.
The index $\vartheta=0,\pm1$ distinguishes the behavior of the beta function
for the NL$\sigma$M in symmetry classes D, C, and A.
        } 
\end{table}

\subsection{
Summary of main results
           }
\label{SummaryResults}

\begin{figure}
\begin{center}
\begin{picture}(220,100)(-120,-80)
\thinlines

\put(-120,-60){\line(1,0){100}}
\put(-10,-60){\line(1,0){100}}

\put(-120,-10){\line(1,0){100}}
\put(-10,-10){\line(1,0){100}}

\thinlines
\put(-120,10){Class C, $d>2$}
\put(-10,10){Class C, $d<2$}

\put(-25,-5){$t$}
\put(85,-5){$t$}

\put(-122,-21){$0$}
\put(-12,-21){$0$}
\put(-72,-21){$t^*$}

\put(-120,-40){Class D, $d>2$}
\put(-10,-40){Class D, $d<2$}

\put(-25,-55){$t$}
\put(85,-55){$t$}

\put(-122,-71){$0$}
\put(-12,-71){$0$}
\put(38,-71){$t^*$}

\put(-120,-10){\circle{6}}
\put(-70,-10){\circle*{6}}

\put(-10,-10){\circle{6}}

\put(-120,-60){\circle{6}}

\put(-10,-60){\circle{6}}
\put(40,-60){\circle*{6}}

\thinlines
\put(-102,-60){\vector(-1,0){10}}
\put(-87,-60){\vector(-1,0){10}}
\put(-72,-60){\vector(-1,0){10}}
\put(-57,-60){\vector(-1,0){10}}
\put(-42,-60){\vector(-1,0){10}}
\put(-27,-60){\vector(-1,0){10}}

\put(-10,-60){\vector(1,0){10}}
\put(5,-60){\vector(1,0){10}}
\put(20,-60){\vector(1,0){10}}
\put(60,-60){\vector(-1,0){10}}
\put(75,-60){\vector(-1,0){10}}
\put(90,-60){\vector(-1,0){10}}

\put(-5,-10){\vector(1,0){10}}
\put(10,-10){\vector(1,0){10}}
\put(25,-10){\vector(1,0){10}}
\put(40,-10){\vector(1,0){10}}
\put(55,-10){\vector(1,0){10}}
\put(70,-10){\vector(1,0){10}}

\put(-105,-10){\vector(-1,0){10}}
\put(-90,-10){\vector(-1,0){10}}
\put(-75,-10){\vector(-1,0){10}}
\put(-65,-10){\vector(1,0){10}}
\put(-50,-10){\vector(1,0){10}}
\put(-35,-10){\vector(1,0){10}}

\end{picture}

\caption{
\label{fig: IR RG flows}
Infrared (IR) renormalization group (RG) flows
for symmetry classes C and D 
after taking the replica limit
depending on whether dimensionality of space $d$
is larger or smaller than 2.
The open circle at $t=0$ is the metallic fixed point.
The filled circle at $t^{*}$ is the metal-insulator critical point
in symmetry class C 
whereas it describes a stable fixed point
for symmetry class D.
Up to one loop, $t^{*}=|\epsilon|$ where
$\epsilon:=d-2$ is positive for symmetry class C while
it is negative for symmetry class D.
        }
\end{center}
\end{figure}
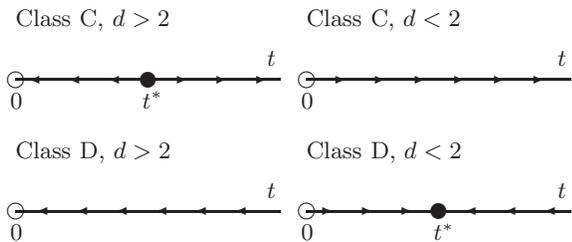

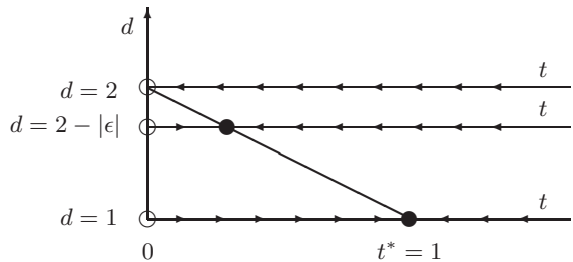
\begin{figure}
\begin{center}
\begin{picture}(220,100)(-120,-80)
\thicklines
\put(-80,-10){\line(2,-1){100}}

\thinlines

\put(-80,-60){\vector(0,1){80}}
\put(-80,-60){\line(1,0){160}}
\put(-80,-25){\line(1,0){160}}
\put(-80,-10){\line(1,0){160}}

\put(- 90, 10){$d$}

\put(-113,-14){$d=2$}
\put(- 80,-10){\circle{6}}
\put(  68,- 6){$t$}

\put(-132,-27){$d=2-|\epsilon|$}
\put(- 80,-25){\circle{6}}
\put(- 50,-25){\circle*{6}}
\put(  68,-21){$t$}

\put(-113,-62){$d=1$}
\put(- 80,-60){\circle{6}}
\put(  19,-60){\circle*{6}}
\put(  68,-56){$t$}

\put(- 82,-75){$0$}
\put(   7,-75){$t^*=1$}

\put(-60,-10){\vector(-1,0){10}}
\put(-45,-10){\vector(-1,0){10}}
\put(-30,-10){\vector(-1,0){10}}
\put(-15,-10){\vector(-1,0){10}}
\put(  0,-10){\vector(-1,0){10}}
\put( 15,-10){\vector(-1,0){10}}
\put( 30,-10){\vector(-1,0){10}}
\put( 45,-10){\vector(-1,0){10}}
\put( 60,-10){\vector(-1,0){10}}

\put(-75,-25){\vector(1,0){10}}
\put(-30,-25){\vector(-1,0){10}}
\put(-15,-25){\vector(-1,0){10}}
\put(  0,-25){\vector(-1,0){10}}
\put( 15,-25){\vector(-1,0){10}}
\put( 30,-25){\vector(-1,0){10}}
\put( 45,-25){\vector(-1,0){10}}
\put( 60,-25){\vector(-1,0){10}}

\put(-75,-60){\vector(1,0){10}}
\put(-60,-60){\vector(1,0){10}}
\put(-45,-60){\vector(1,0){10}}
\put(-30,-60){\vector(1,0){10}}
\put(-15,-60){\vector(1,0){10}}
\put(  0,-60){\vector(1,0){10}}
\put( 40,-60){\vector(-1,0){10}}
\put( 55,-60){\vector(-1,0){10}}
\put( 70,-60){\vector(-1,0){10}}
\end{picture}

\caption{
\label{fig: IR RG flows in Class D}
Possible evolution of the
stable fixed point for symmetry class D in $d=2-|\epsilon|<2$
dimensions between $d=1$ and $d=2$
(after taking the replica limit).
In this sketch it is
assumed that 
the zero $t^*=|\epsilon|+\mathcal{O}(\epsilon^2)$ 
of the beta function, found to 
one-loop  order in the
$\epsilon$ expansion in
$d=2-|\epsilon|$ dimensions, defines
a curve of stable critical points 
which interpolates between the metallic phase
in $d=2$ dimension (at zero coupling $t=0$),
and  the behavior of the corresponding symmetry class D theory
in $d=1$ dimension, which can be solved exactly.
        }
\end{center}
\end{figure}

We begin with the summary of our main results by quoting the
standard one-loop 
result\cite{Hikami81,Friedan85} 
for the (RG) beta function\cite{FootnoteBetafunction}
\begin{eqnarray}
\beta(t)
&=&
-
\epsilon t 
+
(\mathcal{N}-\vartheta) t^2
+
\mathcal{O}(t^3)
\label{eq: meain results: beta fct}
\end{eqnarray}
in $d=2+\epsilon$ spatial dimensions,
obeyed by the dimensionless 
coupling constant $t$ of the relevant NL$\sigma$M 
[defined in (\ref{eq: def NLSM riemann manifold b}) below] 
whose inverse is proportional to 
the mean dimensionless conductance
$\bar{g}$,
once the replica limit 
$\mathcal{N}\to0$
has been taken.
To treat the symmetry classes D, C, and A on equal footing,
we have introduced the replica index $\mathcal{N}$ that takes
the value $N$ for symmetry classes D and C and $p+q$ for symmetry
class A. The index $\vartheta$ is also needed to distinguish the
one-loop beta function for the NL$\sigma$M corresponding to
symmetry classes D ($\vartheta=+1$),
symmetry classes C ($\vartheta=-1$),
and
symmetry classes A ($\vartheta=0$).
The correspondence between the pair 
$(\vartheta,\mathcal{N})$ 
and the symmetry classes D, C, and A
is summarized in Table \ref{tab: RMT and NLSM}. 
The beta function
in Eq.~(\ref{eq: meain results: beta fct})
possesses a trivial fixed point,
\begin{subequations} 
\begin{eqnarray}
\label{eq: M fixed point}
t=0,
\end{eqnarray}
which is infrared (IR) stable for $d=2+\epsilon >2$ (i.e., $\epsilon>0$),
while it is IR unstable when $d=2+\epsilon <2$ (i.e., $\epsilon<0$).
Upon taking the replica limit $\mathcal{N}\to0$
this describes a diffusive metallic phase.
Whenever 
$0 < \epsilon/(\mathcal{N}-\vartheta) <\infty $,
there is a nontrivial zero $t^{*}$ of 
the beta function (\ref{eq: meain results: beta fct})%
\cite{Footnote: zero beta fct for class A,Wegner89,Hikami81,Hikami90-92}
at
\begin{eqnarray}
t^{*}&:=& 
\frac{\epsilon}{\mathcal{N}-\vartheta}
+
\mathcal{O}(\epsilon^2).
\label{eq: MI fixed point}
\end{eqnarray}
\end{subequations}
For symmetry classes C and A this describes in the replica limit
and
in $d=2+\epsilon>2$ dimensions, 
a  metal-insulator transition separating
a metallic from an insulating phase. 
For symmetry class D, on the other hand,
this describes (in the replica limit)
a stable fixed point (``critical phase'') in 
$d=2-|\epsilon|<2$ dimensions.
The IR flows for symmetry classes C and D
are summarized in Fig.~\ref{fig: IR RG flows}.
A possible evolution of the stable fixed point in
symmetry class D for dimensionality
$ d=(2-|\epsilon|)$  between one and two dimensions,
to be discussed in the following  subsection,
is sketched in Fig.~\ref{fig: IR RG flows in Class D}.

In this paper we find the dominant anomalous scaling dimension $x^{(s)}$
among all high-gradient operators containing $2s$ gradients, 
to one-loop order.
These are
\begin{subequations}
\label{eq: scaling dimension around UV fixed points, intro}
\begin{eqnarray}
x^{(s)}&=&
2s- 2(s^2-s)t+\mathcal{O}(t^2)
\nonumber\\
&&
\qquad
\mbox{ for }
\mathrm{U}(p+q)/[\mathrm{U}(p)\times \mathrm{U}(q)],
\label{eq: scaling dimension around UV fixed points A, intro}
\\
x^{(s)}&=&
2s-2(s^2-s)t+\mathcal{O}(t^2)
\nonumber\\
&&
\qquad
\mbox{ for }
\mathrm{Sp}(N)/\mathrm{U}(N),
\label{eq: scaling dimension around UV fixed points C, intro}
\\
x^{(s)}&=&
2s-(s^2-s) t+\mathcal{O}(t^2)
\nonumber\\
&&
\qquad
\mbox{ for } 
\mathrm{SO}(2N)/\mathrm{U}(N),
\label{eq: scaling dimension around UV fixed points D, intro}
\end{eqnarray}
\end{subequations}
when evaluated in the vicinity of the
zero-coupling fixed point
~(\ref{eq: M fixed point}) where $t$ is small.
At the corresponding fixed points in $d=2+\epsilon$ dimensions, 
Eq.\ (\ref{eq: MI fixed point}),
these anomalous scaling dimensions 
become, consequently
[for generic numbers of replicas, $(p,q)$ or $N$]
\begin{subequations}
\label{eq: scaling dimension at IR fixed points, intro}
\begin{eqnarray}
x^{(s)}&=&
2s- 2(s^2-s)\frac{\epsilon}{p+q}
+\mathcal{O}(\epsilon^2)
\nonumber\\
&&
\qquad
\mbox{ for } 
\mathrm{U}(p+q)/[\mathrm{U}(p)\times \mathrm{U}(q)],
\label{eq: scaling dimension at IR fixed points A, intro}
\\
x^{(s)}
&=&
2s-
2(s^2-s)\frac{\epsilon}{N+1}
+\mathcal{O}(\epsilon^2)
\nonumber\\
&&
\qquad
\mbox{ for } 
\mathrm{Sp}(N)/\mathrm{U}(N),
\label{eq: scaling dimension at IR fixed points C, intro}
\\
x^{(s)}
&=&
2s-(s^2-s) \frac{\epsilon}{N-1}
+\mathcal{O}(\epsilon^2)
\nonumber\\
&&
\qquad
\mbox{ for } 
\mathrm{SO}(2N)/\mathrm{U}(N),
\label{eq: scaling dimension at IR fixed points D, intro}
\end{eqnarray}
\end{subequations}
Equations~(\ref{eq: scaling dimension around UV fixed points A, intro})
and (\ref{eq: scaling dimension at IR fixed points A, intro}) 
were derived a long time ago by Lerner and Wegner.\cite{Lerner90,Wegner91}
Equations~(\ref{eq: scaling dimension around UV fixed points, intro})
and (\ref{eq: scaling dimension at IR fixed points, intro}) 
are, otherwise, new and are the main results of this paper.

Taking the replica limit 
of Eq.~(\ref{eq: scaling dimension at IR fixed points, intro}) 
yields the dominant anomalous scaling dimensions
at the nontrivial fixed points of the localization problems
in $d=2+\epsilon$ dimensions.\cite{Footnote: zero beta fct for class A}
In particular, in symmetry class C one obtains, in the replica limit,
the one-loop anomalous scaling dimensions
\begin{subequations}
\label{eq: scaling dimension at IR fixed points in the replica limit around 2d, intro}
\begin{eqnarray}
x^{(s)}&=&
2s
-
2(s^2-s)\epsilon
+
\mathcal{O}(\epsilon^2)
\label{eq: scaling dimension at IR fixed points C in the replica limit 2+dim, intro}
\end{eqnarray}
of dominant $2s$-gradient operators
at the metal-insulator  transition
in $d=2+\epsilon>2$ dimensions ($\epsilon>0$).
On the other hand, in symmetry class D in $d=2-|\epsilon|<2$ dimensions,
the dominant $2s$-gradient operators have the anomalous scaling dimensions
\begin{eqnarray}
x^{(s)}&=&
2s
-
(s^2-s)|\epsilon|
+
\mathcal{O}(\epsilon^2)
\label{eq: scaling dimension at IR fixed points D in the replica limit 2-dim, intro}
\end{eqnarray}
\end{subequations}
at the stable, non-trivial fixed point (in the replica limit).

\subsection{
Discussion
           }

When taken at face value, the one-loop expressions in
Eqs.\ 
(\ref{eq: scaling dimension around UV fixed points, intro}),
(\ref{eq: scaling dimension at IR fixed points, intro}),
and
(\ref{eq: scaling dimension at IR fixed points in the replica limit around 2d, intro})
would appear to imply that the anomalous scaling dimension $x^{(s)}$ 
always turns negative for sufficiently large $s$.%
\cite{footnote: about large N expansion}
If this behavior was to persist when 
all higher orders  in $\epsilon$ are included,
an infinite number of high-gradient operators would be
relevant at the nontrivial fixed points of these NL$\sigma$Ms, 
i.e., infinitely many coupling constants would have to 
be fine-tuned to reach these nontrivial fixed points.
The perturbative treatment of these NL$\sigma$Ms
would then appear to contradict the belief that
the underlying microscopic models undergo continuous phase
transitions whose approach obeys single-parameter scaling laws.
Of course, higher order contributions in $\epsilon$
to these anomalous scaling dimensions,
not computed in  the present or earlier work,
may well change this property and could render the 
actual anomalous scaling dimensions $x^{(s)}$
larger than the spatial dimension $d=2+\epsilon$, 
thereby making them irrelevant.%
~\cite{Wegner90,Ludwig90,Mall93,Castilla93,Castilla97,Brezin97,Chatelain00}
[Indeed, similar behavior is known to occur 
in other random disordered systems 
(see e.g. Refs.~\onlinecite{Ludwig90} and \onlinecite{Chatelain00}).
-- See also the discussion in Ref. \onlinecite{Brezin97}.]

Nonperturbative properties of the metal-insulator transitions in
$d=3$ dimensions have been obtained numerically for the most part.%
~\cite{Slevin99,Slevin01-a,Slevin01-b,Slevin03} 
They support the one-parameter scaling hypothesis for 
metal-insulator transitions in $d=3$ dimensions,
whose critical behaviors are presumed to be described
by the non-trivial fixed point of the replica limit 
of the NL$\sigma$M in the corresponding symmetry classes.
Whether the apparent breakdown
of one-parameter scaling due to the relevance of
high-gradient operators 
at one loop is an artifact of the $2+\epsilon$ expansion
or has a deeper meaning remains an outstanding problem
for the description of Anderson localization using
the NL$\sigma$M approach.

Symmetry classes D and DIII, on the other hand,
may perhaps offer a somewhat different
opportunity for the following reason. 
On the one hand,
there exists a critical phase described by a stable fixed point
in $d=2-|\epsilon| <2$ dimensions whose
properties are accessible perturbatively within the $\epsilon$ expansion.  
These include the properties of the conductance fluctuations
and of the high-gradient operators discussed above.
On the other hand, a critical phase of these models  
is also known to exist in $d=1$ dimensions.
Therefore, one might expect that the results
obtained from the $\epsilon$ expansion in 
$d=2-|\epsilon| <2$ dimensions
would eventually,  when $|\epsilon|=1$,
turn into the results obtained from the
exact solutions of the models in $d=1$ dimensions,
as is schematically depicted in Fig.%
~\ref{fig: IR RG flows in Class D}.
If this was the case,
then one could ``test'' the properties
of the high-gradient operators
and the nature of the conductance fluctuations
obtained from them following the work of AKL,
against exact results in $d=1$ dimensions.
Of course, as with all $\epsilon$ expansions,
one would not be able to predict
exactly the actual numerical values of critical
exponents directly at $|\epsilon|=1$. However,
based on experience with the $\epsilon$ expansion in, 
say, ordinary Landau-Ginzburg theory, one might expect
that the structure of the entire probability distribution
of the conductance would evolve, in some sense, 
``smoothly'' with $\epsilon$, possibly all the way down to
$d=1$ dimensions ($|\epsilon|=1$).
Let us now pursue this possibility.

Using perturbation theory of the NL$\sigma$M,
AKL demonstrate that in general (i.e., for any NL$\sigma$M)
there are two contributions to the $s$th 
cumulant of the (dimensionless) conductance.\cite{Altshuler91}
(Below, we write the corresponding expressions
appearing directly {\it at} the stable fixed point
in $d=2-|\epsilon|$ dimensions.)
The first, ``standard'' contribution 
reads\cite{footnoteNotationCumulant}
\begin{equation}
\label{ConductanceCumulantNormal}
\langle\langle g^s \rangle\rangle^{\ }_{\mathrm{std}}
\propto
(\bar{g})^{2-s}
\end{equation}
where $\bar{g}\equiv\langle g \rangle$ is the mean conductance.
However, when a high-gradient operator with $2s$ 
derivatives is added to the action of the NL$\sigma$M
with coupling constant $z^{\ }_{s}$, 
then there is an additional contribution
\begin{equation}
\label{ConductanceCumulantAdd}
\langle\langle g^{s} \rangle\rangle^{\ }_{\mathrm{add}}
\propto
z^{\ }_{s}
\left(\frac{L}{\ell}\right)^{d-x^{(s)}}
\end{equation}
to this cumulant. Here, $L$ is the linear system size and
$\ell$ the characteristic microscopic length scale 
(the ``mean free path'') for the disorder.
We see that it is when $d-x^{(s)}>0$
that the additional contribution in
Eq.\ (\ref{ConductanceCumulantAdd})
dominates over the standard contribution in
Eq.\ (\ref{ConductanceCumulantNormal}).
This hence happens precisely when the high-gradient operator 
with $2s$ derivatives becomes a relevant  operator in the action 
of the NL$\sigma$M.

Now, AKL argue that, if one was to take the one-loop expressions
for the anomalous scaling dimensions 
of the high-gradient operators at face value,
which for symmetry class D would amount to the result in 
Eq.%
~(\ref{eq: scaling dimension at IR fixed points D in the replica limit 2-dim, intro})
derived by us in this paper, then for  cumulants of 
sufficiently higher order 
$s>\frac{2}{|\epsilon|}[1+\mathcal{O}(\epsilon)]$
the additional contribution would dominate
over the ``standard'' one, yielding
\begin{eqnarray}
\langle\langle g^{s} \rangle\rangle\sim 
A^{\ }_{s}
\left(\frac{L}{\ell}\right)^{d-x^{(s)}},
\qquad 
s>\frac{2}{|\epsilon|}
[1+\mathcal{O}(\epsilon)] 
\end{eqnarray}
as $L/\ell\to\infty$,
where $A^{\ }_{s}$ are certain amplitudes.
Following AKL, the lower cumulants of order
$s <\frac{2}{|\epsilon|}[1+\mathcal{O}(\epsilon)]$,
on the other hand, would be determined by
Eq.~(\ref{ConductanceCumulantNormal}).
In summary, the reasoning of AKL implies, 
when specialized to the stable fixed point in 
symmetry class D
in $d=2-|\epsilon|<2$ spatial dimensions,
the following form of the conductance cumulants
\begin{subequations}
\label{eq: AKL cumulants}
\begin{eqnarray}
\langle\langle g^s \rangle\rangle
\sim
\left\{
\begin{array}{ll}
(\bar{g})^{2-s}, 
&  
\mbox{if $s < s^{\ }_{0}$}, 
\\ 
&
\\
A^{\ }_{s}\,(L/\ell)^{d - x^{(s)}},
&  
\mbox{if $s > s^{\ }_{0}$},
\end{array}
\right.
\end{eqnarray}
where
\begin{eqnarray}
&&
s^{\ }_{0}\approx  
\frac{2}{|\epsilon|}[1 + \mathcal{O}(\epsilon)]
\end{eqnarray}
and 
\begin{eqnarray}
d- x^{(s)}=
|\epsilon| s^2 
- 
(2+|\epsilon|)s 
+ 
(2-|\epsilon|)
+
\mathcal{O}(\epsilon^2).
\end{eqnarray}
\end{subequations}

Let us now discuss the known exact results
for the $d=1$ dimensional case.
Specifically, the leading dependence of 
{\it all} cumulants of the conductance of a very long 
but thick quasi one-dimensional wire 
of length $L$,
believed to be described by a NL$\sigma$M in $d=1$ dimensions,
has been extracted exactly from the DMPK approach
for symmetry classes D and DIII.%
~\cite{Brouwer00,Imamura01,Brouwer05}
For both symmetry classes,
these cumulants decay algebraically with system size,
\begin{eqnarray}
&&
\langle\langle g^s \rangle\rangle
\propto
(L/\ell)^{-1/2},
\quad s=1,2,\ldots,
\label{eq: moments g in quai 1d class D}
\end{eqnarray}
with the {\it same} exponent for all cumulants
(``$\ell$'' is again the mean free path).
Moreover, the logarithm of the conductance exhibits the behavior
\begin{eqnarray}
&&
\langle\langle\ln g\rangle\rangle
\propto
- (L/\ell)^{1/2},
\quad
\mathrm{var}\ln g \sim  L/\ell.
\label{eq: log g in quai 1d class D}
\end{eqnarray}
Thus, the typical conductance exhibits stretched exponential behavior,
in contrast with the mean conductance.
This indicates  a broad conductance distribution.
[For symmetry class DIII the result%
~(\ref{eq: moments g in quai 1d class D})
for the mean conductance has been reproduced 
by Lamacraft, Simons, and Zirnbauer,\cite{Lamacraft04}
using the one-dimensional NL$\sigma$M 
(which is a principal chiral model for symmetry class DIII)
and instanton methods.]

If one assumes that the exact results in $d=1$
describe the end point (at $\epsilon=1$)
of the line of critical points in $2-|\epsilon|$ dimensions
(see Fig.\ref{fig: IR RG flows in Class D}),
there are then two possibilities.
First, the one-loop relevance of high-gradient operators 
is an artifact of the
$\epsilon$-expansion
and these operators are in
fact 
{\it irrelevant}
once all orders in the $\epsilon$-expansion
are taken into account
(see Refs.~\onlinecite{Brezin97,Ludwig90}). 
Second,
this is not the case and
high-gradient operators are truly relevant.
In this case, however, 
deducing the scaling of the cumulants based on the 
one-loop anomalous scaling dimensions 
does not appear to be
sufficient to determine their scaling behavior.
Indeed, in this case
the running coupling constants
$z^{\ }_{s}$ of all the high-gradient operators
in the effective
action of the NL$\sigma$M grow rapidly, more so
the higher the number $2s$ of gradients.
Under these conditions, then, there
is no rationale 
to neglect the nonlinear coupling between 
the running coupling constants
$z^{\ }_{s}$ of the high-gradient operators
as was done to reach the conclusion
~(\ref{eq: AKL cumulants}),
when extracting the scaling of the cumulants of the conductance.
In effect, one needs to implement a functional RG
analysis which can treat
an infinite set of running (and increasing)
coupling constants on equal footing.
We have performed
such an analysis
(on which we will report elsewhere)
for symmetry class A,
but  have so-far
been unable to solve the
resulting RG equations.
Since in the present case the coupling
constants $z^{\ }_{s}$  of the high-gradient
operators all become large,
it is clear (as already mentioned) that they will not
flow independently of each other in the IR limit,
but will strongly mix with each other under the RG flow.
In this situation, their flow does not need to 
result in the breakdown of single parameter scaling.
Indeed, a precisely analogous situation is encountered
when computing the local DOS of the two-dimensional
random hopping problem,\cite{Mudry03}
in which case one-parameter scaling for the local
DOS is recovered using a functional RG  analysis.

\subsection{
Summary of the method of calculation
           }
\label{SummaryMethodCalculation}

The RG scheme that we adopt in this paper is the 
covariant background field method.\cite{Alvarez-Gaume81}
In Refs.~\onlinecite{Castilla93,Castilla97}
this method was used to compute,
up to two loop order,
the RG flows of the high-gradient operators for the 
$\mathrm{O}(N)/\mathrm{O}(N-1)$ NL$\sigma$M,
relevant to statistical models
such as the classical Heisenberg magnet.
The covariant background field method
consists of two key steps.
First, the fields of the NL$\sigma$M are separated into 
short wavelength (``fast'') and 
long wavelength (``slow'') degrees of freedom
where the slow degrees of freedom are defined as the solutions
to the classical equations of motion.
Second, the fast fields are integrated out.
When using this method, no redundant operator~\cite{Ref-Wegner-DombGreenVol-6}
is generated in the second step of
the covariant background field method,%
a great advantage when one needs to consider mixing of 
many scaling operators by the RG transformation (as we do here).
Another advantage is that explicit invariance under a 
coordinate reparametrization of the group manifold is maintained,
allowing for a geometrical  formulation
at all stages of the calculation.

In this paper, we shall also extend
the covariant background field method
to the cases when the target spaces of the NL$\sigma$M 
are Riemannian or K\"ahler supermanifolds.%
\cite{DeWitt92}
We derive a master formula,
Eq.~(\ref{eq: master formula, Kahler}),
in the K\"ahler case
for the RG flow of high-gradient operators,
all of which  possess a vanishing covariant derivative.
[See Eq.~(\ref{eq: master formulae if superriemann})
for the super Riemannian\cite{DeWitt92}
and Eq.~(\ref{eq: master formulae if superkahler}) for the super
K\"ahler\cite{DeWitt92} cases, respectively.
To obtain the corresponding  formula for the non-supersymmetric cases
(for fermionic replicated NL$\sigma$M, i.e., for compact NL$\sigma$M)
we set all factors $(-1)^{p}$ to $+1$,
where $p=0,1$ is here the grade in the supersymmetric expression].
\cite{antisymmetry of riemann tensor} 
This master formula is
expressed solely in terms of the geometrical data of the target manifold
and should apply to all 10 symmetry classes of AZ,\cite{Zirnbauer96,Altland97}
but is not limited to them.
The generality of the master formula suggests that high-gradient operators
are generically relevant to 1-loop order at the nontrivial fixed point of a
NL$\sigma$M defined on a Riemannian target space.

\subsection{
Geometric interpretation of high-gradient operators
           }

All high-gradient operators that enter the computation of
the cumulants for the conductance can be expressed
in terms of tensor fields on the target manifold.
We will show that high-gradient operators
together with the metric and curvature tensors
on the target manifolds of the NL$\sigma$M
form a family of tensors with vanishing covariant derivative.

\subsection{
Outline of the paper
           }

This paper is organized as follows.
Section \ref{sec: NLSM for weakly disordered superconductors}
introduces the NL$\sigma$M on K\"ahler manifolds for 
disordered superconductors with broken time-reversal symmetry
as well as for a disordered metal in the standard unitary symmetry class.
Section \ref{sec: Normal coordinates}
describes how to use the covariant background field method
to deduce perturbative RG flows.
The dominant anomalous scaling dimensions for high-gradient operators are
computed in Sec.\ \ref{sec: One-loop RG for high-gradient operators}
to one-loop order.
We relegate to appendices some key intermediary steps in the derivation
of Eq.~(\ref{eq: scaling dimension around UV fixed points, intro}).
The master formula, Eq.~(\ref{eq: master formula, Kahler}),
which expresses the RG transformation law obeyed
by high-gradient operators solely in terms of the geometrical
data of the target manifold is derived in appendix 
\ref{app: RG of high-gradient operators of NLSMs on supermanifolds}
for any Riemannian or K\"ahlerian supermanifold on which
the high-gradient operators have vanishing covariant derivatives.

\section{
NL$\sigma$M for weakly disordered superconductors
        }
\label{sec: NLSM for weakly disordered superconductors}

\subsection{
NL$\sigma$M on K\"ahler manifolds
           }

We start from the NL$\sigma$M description 
for the symmetry classes D, C and A of Anderson localization
in terms of fermionic replicas.
Alternatively, one can use either bosonic replica or supersymmetric 
methods to perform the disorder average.
These three options are equivalent as far as
perturbative RG computations are concerned.
The field entering the NL$\sigma$M
is a map from the base space,
a $d$-dimensional Euclidean space $\mathbb{R}^{d}$,
to the target manifold $\mathfrak{M}$ defined by 
the pattern of symmetry breaking.
The target manifold has a well-defined natural metric $\mathfrak{g}$.
A classification of all the target manifolds of NL$\sigma$M
relevant to the problem of Anderson localization applied to 
non-interacting quasi-particles
were identified in the seminal work of
Zirnbauer, and Altland and Zirnbauer.\cite{Zirnbauer96}
For the cases of interest in this paper, i.e.,
disordered superconductors with broken time-reversal symmetry,
the target spaces of the relevant NL$\sigma$M turn out 
to be K\"ahler manifolds
which are
the Hermitian symmetric spaces
$\mathrm{SO}(2N)/\mathrm{U}(N)$ for symmetry class D
and 
$\mathrm{Sp}(N)/\mathrm{U}(N)$ for symmetry class C.%
~\cite{CommentDomainWalls}
The target space of the NL$\sigma$M that 
describes a 
disordered metal when time-reversal symmetry is broken
is the Hermitian symmetric space
$\mathrm{U}(2N)/[\mathrm{U}(N)\times\mathrm{U}(N)]$,
which is isomorphic to 
the complex Grassmannian 
$G^{\ }_{2N,N}(\mathbb{C})$.\cite{Nakahara03}

Within the mesoscopic community,
the usual representation of the NL$\sigma$M
in symmetry classes D, C, and A is in terms
of matrices belonging to the relevant symmetric spaces,
``the $Q$-matrix representation'' after the seminal work
by Efetov et al.\ in Ref.~\onlinecite{Efetov80}.
This representation emphasizes the pattern of symmetry breaking
but quickly becomes cumbersome when performing an RG analysis
of the flow of composite operators.
We opt instead to emphasize more directly
the geometrical properties of the target spaces, 
namely the fact that they are examples of K\"ahler manifolds
for symmetry classes D, C, and A. We refer the reader to appendix
\ref{app sec: Group-theory versus geometric representations of the NLsigmaM}
for a relation between the representation of the NL$\sigma$M
that we shall introduce below and the more common $Q$-matrix representation
thereof. 

We now present the definition of a NL$\sigma$M
on a K\"ahler manifold starting from a NL$\sigma$M on 
a $2M$-dimensional Riemannian manifold \cite{Nakahara03,Kobayashi96} 
($\mathfrak{M}$,$\mathfrak{g}$),
\begin{subequations}
\label{eq: def NLSM riemann manifold}
\begin{eqnarray}
Z&:=&
\int
\mathcal{D}[\phi]\,
e^{-S[\phi]},
\label{eq: def NLSM riemann manifold a}
\\
S[\phi]&:=&
\frac{1}{4\pi t}
\int^{\ }_{r}
g^{\ }_{AB}(\phi)\,
\partial^{\ }_{\mu}\phi^{B}
\partial^{\ }_{\mu}\phi^{A}.
\label{eq: def NLSM riemann manifold b} 
\end{eqnarray}
Our conventions are here the following.
We reserve the greek alphabet to label the $d$ coordinates
$r^{\ }_{\mu}$ where $\mu=1,\ldots,d$ of $r\in\mathbb{R}^d$,
and use the short-hand notation 
$\int^{\ }_r\equiv\int_{|r|>\mathfrak{a}} {d}^{d}r/\mathfrak{a}^{d-2}$
for the integration over $\mathbb{R}^{d}$ 
with the short-distance cut-off $\mathfrak{a}$.
The dimensionless coupling constant of the NL$\sigma$M
is here denoted by $t$. The dimensionless conductance
shares with $(\mathfrak{a}^{d-2}t)^{-1}$ the
same engineering dimension.
To each point $r$ from the base space we have assigned
a point $\mathfrak{p}$ on the manifold $\mathfrak{M}$
with the coordinates
$(\phi^{A})\in\mathbb{R}^{2M}$.
In the definition of the NL$\sigma$M on a Riemannian manifold,
Eq.\ (\ref{eq: def NLSM riemann manifold}),
the $2M\times 2M$ components of 
the metric tensor field
$\mathfrak{g}$ of rank $(0,2)$
are represented by  $(g^{\ }_{AB})$.
We reserve the capital latin alphabet to label the $2M$ coordinates
$\phi^{A=1,\ldots,2M}$ of $\mathfrak{p}\in\mathfrak{M}$.

The components of the metric tensor field $g^{\ }_{AB}$ 
on the Riemannian manifold 
$(\mathfrak{M},\mathfrak{g})$
are real and symmetric under the exchange of the indices
\begin{eqnarray}
g^{\ }_{AB}=g^{\ }_{BA},
\qquad A,B=1,\ldots,2M.
\label{eq: def riemann manifold c}
\end{eqnarray}
\end{subequations}
There is a distinction between upper/lower capital latin indices.
The components of $\mathfrak{g}^{-1}$ are denoted by $g^{AB}$
and satisfy 
\begin{eqnarray}
g^{\ }_{A C}\,g^{C B}=
{}^{\ }_{A}\delta^{B},
&&
g^{A C}\,g^{\ }_{C B}=
{}^{A}\delta^{\ }_{B},
\end{eqnarray}
and so on for $A,B,C=1\ldots,2M$. 
On the other hand, there is no distinction between 
upper and lower greek indices and we will
always choose them to be lower indices.

A Hermitian manifold ($\mathfrak{M}$,$\mathfrak{g}$)
is a $2M$-dimensional Riemannian manifold
endowed with a rank $(1,1)$ tensor field $\mathfrak{J}$
that is defined globally on $\mathfrak{M}$
and is called a complex structure.
On a Hermitian manifold, we can always make a coordinate transformation
from $(\phi^A)$ to $(x^a,y^a)$ 
that satisfies, for each point $\mathfrak{p}\in \mathfrak{M}$,
\begin{eqnarray}
\mathfrak{J}_{\mathfrak{p}}
\left(
\frac{\partial}{\partial x^a}
\right)
=
\frac{\partial}{\partial y^a},
\quad
\mathfrak{J}_{\mathfrak{p}}
\left(
\frac{\partial}{\partial y^a}
\right)
=
-
\frac{\partial}{\partial x^a},
\end{eqnarray}
where $a=1,\ldots,M$.
It is then natural to introduce complex coordinates
$(z^{*a},z^a)$ through
\begin{eqnarray}
z^{*a}:=x^{a}-\mathrm{i}y^{a},\quad
z^{a}:=x^{a}+\mathrm{i}y^{a},
\quad
a=1,\ldots,M.
\end{eqnarray}
A NL$\sigma$M on a Hermitian manifold ($\mathfrak{M}$,$\mathfrak{g}$) 
is thus defined by the partition function
\begin{subequations}
\label{eq: def NLSM complex manifold}
\begin{eqnarray}
Z&:=&
\int
\mathcal{D}[z^{*},z]\,
e^{-S[z^{*},z]},
\label{eq: def NLSM complex manifold a}
\\
S[z^{*},z]&:=&
\frac{1}{2\pi t}
\int^{\ }_{r}
g^{\ }_{a^* b^{\ }}(z^{*},z)\,
\partial^{\ }_{\mu}z^{b}
\partial^{\ }_{\mu}z^{*a},
\label{eq: def NLSM complex manifold b} 
\end{eqnarray}
where we have assigned to each point $r$ from the base space 
a point $\mathfrak{p}$ on the manifold $\mathfrak{M}$
with the coordinates
$(z^{*a},z^{a})\in\mathbb{C}^{M}$.
Components of tensor fields on $\mathfrak{M}$ have now
two types of indices, 
    holomorphic ($a,b,\ldots$) and 
antiholomorphic ones ($a^{*},b^{*},\ldots$).
Holomorphic indices can be contracted with 
holomorphic components of the coordinates $z^a$ whereas
antiholomorhic ones with  $z^{*a}$,
the complex conjugate of $z^{a}$.
In the definition of the NL$\sigma$M on a Hermitian manifold,
Eq.\ (\ref{eq: def NLSM complex manifold}),
the $M\times M$ components of 
the Hermitian metric tensor field
$\mathfrak{g}$ of rank $(0,2)$
are represented by  $(g^{\ }_{a^* b})$.
We reserve the latin alphabet to label the $M$
holomorphic and antiholomorphic coordinates
$z^{*a}$ and $z^{a}$ where $a^{*},a=1,\ldots,M$, 
respectively, of $\mathfrak{p}\in\mathfrak{M}$.
The use of a capital latin letter on a Hermitian manifold, 
say $A$, as an index denotes
either an holomorphic ($A=a$) or an antiholomorphic ($A=a^{*}$) index.

The components of the metric tensor field $g^{\ }_{AB}$ 
on the Hermitian manifold 
$(\mathfrak{M},\mathfrak{g})$
are symmetric under the exchange of the indices
and Hermitian
\begin{eqnarray}
g^{\ }_{AB}=g^{\ }_{BA},
\qquad
\left(g^{\ }_{AB}\right)^{*}=g^{\ }_{A^{*}B^{*}}
\label{eq: def complex manifold c}
\end{eqnarray}
\end{subequations}
with $A=a$ or $a^{*}$, $B=b$ or $b^{*}$, $A^{*}=a^{*}$ or $a$, 
and $B^{*}=b^{*}$ or $b$ running from $1$ to $M$.
The components of $\mathfrak{g}^{-1}$ are denoted by $g^{AB}$
and satisfy 
\begin{eqnarray}
g^{\ }_{A C^{*}}\,g^{C^{*} B}=
{}^{\ }_{A}\delta^{B},
&&
g^{A C^{*}}\,g^{\ }_{C^{*} B}=
{}^{A}\delta^{\ }_{B},
\end{eqnarray}
and so on for $A,B,C^{*}=1,\ldots,M$.

A K\"ahler manifold $(\mathfrak{M},\mathfrak{g})$
is a Hermitian manifold whose Hermitian metric can be derived from
a K\"ahler potential $K(z^{*},z)$,
\begin{eqnarray}
&&
g^{\ }_{a b^{*}}(z^{*},z)=
\partial_{a}^{\ }\partial_{b^{*}}^{\ } K(z^{*},z),
\end{eqnarray}
with $a,b^{*}=1\ldots,M$.
Not all Hermitian manifolds are K\"ahlerian.

The  K\"ahler manifolds corresponding to the Hermitian symmetric spaces
$\mathrm{U}(p+q)/[\mathrm{U}(p)\times \mathrm{U}(q)]$ for 
symmetry class A,
$\mathrm{Sp}(N)/\mathrm{U}(N)$ for symmetry class C,
and
$\mathrm{SO}(2N)/\mathrm{U}(N)$ for symmetry class D
are specified once the corresponding K\"ahler potentials have been
constructed. To this end, we begin by parametrizing the complex Grassmannian 
$\mathrm{U}(p+q)/[\mathrm{U}(p)\times \mathrm{U}(q)]$ for symmetry class A
by the stereographic coordinates defined by the independent set
\begin{subequations}
\label{eq: kahler potential grassmannian}
\begin{eqnarray}
\{\varphi^{* }_{Aa},\varphi^{\ }_{Aa},\ A=1,\ldots, p,\ a=1,\ldots, q\}
\end{eqnarray}
of $p\times q$ complex-valued matrices.%
\cite{Zumino79,Higashijima01,Higashijima02b}
With this coordinate system, the K\"ahler potential on 
$\mathrm{U}(p+q)/[\mathrm{U}(p)\times \mathrm{U}(q)]$ 
is given by
\begin{eqnarray}
K(\varphi^{*},\varphi)&=&
c \ln \det
\left(
\mathbb{I}_{q}+\varphi^{\dag}\varphi
\right)
\end{eqnarray}
\end{subequations}
where $c$ is an arbitrary constant (the scale of the metric)
which is chosen to be unity in the following.
(Another choice for $c$ just amounts to a rescaling of $t$.)
The K\"ahler metric induced by 
the K\"ahler potential~(\ref{eq: kahler potential grassmannian})
is
\begin{subequations}
\begin{eqnarray}
g^{\ }_{(Aa)(Bb)^{*}}(\varphi^{*}, \varphi)
&=&
Z^{\ }_{ab} Y^{\ }_{BA}
\label{eq: metric grassmannian}
\end{eqnarray}
where we have introduced 
$q \times q$ and $p\times p$ Hermitian matrices,
\begin{eqnarray}
Z^{\ }_{ab}&:=&
\left(
\mathbb{I}^{\ }_{q}
+
\varphi^{\dagger}\varphi
\right)^{-1}_{ab},
\qquad a,b=1,\dots,q,
\label{eq: def Z}
\\
Y^{\ }_{AB}&:=&
\left(
\mathbb{I}^{\ }_{p}
+
\varphi \varphi^{\dagger} 
\right)^{-1}_{AB},
\quad A,B=1,\dots,p.
\label{eq: def Y}
\end{eqnarray}
\end{subequations}

The cases of 
$\mathrm{Sp}(N)/\mathrm{U}(N)$ for symmetry class C
and 
$\mathrm{SO}(2N)/\mathrm{U}(N)$ for symmetry class D
are obtained 
by setting $p=q=N$
and by imposing the constraint 
\begin{eqnarray}
\varphi+\vartheta \varphi^{\mathrm{T}}=0
\label{eq: stereographic coordinates for Class C and D}
\end{eqnarray}
on the complex Grassmannian 
$\mathrm{U}(p+q)/[\mathrm{U}(p)\times \mathrm{U}(q)]$
where 
$\vartheta=-1$ for $\mathrm{Sp}(N)$ (symmetry class C)
and 
$\vartheta=+1$ for $\mathrm{SO}(2N)$ (symmetry class D).
Due to the constraint (\ref{eq: stereographic coordinates for Class C and D}), 
independent entries of 
$\{\varphi^{* }_{Aa},\varphi^{\ }_{Aa},\ A,a=1,\ldots,N\}$
can be chosen to be those with the indices
$A\le a$ for $\mathrm{Sp}(N)$ (symmetry class C)
and
$A<a$ for $\mathrm{SO}(2N)$ (symmetry class D),
say. 
In turn, the metric for 
$\mathrm{Sp}(N)/\mathrm{U}(N)$ (symmetry  class C)
and the metric for 
$\mathrm{SO}(2N)/\mathrm{U}(N)$ (symmetry class D)
are given by
\begin{subequations}
\begin{eqnarray}
g^{\ }_{(Aa)(Bb)^{*}}(\varphi^*,\varphi)&=&
\alpha^{\ }_{Aa}\alpha^{\ }_{Bb}
\mathcal{S}^{\vartheta}_{(Aa)(Bb)}Z^{\ }_{ab}Z^{\ }_{AB},
\nonumber\\
\end{eqnarray}
where
\begin{eqnarray}
\alpha^{\ }_{Aa}&:=&
1-\delta^{\ }_{Aa}/2,
\\
\mathcal{S}^{\vartheta}_{(Aa)(Bb)\cdots}&:=&
\mathcal{S}^{\vartheta}_{(Aa)}\mathcal{S}^{\vartheta}_{(Bb)}
\cdots,
\\
\mathcal{S}^{\vartheta}_{(Aa)}
T^{\vphantom{\vartheta}}_{\cdots A \cdots a \cdots }&:=&
T^{\vphantom{\vartheta}}_{\cdots A \cdots a \cdots }
-\vartheta 
T^{\vphantom{\vartheta}}_{\cdots a \cdots A \cdots },
\end{eqnarray}
and
\begin{eqnarray}
Z^{\ }_{ab}&:=&
\left(
\mathbb{I}^{\ }_{N}
+
\varphi^{\dagger}\varphi
\right)^{-1}_{ab},
\quad a,b=1,\dots,N.
\end{eqnarray}
\end{subequations}
Observe that
$Z$ in Eq.~(\ref{eq: def Z}) and $Y$ in Eq.~(\ref{eq: def Y}) 
are not independent anymore because $Y=Z^{\mathrm{T}}$
when $p=q=N$ and 
the constraint (\ref{eq: stereographic coordinates for Class C and D})
holds. 
By a slight abuse of notation we will sometimes refer
to the complex Grassmannian case by $\vartheta=0$. 

\subsection{
High-gradient operators           }

The seminal contribution by
Altshuler, Kravtsov, and Lerner,\cite{Altshuler91}
was to observe that the perturbative RG flow obeyed by the cumulants
of the conductance necessarily involves an infinite number of composite
operators and to identify a criterion for the selection of this
infinite family of composite operators.
The selection criterion is that the flow must be closed within a family
of composite operators characterized by the most dominant scaling 
dimensions. Applied to the three cases at hand, we thus seek a family
of composite operators that transform as scalars under reparametrization
of the target manifold and contain as many derivatives as possible 
subject to the condition that their anomalous scaling dimension be dominant.
We are led to the family of high-gradient operators
\begin{eqnarray}
T^{\ }_{IJK\ldots}(z^{*},z)
\underbrace{
\partial z_{\mu}^{I}
\partial z_{\nu}^{J}
\partial z_{\rho}^{K}
\cdots
           }_{2s},
\label{eq: high-gradient opertors in tensor notation}
\end{eqnarray}
where $T^{\ }_{IJK\ldots}(z^{*},z)$ is a tensor 
on the target manifold with $s$ holomorphic and antiholomorphic indices;
we do not yet specify how the greek indices must be contracted.
Although composite operators containing derivatives of the form 
$\partial^n z$ ($n\ge 2$)
are generated by the RG flow for the cumulants of the conductance
they are neither expected to contribute to the dominant RG
anomalous scaling dimensions,\cite{Lerner90} nor needed to close the RG flows
of the high-gradient operators%
~(\ref{eq: high-gradient opertors in tensor notation})
as we shall see below.

As the simplest example of a high-gradient operator belonging to
the family~(\ref{eq: high-gradient opertors in tensor notation})
is the Lagrangian
$g^{\ }_{IJ}\partial^{\ }_{\mu}z^{J}\partial^{\ }_{\mu}z^{I}$
of the NL$\sigma$M action. 
This is in fact the only scalar under both reparametrization
of the target manifold and global rotations in the base space
with $s=1$ that belongs to the family%
~(\ref{eq: high-gradient opertors in tensor notation}).
To see this, observe that the number of independent tensors with two indices 
on the target manifold is related to that of independent bilinear forms 
in the group (coset). For simple Lie groups 
(and their analogues for cosets), there is only one such bilinear form 
(the Casimir), and hence $g^{\ }_{IJ}$ is uniquely fixed.
The scaling flow obeyed by 
$g^{\ }_{IJ}\partial^{\ }_{\mu}z^{J}\partial^{\ }_{\mu}z^{I}$,
controls the scaling flow obeyed by the mean conductance,
i.e., the first cumulant of the conductance
once the replica limit has been taken.

Life already becomes more complicated if we seek all the high-gradient
operators with $s=2$ in 
the family~(\ref{eq: high-gradient opertors in tensor notation}).
On a generic complex manifold more than 2 independent tensors
of rank $(0,4)$ are available,
\begin{eqnarray}
g^{\ }_{IJ}g^{\ }_{KL}, 
\quad 
R^{\ }_{IJKL},
\quad
R^{\ }_{IJMN}R^{MN}{}^{\ }_{KL},
\end{eqnarray} 
and so on, where $R^{\ }_{IJKL}$ is the curvature (Riemann) tensor
on the target manifold. Fortunately, the situation is simpler
for the Grassmannian since we only need to contend
with two independent tensors of rank $(0,4)$,
\begin{subequations}
\begin{eqnarray}
S^{(2)}_{(Aa)(Bb)(Cc)(Dd)}&=&
Z^{\ }_{ab}Y^{\ }_{BC}Z^{\ }_{cd}Y^{\ }_{DA},
\\
S^{(1)}_{(Aa)(Bb)}S^{(1)}_{(Cc)(Dd)}&=&
Z^{\ }_{ab}Y^{\ }_{BA}Z^{\ }_{cd}Y^{\ }_{DC}.
\end{eqnarray}
\end{subequations}
[See Eq.~(\ref{eq: AKL cumulants}).]

Observe that $S^{(1)}_{AB}$
is nothing but the metric tensor $g^{\ }_{AB}$.

For symmetry class A and for given 
$s\in\mathbb{N}$,
all the independent tensors $T^{\ }_{IJK\ldots}$ of rank $(0,2s)$
of the complex Grassmannian 
can be expressed by taking the products
of the building blocks,
\begin{subequations}
\label{eq: construction all tensor of rank (0,2s) if A,C, D}
\begin{eqnarray}
S^{(l)}_{(Aa)(bB)^{*}(Cc)(dD)^{*}\ldots}:=
Z^{\ }_{ab}Y^{\ }_{BC}Z^{\ }_{cd}Y^{\ }_{DE}\cdots Y^{\ }_{ZA}
\nonumber\\
\label{eq: S(l) if Grasmannian}
\end{eqnarray}
made of $l$ pairs of $Z$ and $Y$, 
according to the decomposition
\begin{eqnarray}
T^{\ }_{I^{\ }_1J^{\ }_1K^{\ }_1\ldots}
=
\prod_{i}
S^{(l^{\ }_i)}_{I^{\ }_iJ^{\ }_iK^{\ }_i\ldots},
\qquad
\sum_i l^{\ }_i=s.
\label{eq: tensor T as a product of S's}
\end{eqnarray}
For the cases of 
$\mathrm{Sp}(N)/\mathrm{U}(N)$ (symmetry class C)
and 
$\mathrm{SO}(2N)/\mathrm{U}(N)$ (symmetry class D)
we need to replace Eq.~(\ref{eq: S(l) if Grasmannian}) by
\begin{eqnarray}
S^{(l)}_{\ldots(bB)^{*}(Cc)(dD)^{*}\ldots}:=
\cdots
\alpha^{\ }_{Cc}
\mathcal{S}^{\vartheta}_{Cc}
Y^{\ }_{Bc} Z^{\ }_{Cd}
\cdots.
\label{eq: S(l) if class C or D}
\end{eqnarray}
\end{subequations}

Another simplification attached to the K\"ahler manifolds
for symmetry classes A, C, and D is that
contraction between $S^{(l^{\ }_i)}$ 
and the $\partial z^{I}_{\mu}$'s
reduces to a trace,
\begin{subequations}
\label{eq: def of A{pm}{pm}}
\begin{eqnarray}
S^{(l)}_{(Aa)(bB)^{*}(Cc)\ldots}
\partial^{\ }_{\mu^{\ }_{1}}\varphi^{\   }_{Aa}
\partial^{\ }_{\mu^{\ }_{2}}\varphi^{\dag}_{bB}
\partial^{\ }_{\mu^{\ }_{3}}\varphi^{\   }_{Cc}
\cdots
\qquad\quad
&&
\nonumber \\
=
\mathrm{tr}\,
\big(
A^{+}_{\mu^{\ }_{1}}
A^{-}_{\mu^{\ }_{2}}
A^{+}_{\mu^{\ }_{3}}
\cdots
\big),
\quad\quad
&&
\end{eqnarray}
where we have introduced
\begin{eqnarray}
A^{+}_{\mu}:= 
Y \partial^{\ }_{\mu}\varphi,
\qquad
A^{-}_{\mu}:= 
Z \partial^{\ }_{\mu}\varphi^{\dagger},
\end{eqnarray}
for the complex Grassmannian (symmetry class A).
For $\mathrm{Sp}(N)/\mathrm{U}(N)$ (symmetry class C) 
and
for $\mathrm{SO}(2N)/\mathrm{U}(N)$ (symmetry class D) 
one must impose the additional constraint
\begin{eqnarray}
(A^{+}_{\mu})^{\mathrm{T}}=-\vartheta A^{+}_{\mu},\quad
(A^{-}_{\mu})^{\mathrm{T}}=-\vartheta A^{-}_{\mu}.
\end{eqnarray}
\end{subequations}

Any member $T$ of the family of tensor fields constructed
in Eq.~(\ref{eq: construction all tensor of rank (0,2s) if A,C, D})
has the remarkable property that it vanishes under the action of the
covariant derivative on any one of its indices
\begin{eqnarray}
\nabla_{P}^{\ }T_{I J K L \cdots}^{\ }
&=&
\partial_{P}^{\ } T_{I J K L \cdots}^{ }
\nonumber \\
&&
-
\Gamma^{Q}_{\ P I} T_{Q J K L \cdots}^{\ }
-
\Gamma^{Q}_{\ P J} T_{I Q K L \cdots}^{\ }
-\cdots
\nonumber\\
&=&
0.
\end{eqnarray}
Here, $\Gamma^{A}_{\ BC}$ 
are the components of the Levi-Civita connection
whose non-vanishing entries are given by
\begin{eqnarray}
\Gamma^{a}_{\ bc}
=
g^{a p^*}\,
\partial^{\ }_c g_{p^* b}^{\ },
\quad
\Gamma^{a^*}_{\ b^* c^*}
=
g^{a^* p}\,
\partial_{c}^{*} g_{p b^*}^{\ }.
\label{eq: def Levi-Civita connection}
\end{eqnarray}
This is also true for the NL$\sigma$M on the real Grassmannian
$\mathrm{O}(p+q)/[\mathrm{O}(p)\times \mathrm{O}(q)]$
which is related to the symplectic symmetry class of 
Anderson localization,
and for the cases of simpler 
target manifolds such as $\mathrm{O}(N)/\mathrm{O}(N-1)$
or $\mathbb{C}P^{n+m-1|m}$
for which the tensor fields $T$ are just products of 
the metric tensor. In all these cases,
the fact that all high-gradient operators
have vanishing covariant derivatives
greatly simplifies the computation of
their one-loop RG flows.

For a given number of derivatives, $2s$,
there are many degenerate operators of the type%
~(\ref{eq: high-gradient opertors in tensor notation})
with $T^{\ }_{IJK\ldots}$ given by
Eq.~(\ref{eq: tensor T as a product of S's})
that share the same anomalous scaling dimensions $2s$ 
at the metallic fixed point.
They are specified by the choice of 
$\{ l^{\ }_i\}$ in Eq.\ (\ref{eq: tensor T as a product of S's})
as well as that of the greek indices
in the $\partial^{\ }_{\mu} z^{I}$'s.
This degeneracy is, however, lifted in perturbation theory
as we depart from the metallic fixed point.

We close this section by noting that
it is more convenient to introduce
the conformal indices $+$ and $-$ 
\begin{eqnarray}
\partial^{\ }_{\pm}=\partial^{\ }_{x}\pm \mathrm{i}\partial^{\ }_{y}.
\label{eq: def conformal indices}
\end{eqnarray}
instead of the Euclidean indices $\mu=x,y$ in $2d$.\cite{Altshuler91,Wegner90}

\section{
The covariant background field method
        }
\label{sec: Normal coordinates}

\subsection{
K\"ahler normal coordinates for NL$\sigma$M
           }
\label{subsec: Normal coordinates for NLSM}

To perform the RG program, we divide the fields $(z^{*},z)$
in the NL$\sigma$M into slow
$(\psi^{*},\psi)$,
and fast
$(\pi^{*},\pi)$,
modes. To this end, we write
\begin{eqnarray}
z^{*a}=
\psi^{*a}+\pi^{*a},
\qquad
z^{a}=
\psi^{a}+\pi^{a},
\label{eq: seperation into fast and slow modes}
\end{eqnarray}
where we require that $(\psi^{*},\psi)$ satisfy
the classical equations of motion.
The fast modes $(\pi^{*},\pi)$ represent the
quantum fluctuations around 
the classical background $(\psi^{*},\psi)$.
They are integrated out, thereby
generating an effective action for the slow modes $(\psi^{*},\psi)$.
In practice, this effective action is computed order by order 
within an expansion, the so-called loop expansion, 
which is nothing but the cumulant expansion. 
One drawback with the separation%
~(\ref{eq: seperation into fast and slow modes})
into fast and slow modes is that the
perturbative expansion of the action in powers of the fast modes 
$(\pi^{*},\pi)$
with respect to the action evaluated at 
$(\psi^{*},\psi)$
need not transform covariantly under reparametrization of the manifold.
This danger can be avoided by choosing 
K\"ahler normal coordinates (KNC) 
in terms of which we denote the fast modes by
$(\omega^{*},\omega)$.
The KNC coordinates 
are defined by demanding the manifest covariance 
of the K\"ahler potential when expanded in terms of 
$(\omega^{*},\omega)$.
\cite{Higashijima00,Higashijima02a}
The perturbative expansion of the action in powers of the KNC
about the background field is then manifestly covariant, i.e., 
only covariant objects such as the metric, the Riemann tensor,
the covariant derivative on the manifold, and so on appear in the expansion,
as we shall shortly see.
One advantage of this RG scheme, the covariant background field method,
is that no redundant operator is induced upon renormalization.
This is a very useful property when dealing
with the mixing of the composite operators
induced by an RG transformation.\cite{Castilla97}
A related advantage is that
there is no need to break the symmetry of the NL$\sigma$M 
to tame IR divergences.

The relation between the generic parametrization
$(\pi^{*},\pi)$ and the KNC parametrization $(\omega^{*},\omega)$
of the fast modes is non-linear. For examples,
\begin{subequations}
\label{eq: covariant exp partial mu and metric in normal coo}
\begin{eqnarray}
&&
\partial^{\ }_{\mu}
\left(
\psi^{a}
+
\pi^{a}
\right)=
\partial^{\ }_{\mu}\psi^{a}
+
D^{\ }_{\mu}
w^{a}
\label{eq: covariant exp partial mu and metric in normal coo a}\\
&&
\hphantom{AAAAAA}
+
\frac{1}{2}
R^{a}_{\ c^{\ }_{1}c^{\ }_{2}c^{*}_{3}}
(\psi^{*},\psi)
\partial^{\ }_{\mu}\psi^{* c^{\ }_{3}}
w^{c^{\ }_{2}}w^{c^{\ }_{1}}
+
\cdots,
\nonumber\\
&&
g^{\ }_{i^{*} j}(\psi^{*}+\pi^{*},\psi+\pi)=
g^{\ }_{i^{*} j}(\psi^{*},\psi)
\label{eq: covariant exp partial mu and metric in normal coo b}\\
&&
\hphantom{AAAAAAAAAAAAA}
-
R^{\ }_{i^{*} j k^{\ }_{1} l^{*}_{1}}
w^{* l_{1}^{\ }}
w^{k^{\ }_{1}}
+
\cdots.
\nonumber
\end{eqnarray}
\end{subequations}
As promised, the right-hand side is covariant under reparametrization
as the pair 
$(D^{* }_{\mu},D^{\ }_{\mu})$ 
denotes the holomorphic and antiholomorphic
covariant derivatives
\begin{subequations}
\begin{eqnarray}
D^{\ }_{\mu} w^{j}&=&
\nabla^{\ }_{a} w^{j}\,
\partial^{\ }_{\mu} \psi^{a}
\nonumber\\
&=&
\partial_{\mu}^{\ }w^{j}
+
\Gamma^{j}_{\, ab}\,
w^{b}\,
\partial^{\ }_{\mu}
\psi^{a},
\label{eq: holo cov der}
 \\
D^{* }_{\mu} w^{* j}&=&
\left(
D^{\ }_{\mu} w^{j}
\right)^*
\nonumber\\
&=&
\nabla^{\ }_{a^{*}}
w^{* j}\,
\partial^{\ }_{\mu}\psi^{* a}
\nonumber\\
&=&
\partial_{\mu}^{\ }w^{* j}
+
\Gamma^{j^*}_{\,a^{*}b^{*}}\,
w^{* b}\,
\partial^{\ }_{\mu}\psi^{* a}
,
\label{eq: antiholo cov der}
\end{eqnarray}
\end{subequations}
respectively,
with the components 
\begin{eqnarray}
R^{a}_{\ b c d^*}
=
-\partial_{d^*}^{\ } 
\Gamma^{a}_{\ bc},
\quad
R^{a^*}_{\ b^* c^* d}
=
-
\partial_d^{\ } \Gamma^{a^*}_{\ b^* c^*},
\label{eq: def curvature tensor}
\end{eqnarray}
$R^{\ }_{EDAB}=g^{\ }_{EC}R^{C}_{\, DAB}$ and so on
of the curvature tensor.
The expansion of the action in powers of the KNC is
\begin{eqnarray}
&&
S[\psi^{*}+\pi^{*},\psi+\pi]
-
S[\psi^{*},\psi]
=
\nonumber\\
&&
+
\frac{1}{2\pi t}
\int_r
g^{\ }_{i^{*}j}(\psi^{*},\psi)\,
D^{\ }_{\mu}w^{j}
D^{* }_{\mu}w^{* i}\,
\nonumber\\
&&
-
\frac{1}{2\pi t}
\int_r\,
R^{\ }_{i^{*}jkl^{*}}(\psi^{*},\psi)
\Big(
w^{* l }w^{k}
\partial^{\ }_{\mu}\psi^{j}
\partial^{\ }_{\mu}\psi^{* i }\
\nonumber\\
&&
\qquad\qquad
-
\frac{1}{2}
w^{* l}
\partial^{\ }_{\mu}\psi^{k}
\partial^{\ }_{\mu}\psi^{j}
w^{* i}
-
\frac{1}{2}
\partial^{\ }_{\mu}\psi^{* l}
w^{k}
w^{j}
\partial^{\ }_{\mu}\psi^{* i}
\Big)
\nonumber\\
&&
+\cdots.
\label{eq: Covariant Gaussian expansion action NLSM}
\end{eqnarray}

\subsection{
Canonical form of the metric tensor and vielbeins
           }
\label{subsec: Canonical form of the metric tensor and vielbeins}

The covariant Gaussian expansion
(\ref{eq: Covariant Gaussian expansion action NLSM})
about an extrema $(\psi^{*},\psi)$ of the action 
can still be simplified further by performing a local
$\mathrm{GL}(M,\mathbb{C})$
transformation of the tangent space at $(\psi^{*},\psi)$
that brings the metric tensor at $(\psi^{*},\psi)$
to the canonical form.
To this end, introduce the fast degree of freedom 
$(\zeta^{*},\zeta)$
by the linear transformation%
\cite{Aoyama85,Shore90}
\begin{subequations}
\label{eq: def vielbeins}
\begin{eqnarray}
&&
\zeta^{* \hat{a}}:=
w^{* a}
{}^{\ }_{a^{*}}\hat{e}^{\hat{a}^{*}},
\qquad
w^{* a}=
\zeta^{* \hat{a}}
{}^{\ }_{\hat{a}^{*}}\hat{e}^{\ a^{*}},
\label{eq: def vielbeins a}
\\
&&
\zeta^{\hat{a}}:=
w^{a}\,
{}^{\ }_{a}\hat{e}^{\hat{a}},
\qquad
w^{a}=
\zeta^{\hat{a}}\,
{}^{\ }_{\hat{a}}\hat{e}^{\ a},
\label{eq: def vielbeins b}
\end{eqnarray}
by demanding that
\begin{eqnarray}
g^{\ }_{a^{*}b}(\psi^{*},\psi)
w^{b} w^{* a}=
\eta^{\ }_{\hat{a}^{*}\hat{b}}
\zeta^{\hat{b}}
\zeta^{* \hat{a}}
\label{eq: def vielbeins c}
\end{eqnarray}
where the transformed metric is given by 
\begin{eqnarray}
\eta^{\ }_{\hat{a} \hat{b}^{*}}&=&
\mathrm{diag}
\big(
\mathbb{I}^{\ }_{M^{\ }_{1}},
-
\mathbb{I}^{\ }_{M^{\ }_{2}}
\big),
\quad
M^{\ }_{1} + M^{\ }_{2} = M.
\nonumber\\
&&
\label{eq: def vielbeins d}
\end{eqnarray}
\end{subequations}
The two background dependent matrices
$({}^{\ }_{a^{*}}\hat{e}^{\hat{a}^{*}})$
and
$({}^{\ }_{\hat{a}^{*}}\hat{e}^{a^{*}})$
from $\mathrm{GL}(M,\mathbb{C})$
that implement this transformation in the antiholomorphic sector
are inverse to each other. The same is true of the matrices
$({}^{\ }_{a}\hat{e}^{\hat{a}})$
and
$({}^{\ }_{\hat{a}}\hat{e}^{a})$
in the holomorphic sector.
These four matrices are called the vielbeins and must obey
\begin{eqnarray}
g^{\ }_{a b^{*}}\, 
{}^{b^{*}}\hat{e}^{\ }_{\hat{b}^{*}} 
{}^{a}\hat{e}^{\ }_{\hat{a}}\,=
\eta^{\ }_{\hat{a} \hat{b}^{*}},
&&
\eta^{\ }_{\hat{a} \hat{b}^{*}}\, 
{}^{\hat{b}^{*}}\hat{e}^{\ }_{b^{*}}
{}^{\hat{a}} \hat{e}^{\ }_{a}\,=
g^{\ }_{a b^{*}},
\end{eqnarray}
by condition (\ref{eq: def vielbeins c}).
From now on, latin letters with a hat refer to the coordinates
of the K\"ahler manifold in the vielbein basis
(\ref{eq: def vielbeins}).
Under transformation
(\ref{eq: def vielbeins})
the covariant expansion
(\ref{eq: Covariant Gaussian expansion action NLSM})
becomes
\begin{eqnarray}
&&
S[\psi^{*}+\pi^{*},\psi+\pi]
-
S[\psi^{*},\psi]=
\nonumber\\
&&
+
\frac{1}{2\pi t}
\int_{r}\,
\eta^{\ }_{\hat{a}^{*} \hat{b}}
\hat{D}^{\ }_{\mu}\zeta^{\hat{b}}
\hat{D}^{* }_{\mu}\zeta^{* \hat{a}}
\nonumber\\
&&
-
\frac{1}{2\pi t}
\int_{r}
R^{\ }_{i^{*}jkl^{*}}
\Big[
(\zeta^{* \hat{l}}\, {}_{\hat{l}^{*}}^{\ }\hat{e}^{l^{*}})
(\zeta^{\hat{k}}\, {}_{\hat{k}}^{\ }\hat{e}^{k})
\partial^{\ }_{\mu}\psi^{j}
\partial^{\ }_{\mu}\psi^{* i}
\nonumber\\
&&
\qquad\qquad\qquad\quad
-
\frac{1}{2}
(\zeta^{\hat{l}^{*}}\,{}^{\ }_{\hat{l}^{*}}\hat{e}^{l^{*}})
\partial^{\ }_{\mu}\psi^{k}
\partial^{\ }_{\mu}\psi^{j}
(\zeta^{\hat{i}^{*}}\,{}_{\hat{i}^{*}}^{\ }\hat{e}^{i^{*}})
\nonumber\\
&&
\qquad\qquad\qquad\quad
-
\frac{1}{2}
\partial^{\ }_{\mu}\psi^{* l}
(\zeta^{\hat{k}}\,{}^{\ }_{\hat{k}}\hat{e}^{k})
(\zeta^{\hat{j}}\,{}^{\ }_{\hat{j}}\hat{e}^{j})
\partial^{\ }_{\mu}\psi^{* i}
\Big]
\nonumber\\
&&
+
\cdots
\label{eq: covariant expansion with vielbeins a}
\end{eqnarray}
where yet another pair of covariant derivatives
\begin{subequations}
\label{eq: covariant der with vielbeins}
\begin{eqnarray}
\hat{D}^{* }_{\mu}\zeta^{* \hat{a} }&=&
\partial^{\ }_{\mu}\zeta^{* \hat{a} }
-
\zeta^{* \hat{b} }
{}_{\hat{b}^{*}}(A^{\ }_{\mu})^{\hat{a}^{*}},
\label{eq: covariant der with vielbeins a}
\\
\hat{D}^{\ }_{\mu}\zeta^{\hat{a}}&=&
\partial^{\ }_{\mu}\zeta^{\hat{a}}
-
\zeta^{\hat{b}}\,
{}_{\hat{b}}^{\ }(A^{\ }_{\mu})^{\hat{a}},
\label{eq: covariant der with vielbeins b}
\end{eqnarray}
\end{subequations}
have been introduced.
The field $A^{\ }_{\mu}$ is the 
$\mathrm{U}(M)$ 
gauge field defined through
the $\mathrm{U}(M)$ spin connection 
${}^{\ }_{\hat{a}}\omega^{\hat{b}}{}^{\ }_{c^{*}}$,
${}^{\ }_{\hat{a}}\omega^{\hat{b}}{}^{\ }_{c}$
by
\begin{subequations}
\begin{eqnarray}
{}_{\hat{a}}^{\ }\omega^{\hat{b}}{}_{c^{*}}^{\ }&:=&
{}_{\hat{a}}^{\ }\hat{e}^{a}\,
\big(
\partial_{c^{*}}^{\ }\,
{}_{a}\hat{e}^{\hat{b}}
\big),
\\
{}_{\hat{a}}^{\ }\omega^{\hat{b}}{}_{c}^{\ }
&:=&
{}_{\hat{a}}^{\ }\hat{e}^{a} 
\Big[
(\partial^{\ }_{c}\, {}_{a}^{\ }\hat{e}^{\hat{b}})
-
\Gamma^{r}{}_{ac}^{\ }\,
{}_{r}^{\ }\hat{e}^{\hat{b}}
\Big],
\label{eq: covariant expansion with vielbeins c}
\\
{}_{\hat{a}}^{\ }
\left(
A^{\ }_{\mu}
\right)^{\hat{b}}
&:=&
{}_{\hat{a}}^{\ }\omega^{\hat{b}}{}_{c^{*}}^{\ }
\partial^{\ }_{\mu}\psi^{c^{*}}
+
{}_{\hat{a}}^{\ }\omega^{\hat{b}}{}_{c}^{\ }
\partial^{\ }_{\mu}\psi^{c}.
\label{eq: covariant expansion with vielbeins d}
\end{eqnarray}
[If the metric $\eta$ is pseudo Riemannian,
$\mathrm{U}(M)$ should be replaced by 
$\mathrm{U}(M^{\ }_{1}, M^{\ }_{2})$.]
The emergence of an $\mathrm{U}(M)$ gauge structure
originates from the fact that the vielbeins are not
uniquely specified by condition~(\ref{eq: def vielbeins c}).
As the physical result cannot depend on the specific
choice of gauge, i.e., the specific choice for the vielbeins,
the dependence on the spin connection should occur through
the U($M$) gauge invariant field strength
\begin{eqnarray}
&&
{}_{\hat{a}}^{\ }(F^{\ }_{\mu\nu}){}^{\hat{b}}=
\nonumber\\
&&
\partial^{\ }_{\mu}\,
{}_{\hat{a}}^{\ }(A^{\ }_{\nu})^{\hat{b}}
-
\partial^{\ }_{\nu}\,
{}_{\hat{a}}^{\ }(A^{\ }_{\mu})^{\hat{b}}
+
{}_{\hat{a}}^{\ }(A^{\ }_{\mu}A^{\ }_{\nu})^{\hat{b}}
-
{}_{\hat{a}}^{\ }(A^{\ }_{\nu}A^{\ }_{\mu})^{\hat{b}}=
\nonumber\\
&&
-
{}_{\hat{a}}^{\ }\hat{e}^{a}
R^{r}_{\,ad c^{*}}
\partial^{\ }_{\nu}\psi^{c^{*}}
\partial^{\ }_{\mu}\psi^{d}
{}_{r}^{\ }\hat{e}^{\hat{b}}
+
{}_{\hat{a}}^{\ }\hat{e}^{a}
R^{r}_{\,ad c^{*}}
\partial^{\ }_{\mu}\psi^{c^{*}}
\partial^{\ }_{\nu}\psi^{d}
{}_{r}^{\ }\hat{e}^{\hat{b}}.
\nonumber\\
&&
\label{eq: covariant expansion with vielbeins e}
\end{eqnarray}
\end{subequations}

\subsection{
One-loop beta functions
           }

We are in position to integrate out the fast degrees of freedom
for the complex Grassmannian,
$\mathrm{SO}(2N)/\mathrm{U}(N)$, 
and 
$\mathrm{Sp}(2N)/\mathrm{U}(N)$.
The partition function for slow and fast modes is
\begin{subequations}
\begin{eqnarray}
Z\!\!&:=&\!\!
\int
\mathcal{D}[\psi^{*},\psi]
\int
\mathcal{D}[\zeta^{*},\zeta]\,
e^{-S[\psi^{*},\psi,\zeta^{*},\zeta]}.
\label{eq: Covariant Gaussian quantum theory for slow and fast modes a}
\end{eqnarray}
The action 
$S[\psi^{*},\psi,\zeta^{*},\zeta]$
($=S[\psi^{*}+\pi^{*},\psi+\pi]$)
is separated into three contributions
\begin{eqnarray}
S[\psi^{*},\psi,\zeta^{*},\zeta]\!\!&:=&\!\!
S[\psi^{*},\psi]
+
S^{\ }_{0}[\zeta^{*},\zeta]
+
S^{\ }_{\mathrm{I}}[\psi^{*},\psi,\zeta^{*},\zeta],
\nonumber\\
&&
\label{eq: action for slow and fast modes}
\end{eqnarray}
\end{subequations}
where $S^{\ }_{0}[\zeta^{*},\zeta]$
is the free Gaussian part for the fast modes,
and
$S^{\ }_{\mathrm{I}}[\psi^{*},\psi,\zeta^{*},\zeta]$
represents the interactions between
the slow and fast modes.
Integration over the fast mode $(\zeta^{*},\zeta)$ 
is performed with the help of the cumulant expansion
\begin{subequations}
\label{eq: Covariant cumulant expansion}
\begin{eqnarray}
&&
Z=
\int
\mathcal{D}[\psi^{*},\psi]\,
e^{-S[\psi^{*},\psi]+\delta S[\psi^{*},\psi]},
\label{eq: Covariant cumulant expansion a}
\\
&&
\delta S[\psi^{*},\psi]:=
\left\langle
S_{\mathrm{I}}
\right\rangle^{\zeta}_{0}
-
\frac{1}{2}
\Big[
\big\langle
\left(
S_{\mathrm{I}}
\right)^2
\big\rangle^{\zeta}_{0}
-
\Big(
\big\langle
S_{\mathrm{I}}
\big\rangle^{\zeta}_{0}
\Big)^2
\Big]
+
\cdots,
\nonumber\\
\end{eqnarray}
where
\begin{eqnarray}
\left\langle
(\cdots)
\right\rangle^{\zeta}_{0}
&:=&
\frac{
\int\mathcal{D}[\zeta^{*},\zeta]\,
e^{-S^{\ }_{0}[\zeta^{*},\zeta]}
(\cdots)
     }
     {
\int\mathcal{D}[\zeta^{*},\zeta]\,
e^{-S^{\ }_{0}[\zeta^{*},\zeta]}
\hphantom{(\cdots)}
     }.
\label{eq: def of <>zeta0}
\end{eqnarray}
\end{subequations}
To evaluate the cumulant expansion, 
we introduce the Green function
\begin{eqnarray}
\big\langle
\,
\zeta^{\hat{a}    }(x)
\zeta^{* \hat{b} }(y)
\big\rangle^{\zeta}_{0}
\,
&=&
\eta^{\hat{a}\hat{b}^{*}}\,
2\pi t
\int\nolimits_{e^{-{d}l}\Lambda}^{\Lambda}
 \frac{{d}^dk}{(2\pi)^{d}}\,
\frac{e^{\mathrm{i}k(x-y)}}{k^2}
\nonumber\\
&=:&
\eta^{\hat{a}\hat{b}^{*}}\,
2\pi t
G^{\ }_{0}(x-y),
\label{eq: def bare Green fct}
\end{eqnarray}
where $\Lambda=\mathfrak{a}^{-1}$ is the UV cutoff.
The leading contribution to the increment of the
action of the slow mode $(\psi^{*},\psi)$ 
resulting from integrating out the fast
mode $(\zeta^{*},\zeta)$ 
within the second-order cumulant expansion is
\begin{eqnarray}
\delta S[\psi^{*},\psi]&=&
-
G^{\ }_{0}(0)
\int_{r}
R^{\ }_{a^{*} b}
\partial^{\ }_{\mu}\psi^{b}\,
\partial^{\ }_{\mu}\psi^{* a},
\label{eq: one-loop effective action for slow modes}
\end{eqnarray}
with
\begin{eqnarray}
R_{a^* b}^{\ }&:=& R^{c^*}{}^{\ }_{a^* c^* b}=R^{c^*}{}^{\ }_{c^* a^* b},
\end{eqnarray}
the Ricci tensor.

For symmetry classes C and D, the Ricci tensor is 
proportional to the metric (the target space is an Einstein manifold)
and given by
\begin{eqnarray}
R^{\ }_{(Cc) (Ee)^{*}}&=&
(\mathcal{N}-\vartheta)
g^{\ }_{(Cc) (Ee)^{*}}.
\end{eqnarray}
Thus, 
from Eq.~(\ref{eq: one-loop effective action for slow modes}),
the one-loop beta functions in $(2+\epsilon)d$
for the NL$\sigma$M coupling 
$\beta(t)=
{d}t/{d}l$
is
\begin{eqnarray}
\beta(t)
&=&
-
\epsilon t 
+
(\mathcal{N}-\vartheta) t^2
+
\mathcal{O}(t^3).
\label{eq: derivation beta fct}
\end{eqnarray}
Here, ${d}l$ is the infinitesimal 
rescaling of the ultraviolet cutoff, 
$\mathfrak{a}\rightarrow \mathfrak{a}e^{+{d}l}$
for which $G^{\ }_0(0)= {d}l/2\pi$.
This one-loop result agrees with
the known three-, four-, and five-loop results.
\cite{Wegner89,Hikami81,Hikami90-92}
Another check comes from
the group isomorphisms%
~\cite{Hikami81, Higashijima01}
\begin{subequations}
\begin{eqnarray}
\mathrm{U}(2)/[\mathrm{U}(1)\times\mathrm{U}(1)]
&\simeq& \mathrm{SO}(4)/\mathrm{U}(2)
\nonumber\\
&\simeq& \mathrm{Sp}(1)/\mathrm{U}(1),
\label{eq: group isomorphism 1}\\
\mathrm{U}(4)/[\mathrm{U}(1)\times\mathrm{U}(3)]
&\simeq& \mathrm{SO}(6)/\mathrm{U}(3),
\label{eq: group isomorphism 2}\\
\mathrm{Sp}(N)/\mathrm{U}(N)
&\simeq&
\mathrm{SO}(-2N)/\mathrm{U}(-N).
\qquad
\label{eq: group isomorphism 3}
\end{eqnarray}
\end{subequations}
One verifies that the beta functions 
match if the couplings are appropriately rescaled 
for isomorphisms~(\ref{eq: group isomorphism 1}) 
and~(\ref{eq: group isomorphism 2})
and with the additional substitution $t\to-t$
for isomorphisms~(\ref{eq: group isomorphism 3}). 
The nontrivial zeros of the beta function in 
symmetry classes A, C, and D are%
\begin{eqnarray}
0<t^{*}&=& 
\frac{\epsilon}{\mathcal{N}-\vartheta}+\mathcal{O}(\epsilon^2).
\end{eqnarray}

\section{
One-loop RG for high-gradient operators
        }
\label{sec: One-loop RG for high-gradient operators}

\subsection{
Strategy
           }

We now turn to the computation of the 
dominant anomalous scaling dimension $x^{(s)}_{i}$
associated to any high-gradient operator $\mathcal{O}^{(s)}_{i}$ 
of type~(\ref{eq: high-gradient opertors in tensor notation}),
where the collective index $i$ runs over any allowed choice 
for $\{l^{\ }_{i}\}$
in Eq.~(\ref{eq: tensor T as a product of S's}) 
and over the greek indices in
$\partial^{\ }_{\mu} z^{I}
 \partial^{\ }_{\nu} z^{J}
 \partial^{\ }_{\rho}z^{K}\cdots$.
The one-loop anomalous scaling dimension $x^{(s)}_{i}$
of $\mathcal{O}^{(s)}_{i}$ is extracted from the one-loop
RG transformation law
\begin{subequations}
\label{eq: def mixing matrix}
\begin{eqnarray}
\left\langle
\left[ 
\mathcal{O}^{(s)}_{i}(z^{*},z) 
\right]^{\ }_{\zeta^2} 
\right\rangle
&=&
t {d}l\widehat{\Delta}^{(s)}_{ij}\mathcal{O}^{(s)}_{j}(\psi^{*},\psi).
\end{eqnarray}
Here, 
$\left[\mathcal{O}^{(s)}_{i}(z^{*},z)\right]^{\ }_{\zeta^k}$ 
denotes all the terms that are of order $k$ in $\zeta^{A}$
in the KNC expansion of $\mathcal{O}^{(s)}_{i}(z^{*},z)$,
and
\begin{eqnarray}
\langle
(\cdots)
\rangle
&=&
\left\langle
e^{-S^{\ }_{\mathrm{I}}}
(\cdots)
\right\rangle^{\zeta}_{0}
\Big/
\left\langle
e^{-S^{\ }_{\mathrm{I}}}
\right\rangle^{\zeta}_{0}
\end{eqnarray}
\end{subequations}
is evaluated up to first order in $t{d}l$ 
[recall Eq.~(\ref{eq: def of <>zeta0})].

Once the mixing matrix
\begin{subequations} 
\begin{eqnarray}
\widehat{\Delta}^{(s)}=\left(\widehat{\Delta}^{(s)}_{ij}\right)
\end{eqnarray} 
is known, its eigenvalues $\left\{\alpha^{(s)}_{i}\right\}$ 
yield the spectrum of anomalous scaling dimensions
\begin{eqnarray}
x^{(s)}_{i}
&=&
2s-\alpha^{(s)}_{i} t
\label{eq: scaling dimensions x and eigenvalues alpha}
\end{eqnarray}
and the spectrum of anomalous scaling dimensions
\begin{eqnarray}
y^{(s)}_{i}:=
d-x^{(s)}_{i}
\end{eqnarray}
\end{subequations} 
associated to the high-gradient operator $\mathcal{O}^{(s)}_{i}$.
Any eigenoperator of Eq.~(\ref{eq: def mixing matrix})
is relevant (marginal, irrelevant)
when its eigenvalue $y^{(s)}_{i}$ 
is positive (zero, negative). 
We are going to construct the eigenoperator
of Eq.~(\ref{eq: def mixing matrix})
with the largest eigenvalue of the mixing matrix $\widehat{\Delta}^{(s)}$
up to one-loop accuracy close to the metallic fixed point.

To this end, we need to expand 
$\mathcal{O}^{(s)}_i(z^*,z)$ in the KNC.
By taking advantage of the fact that all the covariant derivatives
of the tensor $T^{\ }_{IJ\ldots} (z^{*},z)$
vanish and that 
$\langle [\partial z^{I}]_{\zeta^2} \rangle =0$,
we infer that
\begin{widetext}
\begin{subequations}
\label{eq: master formula, Kahler}
\begin{eqnarray}
\left\langle 
\left[
T^{\ }_{IJ\ldots} (z^{*},z)
\partial^{\ }_{\mu} z^{I}
\partial^{\ }_{\nu} z^{J}
\cdots
\right]^{\ }_{\zeta^2}
\right\rangle
&=&
\left\langle
\left[
T^{\ }_{IJ\ldots}
\right]^{\ }_{\zeta^2}
\right\rangle
\partial^{\ }_{\mu}\psi^{I}
\partial^{\ }_{\nu}\psi^{J}
\cdots
+
T^{\ }_{IJ\ldots}(\psi,\psi^{*})
\left\langle
\left[
\partial^{\ }_{\mu} z^{I}
\right]^{\ }_{\zeta^1}
\left[
\partial^{\ }_{\nu} z^{J}
\right]^{\ }_{\zeta^1}
\right\rangle
\cdots
+
\cdots.
\nonumber\\
&&
\label{eq: master formula 0, Kahler}
\end{eqnarray}
Thus, all we need are
\begin{eqnarray}
\big\langle
\big[\partial^{\ }_{\mu} z^{*i}\big]_{\zeta^1}
\big[\partial^{\ }_{\nu} z^{*j}\big]_{\zeta^1}
\big\rangle
&=&
+
t{d}l \delta^{\ }_{\mu,-\nu}
\partial^{\ }_{\mu} \psi^{*k}
\partial^{\ }_{\nu} \psi^{*l}
R_{k^{*}}{}^{i^{*}}{}_{l^{*}}{}^{j^{*}},
\\
\big\langle
\big[ \partial^{\ }_{\mu}z^{i} \big]_{\zeta^1}
\big[ \partial^{\ }_{\nu}z^{*j} \big]_{\zeta^1}
\big\rangle
&=&
-t{d}l \delta^{\ }_{\mu,-\nu}
\partial^{\ }_{\mu} \psi^{k} 
\partial^{\ }_{\nu} \psi^{*l}
R^{i j^{*}}{}_{l^{*} k},
\\
\big\langle
\big[\partial^{\ }_{\mu}z^{i} \big]_{\zeta^1}
\big[\partial^{\ }_{\nu}z^{j} \big]_{\zeta^1}
\big\rangle
&=&
+
t{d}l \delta^{\ }_{\mu,-\nu}
\partial^{\ }_{\mu} \psi^{k}
\partial^{\ }_{\nu} \psi^{l}
R^{i}{}_{k}{}^{j}{}_{l},
\label{eq: master formulae 1, Kahler}
\end{eqnarray}
on the one hand and
\begin{eqnarray}
\left\langle 
\left[
T^{\ }_{b^{\ }_{1}\cdots b^{\ }_{r}b^{*}_{1}\cdots b^{*}_{s}}
\right]^{\ }_{\zeta^2}
(z^{*},z)
\right\rangle
&=&
t{d}l
\sum_{i=1}^{s}
R^{l^{*}}{}_{l^{*}}{}^{a^{*}}{}_{b^{*}_{i}}
T^{\ }_{
b^{\ }_{1}
\cdots  
b^{\ }_{r}
b^{*}_{1}
\cdots 
b^{*}_{i-1}
a^{*}
b^{*}_{i+1} 
\cdots 
b^{*}_{s}
       }
(\psi^{*},\psi)
\nonumber\\
&=&
t{d}l
\sum_{i=1}^{r}
R^{k}{}_{k}{}^{a}{}_{b^{\ }_{i}}
T^{\ }_{
b^{\ }_{1}
\cdots 
b^{\ }_{i-1}
a
b^{\ }_{i+1} 
\cdots 
b^{\ }_{r}
b^{*}_{1}
\cdots 
b^{*}_{s}
       }
(\psi^{*},\psi),
\label{eq: master formulae 2, Kahler}
\end{eqnarray}
\end{subequations}
\end{widetext}
on the other hand.
Here, we are using the conformal indices defined in
Eq.\ (\ref{eq: def conformal indices}).
Note also that since holomorphic 
$\nabla^{\ }_{i}$
and antiholomorphic 
$\nabla^{\ }_{j^{*}}$
covariant derivatives 
do not commute in general,
we have two distinct representations for 
$\left\langle\left[T^{\ }_{IJK\cdots}\right]^{\ }_{\zeta^2}\right\rangle$.

\subsection{
RG transformation
           }
\begin{widetext}
The RG transformation law
(\ref{eq: master formula, Kahler})
obeyed by any high-gradient operator
for a NL$\sigma$M on the Hermitian symmetric spaces
$\mathrm{U}(p+q)/[ \mathrm{U}(p)\times \mathrm{U}(q) ]$,
$\mathrm{SO}(2N)/\mathrm{U}(N)$,
and
$\mathrm{Sp}(N)/\mathrm{U}(N)$
becomes explicit with the help of the

\noindent
(a) intra-trace formulae
\begin{subequations}
\label{eq: RG transformation, intra-trace}
\begin{eqnarray}
\left\langle
\mathrm{tr}\,
\left( 
\left[ 
A^{\tau}_{\mu} 
\right]^{\ }_{\zeta^1}
\mathcal{M} 
\left[ 
A^{\tau}_{\nu} 
\right]^{\ }_{\zeta^1}
\mathcal{N}
\right)
\right\rangle
&=&
-
\frac{2-\vartheta^{2}}{2}
t{d}l \delta^{\ }_{\mu,-\nu}
\left[
\vartheta^{2}
\mathrm{tr}\,
\left(
A^{\tau}_{\mu} 
\mathcal{N} 
A^{\tau}_{\nu} 
\mathcal{M}^{\mathrm{T}}
\right)  
+
\mathrm{tr}\,
\left( 
A^{\tau}_{\mu} 
\mathcal{N} 
\right)
\mathrm{tr}\,
\left( 
A^{\tau}_{\nu} 
\mathcal{M}  
\right)  
\right.
\nonumber\\
&&
\qquad\qquad
+
\mathrm{tr}\,
\left.
\left( 
A^{\tau}_{\mu}  
\mathcal{M} 
\right)
\mathrm{tr}\,
\left( 
A^{\tau}_{\nu} 
\mathcal{N} 
\right)  
+
\vartheta^{2} 
\mathrm{tr}\,
\left( 
A^{\tau}_{\mu}  
\mathcal{M}^{\mathrm{T}} 
A^{\tau}_{\nu} 
\mathcal{N}
\right) 
\right],
\\
\left\langle
\mathrm{tr}\,
\left( 
\left[ 
A^{+\tau}_{\mu}
\right]^{\ }_{\zeta^1}  
\mathcal{M}  
\left[ 
A^{-\tau}_{\nu}
\right]^{\ }_{\zeta^1} 
\mathcal{N} 
\right)
\right\rangle
&=&
+
\frac{2-\vartheta^{2}}{2} 
t{d}l 
\delta^{\ }_{\mu,-\nu}
\left[
\mathrm{tr}\,
\left(
A^{+\tau}_{\mu}
A^{-\tau}_{\nu}
\mathcal{N} 
\right) 
\mathrm{tr}\,
\left( 
\mathcal{M} 
\right)
- 
\vartheta 
\mathrm{tr}\,
\left( 
A^{+\tau}_{\mu}
A^{-\tau}_{\nu}  
\mathcal{N}  
\mathcal{M}^{\mathrm{T}} 
\right)
\right.
\nonumber\\
&&
\qquad\qquad
- 
\vartheta 
\mathrm{tr}\,
\left.
\left( 
A^{+\tau}_{\mu}
A^{-\tau}_{\nu} 
\mathcal{M}^{\mathrm{T}}  
\mathcal{N} 
\right)
+ 
\mathrm{tr}\,
\left( 
\mathcal{N}
\right) 
\mathrm{tr}\,
\left(  
\mathcal{M}
A^{-\tau}_{\nu} 
A^{+\tau}_{\mu}
\right)
\right],
\end{eqnarray}
\end{subequations}
(b) inter-trace formulae
\begin{subequations}
\label{eq: RG transformation, inter-trace}
\begin{eqnarray}
\left\langle
\mathrm{tr}\,
\left(
\left[
A^{\tau}_{\mu}
\right]^{\ }_{\zeta^1} 
\mathcal{M}
\right) 
\mathrm{tr}\,
\left(
\left[
A^{\tau}_{\nu}
\right]^{\ }_{\zeta^1} 
\mathcal{N}
\right)
\right\rangle
&=&
\frac{2-\vartheta^{2}}{2} t{d}l\delta^{\ }_{\mu,-\nu}
\left[
\vartheta
\mathrm{tr}\,
\left(
A^{\tau}_{\nu} 
\mathcal{N}^{\mathrm{T}} 
A^{\tau}_{\mu} 
\mathcal{M}
\right)  
-
\mathrm{tr}\,
\left(
A^{\tau}_{\mu} 
\mathcal{N} 
A^{\tau}_{\nu} 
\mathcal{M}
\right)
\right.  
\nonumber\\
&&
\qquad \qquad
- 
\mathrm{tr}\,
\left.
\left(
A^{\tau}_{\nu} 
\mathcal{N} 
A^{\tau}_{\mu} 
\mathcal{M}
\right) 
+
\vartheta 
\mathrm{tr}\,
\left(
A^{\tau}_{\mu} 
\mathcal{N}^{\mathrm{T}} 
A^{\tau}_{\nu} 
\mathcal{M}
\right)
\right],
\label{eq: RG transformation, inter-trace a}
\\
\left\langle
\mathrm{tr}\, 
\left(
\left[
A^{+\tau}_{\mu}
\right]^{\ }_{\zeta^1} 
\mathcal{M}
\right) 
\mathrm{tr}\, 
\left(
\left[
A^{-\tau}_{\nu}
\right]^{\ }_{\zeta^1} 
\mathcal{N}
\right)
\right\rangle
&=&
\frac{2-\vartheta^{2}}{2} t{d}l
\delta^{\ }_{\mu,-\nu}
\left[
\mathrm{tr}\,
\left(
A^{+\tau}_{\mu} 
A^{-\tau}_{\nu} 
\mathcal{N} \mathcal{M}
\right)
- 
\vartheta 
\mathrm{tr}\,
\left( \
A^{-\tau}_{\nu} 
A^{+\tau}_{\mu}  
\mathcal{M} 
\mathcal{N}^{\mathrm{T}} 
\right)
\right.
\nonumber\\
&&
\qquad\qquad
- 
\vartheta 
\mathrm{tr}\,
\left.
\left(
A^{+\tau}_{\mu} 
A^{-\tau}_{\nu} 
\mathcal{N}^{\mathrm{T}} 
\mathcal{M}
\right) 
+ 
\mathrm{tr}\,
\left( 
A^{-\tau}_{\nu} 
A^{+\tau}_{\mu} 
\mathcal{M}  
\mathcal{N} 
\right)
\right],
\label{eq: RG transformation, inter-trace b}
\end{eqnarray}
\end{subequations}
and (c) diagonal contributions
\begin{eqnarray}
\left\langle
\left[
T^{\ }_{
a^{\ }_1
\cdots 
a^{\ }_s
b_1^{*}
\cdots 
b^{* }_s
       }
(z^{*},z)
\right]^{\ }_{\zeta^2}
\right\rangle
&=&
- t{d}l s 
\left(
\mathcal{N}
-
\vartheta 
\right)
T^{\ }_{
a^{\ }_1 
\cdots 
a^{\ }_s 
b^{* }_1
\cdots 
b^{* }_s
       }
(\psi^{*},\psi),
\label{eq: RG transformation, dia cont}
\end{eqnarray}
where $\mathcal{M}$ and $\mathcal{N}$ represent arbitrary 
rectangular matrices of the proper size
and the matrix-valued $A^{\pm}_{\pm}$
were defined in Eq.~(\ref{eq: def of A{pm}{pm}}).
This result, that follows from substituting the components of 
the curvature tensor~(\ref{eq: def curvature tensor}) 
when expressed in the 
stereographic coordinates~(\ref{eq: kahler potential grassmannian})
into the master formula~(\ref{eq: master formula, Kahler}),
agrees with the one-loop RG result for symmetry class A ($\vartheta=0$)
in Ref.\ \onlinecite{Lerner90}. 
\end{widetext}

\subsection{
Diagonalization of the RG equation within the subspace of
maximal number of switches
           }

As is shown in Appendix 
\ref{app: Upper triangular structure of the RG transformation},
the RG transformation law~(\ref{eq: master formula, Kahler})
has an upper (or lower) triangular structure when
appropriate quantum numbers (``number of switches'') are defined.
Since the number of switches
never increases under an RG transformation,
we can solve the eigenvalue and eigenvector problem%
~(\ref{eq: master formula, Kahler})
for each subspace with a fixed number of switches.
We will limit ourselves to 
operators with the maximal number of switches,
as the eigenscaling operator with the most dominant 
RG anomalous scaling dimension is known to be among them for 
symmetry class A.\cite{Wegner91}
In this subspace, any operator can be expressed 
as a polynomial of the objects
\begin{eqnarray}
\Big\{
\mathrm{tr}\,\left[(A^{+}_{+}A^{-}_{-})^{l}\right],\,
\mathrm{tr}\,\left[(A^{+}_{-}A^{-}_{+})^{l}\right]
\Big\}_{l=1,2,\ldots},
\end{eqnarray}
i.e., conformal indices in a single trace 
are completely alternating. It can also be shown that
the spaces spanned by
$
\big\{\mathrm{tr}\,\big[(A^{+}_{+}A^{-}_{-})^{l}\big] \big\}_{l=1,2,\ldots}
$
and
$
\big\{\mathrm{tr}\,\big[(A^{+}_{-}A^{-}_{+})^{l}\big]\big\}_{l=1,2,\ldots}
$
are not mixed by the RG transformation.
We can thus consider these two subspaces separately as long as we do not
require composite operators to be Hermitian.

We now compute the action 
of the RG transformation~(\ref{eq: master formula, Kahler})
on a composite operator of the form
\begin{eqnarray}
\mathcal{O}^{(s)}_{\{r\}}
&=&
\Omega^{r_{1}}_{1}
\Omega^{r_{2}}_{2}
\cdots
\Omega^{r_{L}}_{L},
\qquad
\sum_{l=1}^{L} l r^{\ }_l =s
\end{eqnarray}
where all $\Omega_{l}$ on the right-hand side
are a short-hand notation for objects of type
$
\mathrm{tr}\,\big[(A^{+}_{+}A^{-}_{-})^{l}\big]
$
without loss of generality.
Since the diagonal contributions (c) to the RG eigenvalues
are trivial, we focus on the intra-trace contributions (a) and 
inter-trace contributions (b) for the moment. 
If we define the off-diagonal mixing matrix 
$\widehat{R}^{(s)}=\left(\widehat{R}^{(s)}_{\{r\}\{r'\}}\right)$ by
\begin{eqnarray}
\left.
\left\langle
\mathcal{O}^{(s)}_{\{r\}} (z^{*},z)
\right\rangle
\right|^{\ }_{(a),(b)}
=:
t{d}l 
\widehat{R}^{(s)}_{\{r\}\{r'\}}
\mathcal{O}^{(s)}_{\{r'\}}(\psi^{*},\psi),
\end{eqnarray}
i.e., we have only collected the off-diagonal contributions
(\ref{eq: RG transformation, intra-trace})
and
(\ref{eq: RG transformation, inter-trace})
to the RG transformation law obeyed by
$\mathcal{O}^{(s)}_{\{r\}} (z^{*},z)$,
the action of $\widehat{R}^{(s)}$ can be cast into the form
\begin{eqnarray}
\widehat{R}^{(s)}&=&
-\vartheta\sum_{k}
k^{2}\Omega^{\ }_{k}
\frac{\partial}{\partial\Omega^{\ }_{k}}
+
\mathcal{N}
\sum_{k}
k\Omega_{k}\frac{\partial}{\partial\Omega^{\ }_{k}}
\nonumber\\
&&
+
\sum_{l,n}
\Bigg[
2
ln\Omega^{\ }_{l+n}
\frac{\partial^2}{\partial \Omega^{\ }_l\partial\Omega^{\ }_n}
\nonumber\\
&&
\hphantom{+\sum_{l,n}\Big[}
+
(2-\vartheta^2)
(l+n)
\Omega^{\ }_{l} 
\Omega^{\ }_{n}
\frac{\partial}{\partial\Omega^{\ }_{l+n}}
\Bigg].
\nonumber\\
&&
\label{eq: rep widehat R}
\end{eqnarray}

The eigenvalue problem~(\ref{eq: rep widehat R})
can be solved by brute force for small $s$ in which cases we have verified 
that the sign of the eigenvalues 
of the off-diagonal mixing matrix 
$\widehat{R}^{(s)}$ for $\mathrm{SO}(2N)/\mathrm{U}(N)$
are opposite to those for $\mathrm{Sp}(N)/\mathrm{U}(N)$.
In particular, the largest eigenvalue of $\widehat{R}^{(s)}$
for $\mathrm{SO}(2N)/\mathrm{U}(N)$
is given by minus the smallest eigenvalue for 
$\mathrm{Sp}(N)/\mathrm{U}(N)$ and vice et versa.
This fact is consistent with the group isomorphism
(\ref{eq: group isomorphism 2}),
and the fact that the beta functions for the corresponding NL$\sigma$M
match upon the replacement $t\to-t$.

More generally, one verifies that
\begin{subequations}
\begin{eqnarray}
\lambda^{(s)}&=&
2(s^2-s)
+ s(p+q),
\quad
(\vartheta=0),
\\
\lambda^{(s)}
&=&
s^2-2s
+ sN,
\quad
(\vartheta=+1),
\\
\lambda^{(s)}
&=&
2s^2-s
+ sN,
\quad
(\vartheta=-1),
\end{eqnarray}
\end{subequations}
is an eigenvalue of the off-diagonal mixing operator 
$\widehat{R}^{(s)}$ for any given $s$
with the eigenvector
\begin{subequations}
\begin{eqnarray}
v^{(s)}
&=&
\sum_{\{r \}}
\prod_{j}
\frac{1}{r_j ! j^{r_j}} (\Omega_{j})^{r_j},
\quad
(\vartheta=0),
\\
v^{(s)}
&=&
\sum_{\{r \}}
\prod_{j}
\frac{1}{r_j ! j^{r_j}} (\Omega_{j})^{r_j},
\quad
(\vartheta=+1),
\\
v^{(s)}
&=&
\sum_{\{r \}}
\prod_{j}
\frac{2^{j-1}}{r_j ! j^{r_j}} (\Omega_{j})^{r_j},
\quad
(\vartheta=-1).
\end{eqnarray}
\end{subequations}
As it is known that $\lambda^{(s)}$ is the largest eigenvalue
of $\widehat{R}^{(s)}$ in symmetry class A, 
we believe this to be also true for symmetry classes C and D. 
If so, the largest RG eigenvalues
for a given number $s$ of gradients is given by,
after adding the diagonal contributions (c),
\begin{subequations}
\label{eq: one-loop RG, alpha}
\begin{eqnarray}
\alpha^{(s)}
&=&
\lambda^{(s)}
-s (p+q)
=
2(s^2-s),
\quad
(\vartheta=0),
\nonumber\\
&&
\\
\alpha^{(s)}
&=&
\lambda^{(s)}
-s (N-1)
=
s^2-s,
\quad
(\vartheta=+1),
\nonumber\\
&&
\\
\alpha^{(s)}
&=&
\lambda^{(s)}
-s(N+1)
=
2(s^2-s),
\quad
(\vartheta=-1).
\nonumber\\
&&
\end{eqnarray}
\end{subequations}
Note that these results are valid up to one loop before
taking the replica limit.
Equation (\ref{eq: scaling dimension around UV fixed points, intro})
then follows by combining
Eq.~(\ref{eq: one-loop RG, alpha})
with Eq.~(\ref{eq: scaling dimensions x and eigenvalues alpha}).

\section*{Acknowledgements}

C.M.\ and A.F.\ acknowledge hospitality of the Kavli Institute for 
Theoretical Physics at Santa Barbara during the completion
of the manuscript. This research was supported in part
by the National Science Foundation under
Grants No. DMR-00-75064 (A.W.W.L.)
and No.\ PHY99-07949 (S.R., A.F., C.M.).

\appendix

\section{
Group-theory versus geometric representations of the NL$\sigma$M
        }
\label{app sec: Group-theory versus geometric representations of the NLsigmaM}

The purpose of this appendix
is to bridge 
the geometrical representation of the NL$\sigma$M
used in this paper 
and the one in terms of the so-called ``$Q$-matrix''.

First, recall that an element $Q$ in 
$\mathrm{U}(p+q)/[\mathrm{U}(p)\times \mathrm{U}(q)$]
can be written as\cite{Efetov80}
\begin{eqnarray}
Q=
U^{-1}\Lambda U,
\quad 
\Lambda = 
\mathrm{diag}(\mathbb{I}_{p}, -\mathbb{I}_{q}),
\quad
U\in \mathrm{U}(p+q).
\end{eqnarray}
As noticed by Lerner and Wegner,\cite{Lerner90}
high-gradient operators in the $Q$-matrix representation
as well as the action of the NL$\sigma$M 
can be conveniently expressed in terms of the objects
\begin{subequations}
\label{eq: Q matrix rep}
\begin{eqnarray}
A^{\tau}_{\mu}&:=&
\frac{1}{2}
P^{+\tau}
\left(\partial^{\ }_{\mu}Q\right) 
P^{-\tau}
\end{eqnarray}
where $\tau=\pm$ and the projectors 
$P^{\pm}$ 
\begin{eqnarray}
&&
P^{\pm }:=\frac{1}{2}\left(1\pm Q\right),
\qquad
(P^{\pm})^{2}=P^{\pm}
\label{eq: def projectprs Ppm}
\end{eqnarray}
\end{subequations}
were introduced.
A generic high-gradient operator is then built by taking 
a product of 
\begin{eqnarray}
\mathrm{tr}\,
\big(
A^{+}_{\mu}A^{-}_{\nu}A^{+}_{\rho}\cdots
\big).
\label{eq: def high gradient operators in terms product A's}
\end{eqnarray}
For example, the action of the NL$\sigma$M is 
\begin{eqnarray}
S&=&
\frac{1}{16\pi t}
\int_r
\mathrm{tr}\,
\left[
\left(\partial^{\ }_{\mu}Q\right)
\left(\partial^{\ }_{\mu}Q\right)
\right]
\nonumber\\
&=&
\frac{1}{4 \pi t}
\int_r
\mathrm{tr}\,
\left(A_{\mu}^{+} A^{-}_{\mu}\right).
\label{eq: NLSM action in terms A}
\end{eqnarray}

In order to establish explicitly
the connection between the 
$Q$-matrix representation%
~(\ref{eq: Q matrix rep})%
--(\ref{eq: NLSM action in terms A})
and the geometrical representation in the 
stereographic coordinates~(\ref{eq: kahler potential grassmannian}),
we parametrize $P^{+}$ and $P^{-}=P^{+}-Q$
in Eq.~(\ref{eq: def projectprs Ppm})
through a $(p+q) \times q$ matrix 
$\Phi^{\ }_{\alpha i}$ according to\cite{Brezin80}
\begin{subequations}
\label{eq: P in terms Phi}
\begin{eqnarray}
P^{+}_{\alpha\beta}=
\left(\Phi \Phi^{\dag}\right)^{\ }_{\alpha\beta}=
\sum_{i=1}^{q} 
\Phi^{\ }_{\alpha i}
\left(\Phi^{\dag }\right)^{\ }_{i\beta}=
\sum_{i=1}^{q} \Phi^{\ }_{\alpha i}\Phi^{*}_{\beta i}
\label{eq: P in terms Phi a}
\end{eqnarray}
where $\Phi$ satisfies the constraint
\begin{eqnarray}
\Phi^{\dag}\Phi= \mathbb{I}^{\ }_{q}.
\label{eq: P in terms Phi b}
\end{eqnarray}
\end{subequations}
The matrix $\Phi$ is thus a collection of 
$q$ orthonormal vectors 
$\{\Phi^{\ }_{a}\in\mathbb{C}^{p+q}\}_{a=1,\ldots,q}$.
We can use this constraint to fix the upper $q \times q$ block of $\Phi$.
We write
\begin{subequations}
\label{eq: solving for upper qxq block Phi}
\begin{eqnarray}
\Phi=
\left(
\begin{array}{c}
\overbrace{X}^{q} \, \}^{\ }_{q} \\ \hline
\varphi X \,\, \}^{\ }_{p}
\end{array}
\right)
\label{eq: Phi in terms X varphi}
\end{eqnarray}
where
$\Phi^{\ }_{\alpha i}$,
$\varphi^{\ }_{I j}$,
$X^{\ }_{ij}$,
and
$\varphi^{\ }_{Ij}X^{\ }_{jk}$,
are
$(p+q) \times q$,
$p \times q$,
$q \times q$,
and
$p \times q$,
matrices, respectively,
and the indices
$i$, $I$, 
and $\alpha$
run over the sets
$i=1,\ldots, q$,
$I=1,\ldots, p$,
$\alpha=1,\ldots, p+q$,
respectively.
Next, we express $X$ in terms of $\varphi$ by making use of the constraint%
~(\ref{eq: P in terms Phi b}),
\begin{eqnarray}
X^{*}_{ni}
Z^{-1}_{nm}
X^{\ }_{mk}
=
\delta^{\ }_{ik}
\qquad
(X^{\dag}Z^{-1}X=\mathbb{I}_{q}).
\label{eq: constraint}
\end{eqnarray}
Here the Hermitian $q\times q$ matrix $Z$
was introduced in 
Eqs.\ (\ref{eq: def Z}) and (\ref{eq: def Y}).
Constraint~(\ref{eq: constraint})
is satisfied if $X$ is chosen to be
\begin{eqnarray}
X^{\dag}=X= Z^{1/2}.
\label{eq: solution to constraint}
\end{eqnarray}
\end{subequations}
A generic high-gradient operators%
~(\ref{eq: def high gradient operators in terms product A's})
in the $Q$-matrix representation
is thus expressed in terms
of the stereographic coordinate $(\varphi^{*},\varphi)$
from Eq.~(\ref{eq: kahler potential grassmannian}).
For example, one verifies that the action of the NL$\sigma$M%
~(\ref{eq: NLSM action in terms A})
reduces to
Eq.\ (\ref{eq: def NLSM complex manifold})
with the metric~(\ref{eq: metric grassmannian})
derived from the K\"ahler potential%
~(\ref{eq: kahler potential grassmannian})
once the parametrization~(\ref{eq: P in terms Phi})
with Eq.~(\ref{eq: solving for upper qxq block Phi})
is used. Similarly,
\begin{eqnarray}
&&
\mathrm{tr}\,
\big(
A^{+}_{\mu}A^{-}_{\nu}A^{+}_{\rho}A^{-}_{\sigma}
\big)=
\nonumber\\
&&
\hphantom{A}
\mathrm{tr}\, 
\big[
Z 
\left(\partial^{\ }_{\mu}\varphi^{\dag}\right) 
Y 
\left(\partial^{\ }_{\nu}\varphi\right)
Z 
\left(\partial^{\ }_{\rho}\varphi^{\dag}\right)
Y 
\left(\partial^{\ }_{\sigma }\varphi\right)
\big].
\end{eqnarray}
Observe that the alternating structure in $A^+$ and $A^-$
in the high-gradient operators
has a one-to-one correspondence to that in 
$Z$ and $Y$.

\section{Upper triangular structure
of the RG transformation}
\label{app: Upper triangular structure of the RG transformation}

We prove in this appendix that the RG transformation law
obeyed by high-gradient operators
has an upper (or lower) triangular structure when
appropriate ``quantum'' numbers (``number of switches'')
are defined. A trace made of an even product of $A$'s, 
\begin{eqnarray}
\mathrm{tr}\,
\left(
A^{+\tau}_{\sigma_{1}}
A^{-\tau}_{\sigma_{2}}
A^{+\tau}_{\sigma_{3}}
A^{-\tau}_{\sigma_{4}}
\cdots
\right),
\end{eqnarray}
is said to have $n^{\tau}_{\sigma}$ number of switches 
from $A^{\tau}_{\sigma}$ to $A^{-\tau}_{-\sigma}$
when, dividing the sequence into pairs according to
\begin{eqnarray}
\mathrm{tr}\,
\left[
\left(
A^{+\tau}_{\sigma_{1}}
A^{-\tau}_{\sigma_{2}}
\right)
\left(
A^{+\tau}_{\sigma_{3}}
A^{-\tau}_{\sigma_{4}}
\right)
\cdots
\right],
\label{eq: given trace}
\end{eqnarray}
there are $n^{\tau}_{\sigma}$ pairs
$\left(A^{\tau}_{\sigma}A^{-\tau}_{-\sigma}\right)$
in the sequence. The number $n^{\tau}_{\sigma}$
does not depend on the cyclic permutations of the pairs
$\left(A^{\tau}_{\sigma}A^{-\tau}_{-\sigma}\right)$.
For a given trace of the form~(\ref{eq: given trace})
we can define
\begin{eqnarray}
n^{\ }_{\mathrm{tr}}:=
\sum_{\tau,\sigma=\pm}n^{\tau}_{\sigma}.
\end{eqnarray}

We are going to show that,
for the symmetry classes A, C, and D, 
the total number 
\begin{eqnarray}
n:=
\sum_{\mathrm{tr}}n^{\ }_{\mathrm{tr}}
\label{eq: def quantum number}
\end{eqnarray}
of switches in a composite operator
of type~(\ref{eq: high-gradient opertors in tensor notation})
never increases under 
the RG transformation law~(\ref{eq: master formula, Kahler}).
Here, the summation extends over the product over all traces 
of the form~(\ref{eq: given trace}) 
making up the high-gradient operator.
In other words,
the RG transformation law~(\ref{eq: master formula, Kahler})
is triangular with respect to the quantum number%
~(\ref{eq: def quantum number}).

\subsection{
Symmetry class A
           }

We begin with symmetry class A. This case ($\vartheta=0$)
was already treated in 
Refs.~\onlinecite{Lerner90,Wegner91,Mall93}
and is needed for symmetry classes C ($\vartheta=-1$) and D ($\vartheta=+1$). 
Consider any string
$
L^{\ }_{1}
A^{ \tau}_{\sigma^{\ }_{1}}
A^{-\tau}_{\sigma^{\prime}_{1}}
R^{\ }_{1}
$
and any string
$
L^{\ }_{2}
A^{ \tau}_{\sigma^{\ }_{2}}
A^{-\tau}_{\sigma^{\prime}_{2}}
R^{\ }_{2}
$
of 4 matrices each that enter a high-gradient operator.

First, we assume that each string belongs to two distinct traces.
We seek the one-loop RG transformation law obeyed by the number of
switches 
\begin{subequations}
\begin{eqnarray}
\delta^{\ }_{\sigma^{\ }_{1},-\sigma^{\prime}_{1}}
+
\delta^{\ }_{\sigma^{\ }_{2},-\sigma^{\prime}_{2}}
\label{eq: number of switces for tr tr before RG}
\end{eqnarray}
contained in 
\begin{eqnarray}
\mathrm{tr}\,
\left(
L^{\ }_{1}
A^{ \tau}_{\sigma^{\     }_{1}}
A^{-\tau}_{\sigma^{\prime}_{1}}
R^{\ }_{1}
\right)
\mathrm{tr}\,
\left(
L^{\ }_{2}
A^{ \tau}_{\sigma^{\ }_{2}}
A^{-\tau}_{\sigma^{\prime}_{2}}
R^{\ }_{2}
\right).
\label{eq: intertrace 2 string}
\end{eqnarray}
\end{subequations}
By making use of Eq.~(\ref{eq: RG transformation, inter-trace})
with $\vartheta=0$ we deduce that the
one-loop RG transformation law obeyed by 
Eq.~(\ref{eq: intertrace 2 string}) is
\begin{subequations}
\label{eq: RG trsf tr string tr string}
\begin{eqnarray}
&&
\left\langle
\mathrm{tr}
\left(
\left[
A^{ \tau}_{\sigma^{\     }_{1}}
A^{-\tau}_{\sigma^{\prime}_{1}}
\right]^{\ }_{\zeta^1}\!
R^{\ }_{1}
L^{\ }_{1}
\right)
\mathrm{tr}
\left(
\left[
A^{ \tau}_{\sigma^{\     }_{2}}
A^{-\tau}_{\sigma^{\prime}_{2}}
\right]^{\ }_{\zeta^1}\!
R^{\ }_{2}
L^{\ }_{2}
\right)
\right\rangle=
\nonumber\\
&&
t{d}l\left(
-\delta^{\ }_{\sigma_{1},-\sigma_{2}}
+\delta^{\ }_{\sigma_{1},-\sigma_{2}^{\prime}}
+\delta^{\ }_{\sigma_{1}^{\prime},-\sigma_{2}}
-\delta^{\ }_{\sigma_{1}^{\prime},-\sigma_{2}^{\prime}}
\right)
\nonumber\\
&&
\hphantom{\delta^{\ }_{\sigma_{1},-\sigma_{2}^{\prime}}}
\times
\mathrm{tr}
\left(
A^{+\tau}_{\sigma^{\     }_{1}}
A^{-\tau}_{\sigma^{\prime}_{2}}
R^{\ }_{2}
L^{\ }_{2}
A^{+\tau}_{\sigma^{\     }_{2}}
A^{-\tau}_{\sigma^{\prime}_{1}}
R^{\ }_{1}
L^{\ }_{1}
\right)
\label{eq: RG trsf tr string tr string a}
\\
&&
-t{d}l\delta^{\ }_{\sigma^{\     }_{1},-\sigma^{\     }_{2}}
\mathrm{tr}
\left(
A^{+\tau}_{\sigma^{\     }_{2}}
A^{-\tau}_{\sigma^{\prime}_{2}}
R^{\ }_{2}
L^{\ }_{2}
A^{+\tau}_{\sigma^{\     }_{1}}
A^{-\tau}_{\sigma^{\prime}_{1}}
R^{\ }_{1}
L^{\ }_{1}
\right)
\label{eq: RG trsf tr string tr string b}
\\
&&
+t{d}l\delta^{\ }_{\sigma^{\     }_{1},-\sigma^{\prime}_{2}}
\mathrm{tr}
\left(
R^{\ }_{2}
L^{\ }_{2}
A^{+\tau}_{\sigma^{\     }_{2}}
A^{-\tau}_{\sigma^{\prime}_{2}}
A^{+\tau}_{\sigma^{\     }_{1}}
A^{-\tau}_{\sigma^{\prime}_{1}}
R^{\ }_{1}
L^{\ }_{1}
\right)
\label{eq: RG trsf tr string tr string c}
\\
&&
+
t{d}l\delta^{\ }_{\sigma^{\prime}_{1},-\sigma^{\     }_{2}}
\mathrm{tr}
\left(
A^{+\tau}_{\sigma^{\     }_{1}}
A^{-\tau}_{\sigma^{\prime}_{1}}
A^{+\tau}_{\sigma^{\     }_{2}}
A^{-\tau}_{\sigma^{\prime}_{2}}
R^{\ }_{2}
L^{\ }_{2}
R^{\ }_{1}
L^{\ }_{1}
\right)
\label{eq: RG trsf tr string tr string d}
\\
&&
-
t{d}l\delta^{\ }_{\sigma^{\prime}_{1},-\sigma^{\prime}_{2}}
\mathrm{tr}
\left(
A^{+\tau}_{\sigma^{\     }_{1}}
A^{-\tau}_{\sigma^{\prime}_{1}}
R^{\ }_{2}
L^{\ }_{2}
A^{+\tau}_{\sigma^{\     }_{2}}
A^{-\tau}_{\sigma^{\prime}_{2}}
R^{\ }_{1}
L^{\ }_{1}
\right).
\label{eq: RG trsf tr string tr string e}
\end{eqnarray}
\end{subequations}
Line~(\ref{eq: RG trsf tr string tr string a})
contains 
$
\delta^{\ }_{\sigma^{\ }_{1},-\sigma^{\prime}_{2}}
+
\delta^{\ }_{\sigma^{\ }_{2},-\sigma^{\prime}_{1}}
$
switches.
Lines%
~(\ref{eq: RG trsf tr string tr string b})
to%
~(\ref{eq: RG trsf tr string tr string e})
contain
$
\delta^{\ }_{\sigma^{\ }_{1},-\sigma^{\prime}_{1}}
+
\delta^{\ }_{\sigma^{\ }_{2},-\sigma^{\prime}_{2}}
$
switches. Hence, only
line~(\ref{eq: RG trsf tr string tr string a})
has the potential to increase the initial number%
~(\ref{eq: number of switces for tr tr before RG})
of switches. However, an increase over the
initial number%
~(\ref{eq: number of switces for tr tr before RG})
of switches is not possible in view of the identity
$
-\delta^{\ }_{\sigma^{\     }_{1},-\sigma^{\     }_{2}}
+\delta^{\ }_{\sigma^{\     }_{1},-\sigma^{\prime}_{2}}
+\delta^{\ }_{\sigma^{\prime}_{1},-\sigma^{\     }_{2}}
-\delta^{\ }_{\sigma^{\prime}_{1},-\sigma^{\prime}_{2}}
=
2\sigma^{\ }_{1}\sigma^{\ }_{2}
\delta^{\ }_{\sigma^{\ }_{1},-\sigma^{\prime}_{1}}
\delta^{\ }_{\sigma^{\ }_{2},-\sigma^{\prime}_{2}}
$.
Line~(\ref{eq: RG trsf tr string tr string a})
only contributes if
$\sigma^{\prime}_{1}=-\sigma^{\ }_{1}$
and
$\sigma^{\prime}_{2}=-\sigma^{\ }_{2}$
both hold, i.e., when the initial number of switches%
~(\ref{eq: number of switces for tr tr before RG})
is maximum (2). Consequently, the number of switches 
$2\delta^{\ }_{\sigma^{\ }_{1},\sigma^{\ }_{2}}$
induced by line~(\ref{eq: RG trsf tr string tr string a}) 
equals the initial number of switches 2
when $\sigma^{\ }_{1}=\sigma^{\ }_{2}$ 
while it decreases otherwise.

Second, we assume that each string belongs to the same trace.
We seek the one-loop RG transformation law obeyed by the number of
switches 
\begin{subequations}
\begin{eqnarray}
\delta^{\ }_{\sigma^{\ }_{1},-\sigma^{\prime}_{1}}
+
\delta^{\ }_{\sigma^{\ }_{2},-\sigma^{\prime}_{2}}
\label{eq: number of switces for tr before RG}
\end{eqnarray}
contained in 
\begin{eqnarray}
\mathrm{tr}\,
\left(
L^{\ }_{1}
A^{ \tau}_{\sigma^{\     }_{1}}
A^{-\tau}_{\sigma^{\prime}_{1}}
R^{\ }_{1}
L^{\ }_{2}
A^{ \tau}_{\sigma^{\ }_{2}}
A^{-\tau}_{\sigma^{\prime}_{2}}
R^{\ }_{2}
\right).
\label{eq: intratrace 2 string}
\end{eqnarray}
\end{subequations}
By making use of Eq.~(\ref{eq: RG transformation, intra-trace})
with $\vartheta=0$
one verifies again that the number of switches $n^{\tau}_{\sigma}$
cannot increase after integrating over the fast modes up to one loop.

We have proven that
the RG flows of high-gradient operators made of
any given number of switches decouple from the RG flows of high-gradient
operators made of a lesser number of switches in symmetry class A. 
As a corollary, the mixing matrix~(\ref{eq: def mixing matrix})
is triangular in symmetry class A.

\subsection{
Symmetry classes C and D
           }

We turn our attention to symmetry classes $|\vartheta|=1$,
i.e., $\mathrm{SO}(2N)/\mathrm{U}(N)$ 
and $\mathrm{Sp}(N)/\mathrm{U}(N)$.
We first consider the inter-trace formula of the RG transformation%
~(\ref{eq: RG transformation, inter-trace})
applied to
\begin{eqnarray}
\cdots
\times
\mathrm{tr}\,
\left(
A^{ \tau}_{\sigma^{\ }_{1}}
A^{-\tau}_{\sigma^{\prime}_{1}}
\mathcal{M}
\right)
\times
\mathrm{tr}\,
\left(
A^{ \tau}_{\sigma^{\ }_{2}}
A^{-\tau}_{\sigma^{\prime}_{2}}
\mathcal{N}
\right)
\times
\cdots
\end{eqnarray}
where $\mathcal{M}:=R^{\ }_1 L^{\ }_1$ 
and $\mathcal{N}:=R^{\ }_2 L^{\ }_1$.
It is sufficient to consider the contributions
proportional to $\vartheta$
in
Eq.~(\ref{eq: RG transformation, inter-trace})
as the remaining terms can be treated along the lines of 
symmetry class A.
There are four such contributions $(\mathrm{a})-({d})$:
\begin{subequations}
\label{eq: trsf law mathcal of A if classes D and C}
\begin{eqnarray}
&&
\delta^{\ }_{\sigma_{2},-\sigma_{2}^{\prime}}
\big(
\delta^{\ }_{\sigma_{1},\sigma_{2}} 
-
\delta^{\ }_{\sigma_{1},-\sigma_{2}}
\big)
\mathrm{tr}
(
A^{+\tau}_{\sigma_{2}} 
\mathcal{N}^{\mathrm{T}} 
A^{-\tau}_{\sigma_{2}^{\prime}} 
A^{+\tau}_{\sigma_{1}} 
A^{-\tau}_{\sigma_{1}^{\prime}} 
\mathcal{M}
),
\nonumber\\
&&
\label{eq: trsf law mathcal of A if classes D and C a}
\\
&&
\delta^{\ }_{\sigma_{1},-\sigma_{1}^{\prime}}
\big(
\delta^{\ }_{\sigma_{1},\sigma_{2}}
-
\delta^{\ }_{\sigma_{1},-\sigma_{2}} 
\big)
\mathrm{tr}
(
A^{+\tau}_{\sigma_{1}} 
\mathcal{N}^{\mathrm{T}} 
A^{-\tau}_{\sigma_{2}^{\prime}} 
A^{+\tau}_{\sigma_{2}} 
A^{-\tau}_{\sigma_{1}^{\prime}} 
\mathcal{M}
),
\nonumber\\
&&
\\
&&
\delta^{\ }_{\sigma_{1},-\sigma_{1}^{\prime}}
\big(
\delta^{\ }_{\sigma_{1}^{\prime},\sigma_{2}^{\prime}}
-
\delta^{\ }_{\sigma_{1}^{\prime},-\sigma_{2}^{\prime}}
\big)
\mathrm{tr}
( 
A^{-\tau}_{\sigma_{2}^{\prime}} 
A_{\sigma_{2}}^{+\tau} 
\mathcal{N}^{\mathrm{T}} 
A_{\sigma_{1}^{\prime}}^{-\tau} 
\mathcal{M}
A^{+\tau}_{\sigma_{1}}
), 
\nonumber\\
&&
\\
&&
\delta^{\ }_{\sigma_{2},-\sigma_{2}^{\prime}}
\big(
\delta^{\ }_{\sigma_{1}^{\prime},\sigma_{2}^{\prime}}
-
\delta^{\ }_{\sigma_{1}^{\prime},-\sigma_{2}^{\prime}}
\big)
\mathrm{tr}
(
A^{-\tau}_{\sigma_{1}^{\prime}} 
A^{+\tau}_{\sigma_{2}} 
\mathcal{N}^{\mathrm{T}} 
A_{\sigma_{2}^{\prime}}^{-\tau}
\mathcal{M} 
A_{\sigma_{1}}^{+\tau}
).
\nonumber\\
&&
\end{eqnarray}
\end{subequations}

We assign to 
\begin{eqnarray}
\mathcal{A}:=
\mathrm{tr}\,
(A^{+\tau}_{\sigma_{1}}A^{-\tau}_{\sigma_{1}^{\prime}}\mathcal{M})
\times
\mathrm{tr}\,
(A^{+\tau}_{\sigma_{2}}A^{-\tau}_{\sigma_{2}^{\prime}}\mathcal{N})
\end{eqnarray}
the quartet of numbers 
\begin{eqnarray}
(n_{+}^{+},n_{-}^{+},n_{+}^{-},n_{-}^{-})
\end{eqnarray}
where
$n^{\tau}_{\sigma}$
counts the total number of switches of type
$A^{\tau}_{\sigma}A^{-\tau}_{-\sigma}$
in $\mathcal{A}$.
In symmetry classes D and C, we can combine
$\mathrm{tr}\, \mathcal{O}=\mathrm{tr}\,\mathcal{O}^{\mathrm{T}}$ 
and the symmetry
$
(A_{\mu}^{\pm })^{\mathrm{T}}
=
-\vartheta A_{\mu}^{\pm }
$,
to infer that
\begin{eqnarray}
(n_{+}^{+},n_{-}^{+},n_{+}^{-},n_{-}^{-})=
(n_{-}^{-},n_{+}^{-},n_{-}^{+},n_{+}^{+}).
\end{eqnarray}
As opposed to symmetry class A, individual switches 
$n^{\tau}_{\sigma}$ are not separately conserved
but the total number of switches 
\begin{eqnarray}
n=\sum_{\sigma,\tau}n_{\tau}^{\sigma}
\end{eqnarray}
will be shown to be conserved.

Without loss of generality we choose $\tau=+$ in $\mathcal{A}$.
The total number of switches $n^{+}_{+}$ and 
$n^{+}_{-}$ are
\begin{subequations}
\label{eq: computing npmpm for mathcal{A} before RG}
\begin{eqnarray}
n_{+}^{+}&=&
n_{+}^{+}[A^{+}_{\sigma_{1}}A^{-}_{\sigma_{1}^{\prime}}]
+
n^{+}_{+}[\mathcal{M}]
+
n_{+}^{+}[A^{+}_{\sigma_{2}}A^{-}_{\sigma_{2}^{\prime}}]
+
n^{+}_{+}[\mathcal{N}],
\nonumber\\
&&\\
n_{-}^{+}&=&
n_{-}^{+}[A^{+}_{\sigma_{1}}A^{-}_{\sigma_{1}^{\prime}}]
+
n^{+}_{-}[\mathcal{M}]
+
n_{-}^{+}[A^{+}_{\sigma_{2}}A^{-}_{\sigma_{2}^{\prime}}]
+
n^{+}_{-}[\mathcal{N}],
\nonumber\\
\end{eqnarray}
respectively.
The total number of switches $n^{-}_{+}$ and $n^{-}_{-}$ 
follow from using
$
\mathcal{A}=
\mathrm{tr}\,
(A^{-}_{\sigma_{1}^{\prime}}\mathcal{M}A^{+}_{\sigma_{1}})
\times
\mathrm{tr}\,
(A^{-}_{\sigma_{2}^{\prime}}\mathcal{N}A^{+}_{\sigma_{2}})
$
and are given by
\begin{eqnarray}
n_{+}^{-}&=&
n^{-}_{+}[A^{-}_{\sigma_{1}^{\prime}}\mathcal{M}A^{+}_{\sigma_{1}}]
+
n^{-}_{+}[A^{-}_{\sigma_{2}^{\prime}}\mathcal{N}A^{+}_{\sigma_{2}}],
\nonumber\\
&&\\
n_{-}^{-}&=&
n^{-}_{-}[A^{-}_{\sigma_{1}^{\prime}}\mathcal{M}A^{+}_{\sigma_{1}}]
+
n^{-}_{-}[A^{-}_{\sigma_{2}^{\prime}}\mathcal{N}A^{+}_{\sigma_{2}}],
\nonumber\\
&&
\end{eqnarray}
\end{subequations}
respectively.

The number of switches~(\ref{eq: computing npmpm for mathcal{A} before RG})
in $\mathcal{A}$ changes under the RG transformation%
~(\ref{eq: trsf law mathcal of A if classes D and C})
as follows.
The total number of switches $n^{+}_{+}$ and $n^{+}_{-}$ 
from line~(\ref{eq: trsf law mathcal of A if classes D and C a})
is given by
\begin{subequations}
\begin{eqnarray}
n_{+}^{+}[(a)]&=&
n^{+}_{+}
[A^{+}_{\sigma_{2}} \mathcal{N}^{\mathrm{T}} A^{-}_{\sigma_{2}^{\prime}}]
+
n^{+}_{+}[A^{+}_{\sigma_{1}} A^{-}_{\sigma_{1}^{\prime}}]
+
n^{+}_{+}[\mathcal{M}]
\nonumber\\
&=&
n^{-}_{-}[A^{-}_{\sigma_{2}^{\prime}} \mathcal{N} A^{+}_{\sigma_{2}}]
+
n^{+}_{+}[A^{+}_{\sigma_{1}} A^{-}_{\sigma_{1}^{\prime}}]
+
n^{+}_{+}[\mathcal{M}],
\nonumber\\
&&\\
n_{-}^{+}[(a)]&=&
n^{+}_{-}[A^{+}_{\sigma_{2}} \mathcal{N}^{\mathrm{T}} 
A^{-}_{\sigma_{2}^{\prime}}]
+
n^{+}_{-}[A^{+}_{\sigma_{1}} A^{-}_{\sigma_{1}^{\prime}}]
+
n^{+}_{-}[\mathcal{M}]
\nonumber\\
&=&
n^{-}_{+}[A^{-}_{\sigma_{2}^{\prime}} \mathcal{N} A^{+}_{\sigma_{2}}]
+
n^{+}_{-}[A^{+}_{\sigma_{1}} A^{-}_{\sigma_{1}^{\prime}}]
+
n^{+}_{-}[\mathcal{M}],
\nonumber\\
&&
\end{eqnarray}
respectively,
where we again made use of the symmetry condition
$
(A_{\mu}^{\pm })^{\mathrm{T}}=
-\vartheta A_{\mu}^{\pm }
$,
and the fact that the number of $A_{\mu}^{\pm}$ in a trace is always even.
Similarly, the total number of switches $n^{-}_{+}$ and $n^{-}_{-}$ 
from line~(\ref{eq: trsf law mathcal of A if classes D and C a})
is given by
\begin{eqnarray}
n^{-}_{+}[(a)]&=&
n^{-}_{+}[\mathcal{N}^{\mathrm{T}}]
+
n^{-}_{+}[A^{-}_{\sigma_{2}^{\prime}} A^{+}_{\sigma_{1}} ]
+
n^{-}_{+}[A^{-}_{\sigma_{1}^{\prime}} \mathcal{M}A^{+}_{\sigma_{2}}]
\nonumber \\
&=&
n^{+}_{-}[\mathcal{N}]
+
n^{-}_{+}[A^{-}_{\sigma_{2}^{\prime}} A^{+}_{\sigma_{1}} ]
+
n^{-}_{+}[A^{-}_{\sigma_{1}^{\prime}} \mathcal{M}A^{+}_{\sigma_{2}}],
\nonumber \\
&&\\
n^{-}_{-}[(a)]&=&
(\mbox{as above with $\sigma\to -\sigma$}),
\end{eqnarray}
\end{subequations}
respectively. Thus, the total number of switches changes from
\begin{eqnarray}
n&=&
\sum_{\sigma}
\Big\{
 n_{\sigma}^{+}[A^{+}_{\sigma_{1}}A^{-}_{\sigma_{1}^{\prime}}]
+n^{+}_{\sigma}[\mathcal{M}]
+n_{\sigma}^{+}[A^{+}_{\sigma_{2}}A^{-}_{\sigma_{2}^{\prime}}]
\nonumber\\
&&
+n^{+}_{\sigma}[\mathcal{N}]
+n^{-}_{\sigma}[A^{-}_{\sigma_{1}^{\prime}}\mathcal{M}A^{+}_{\sigma_{1}}]
+
n^{-}_{\sigma}[A^{-}_{\sigma_{2}^{\prime}}\mathcal{N}A^{+}_{\sigma_{2}}]
\Big\}
\nonumber\\
&&
\end{eqnarray}
to
\begin{eqnarray}
n[(a)]&=&
\sum_{\sigma}
\Big\{
n^{-}_{-\sigma}[A_{\sigma_{2}^{\prime}}^{-} \mathcal{N} A_{\sigma_{2}}^{+}]
+
n^{+}_{\sigma}[A^{+}_{\sigma_{1}}A^{-}_{\sigma_{1}^{\prime}}]
+
n^{+}_{\sigma}[\mathcal{M}]
\nonumber\\
&&
+
n^{+}_{-\sigma}[\mathcal{N}]
+
n^{-}_{\sigma}[A^{-}_{\sigma_{2}^{\prime}} A^{+}_{\sigma_{1}} ]
+
n^{-}_{\sigma}[A^{-}_{\sigma_{1}^{\prime}} \mathcal{M}A^{+}_{\sigma_{2}}]
\Big\}
\nonumber\\
&&
\end{eqnarray}
for the contribution from 
line~(\ref{eq: trsf law mathcal of A if classes D and C a}).
The net change 
$\Delta n|_{(a)}=n-n[(a)]$ is
\begin{eqnarray}
\Delta n|_{(a)}&=&
\sum_{\sigma}
\Big\{
n_{\sigma}^{+}[A^{+}_{\sigma_{2}}A^{-}_{\sigma_{2}^{\prime}}]
-
n^{-}_{\sigma}[A^{-}_{\sigma_{2}^{\prime}} A^{+}_{\sigma_{1}} ]
\nonumber\\
&&
+
n^{-}_{\sigma}[A^{-}_{\sigma_{1}^{\prime}}\mathcal{M}A^{+}_{\sigma_{1}}]
-
n^{-}_{\sigma}[A^{-}_{\sigma_{1}^{\prime}} \mathcal{M}A^{+}_{\sigma_{2}}]
\Big\}.
\nonumber\\
&&
\end{eqnarray}
When
$\delta^{\ }_{\sigma_{2},-\sigma_{2}^{\prime}}
\delta^{\ }_{\sigma_{1},\sigma_{2}} =1 $,
we find
\begin{eqnarray}
\Delta n|_{(a)}&=&
\sum_{\sigma}
\Big\{
n^{+}_{\sigma}[A^{+}_{\sigma_{2}}A^{-}_{\sigma_{2}^{\prime}}]
-
n^{-}_{\sigma}[A^{-}_{\sigma_{2}^{\prime}} A^{+}_{\sigma_{2}} ]
\nonumber\\
&&
+
n^{-}_{\sigma}[A^{-}_{\sigma_{1}^{\prime}}\mathcal{M}A^{+}_{\sigma_{1}}]
-
n^{-}_{\sigma}[A^{-}_{\sigma_{1}^{\prime}} \mathcal{M}A^{+}_{\sigma_{1}}]
\Big\}
\nonumber\\
&=&
0
\end{eqnarray}
with the help of the prefactor
$
\delta^{\ }_{\sigma_{2},-\sigma_{2}^{\prime}}\big(
\delta^{\ }_{\sigma_{1},\sigma_{2}} -\delta^{\ }_{\sigma_{1},-\sigma_{2}}
\big)
$.
When
$\delta^{\ }_{\sigma_{2},-\sigma_{2}^{\prime}}
\delta^{\ }_{\sigma_{1},-\sigma_{2}}=1$,
we find
\begin{eqnarray}
\Delta n|_{(a)}&=&
\sum_{\sigma}
\Big\{
n_{\sigma}^{+}[A^{+}_{\sigma_{2}}A^{-}_{\sigma_{2}^{\prime}}]
-
n^{-}_{\sigma}[A^{-}_{\sigma_{2}^{\prime}} A^{+}_{\sigma_{2}^{\prime}} ]
\nonumber\\
&&
+
n^{-}_{\sigma}[A^{-}_{\sigma_{1}^{\prime}}\mathcal{M}A^{+}_{\sigma_{1}}]
-
n^{-}_{\sigma}[A^{-}_{\sigma_{1}^{\prime}} \mathcal{M}A^{+}_{-\sigma_{1}}]
\Big\}
\nonumber\\
&=&
1
+
\sum_{\sigma}
\Big\{
n^{-}_{\sigma}[A^{-}_{\sigma_{1}^{\prime}}\mathcal{M}A^{+}_{\sigma_{1}}]
-
n^{-}_{\sigma}[A^{-}_{\sigma_{1}^{\prime}} \mathcal{M}A^{+}_{-\sigma_{1}}]
\Big\}
\nonumber\\
&\ge&  0.
\end{eqnarray}
We conclude that the total number of switches 
does not increase under the RG 
transformation~(\ref{eq: trsf law mathcal of A if classes D and C a}).

Repeating the same analysis for the contributions $(b)-(d)$
in Eq.~(\ref{eq: trsf law mathcal of A if classes D and C})
yields that the change in the total number of switches is $\ge 0$.

We leave it to the reader to verify that the same conclusion holds for
the intra-trace formula~(\ref{eq: RG transformation, intra-trace}).

This completes the proof for the upper triangular structure of the RG equation 
with respect to the total number of switches in symmetry classes D and C.

\section{
RG flows of high-gradient operators for NL$\sigma$M on supermanifolds
        }
\label{app: RG of high-gradient operators of NLSMs on supermanifolds}

We derive in this appendix two ``master formulae" 
needed to perform the one-loop 
RG analysis of high-gradient operators for
NL$\sigma$M defined on an arbitrary Riemannian or
K\"ahlerian supermanifold.
Master formulae 
~(\ref{eq: master formulae if superriemann})
and%
~(\ref{eq: master formulae if superkahler})
are supersymmetric versions of 
Eq.~(\ref{eq: master formula, Kahler}) and
can be used for any target supermanifold once
the geometrical data of the manifold are known.
An expansion in terms of normal coordinate of tensor fields plays
again an essential role.
For the non-supersymmetric Riemannian case,
an expansion in terms of Riemann normal coordinate (RNC)
based on the notion of geodesics
was used for the computation of the beta function
in all generality%
~\cite{Friedan85,Alvarez-Gaume81}
and the RG analysis of high-gradient operators 
for the special case of the O($N$)$/$O($N-1$) NL$\sigma$M%
~\cite{Castilla93,Castilla97}.
The RNC expansion can also be applied to Hermitian manifolds.
For the (non-supersymmetric) K\"ahlerian case, however,
one can also use the K\"ahler potential 
to find a normal coordinate system,
the K\"ahler normal coordinates (KNC) 
introduced in Refs.\ \onlinecite{Higashijima00}
and
\onlinecite{Higashijima02a}.
In this Appendix,
we will generalize these normal coordinate expansions
to the case of supersymmetric target manifolds.
All the results that follow
reduce to the non-supersymmetric ones
if we ignore the commuting-anticommuting nature of the objects
and fermionic sign factors.

\subsection{
NL$\sigma$M on Riemannian supermanifolds
           }

\paragraph{%
Notations and conventions for Riemannian supermanifolds%
          }

We begin with a quick review of concepts and notation. 
We refer the reader 
to Ref.\ \onlinecite{DeWitt92}
for a textbook on the  geometry of supermanifolds.

\begin{itemize}
\item
A tensor field of rank $(r,s)$ on a supermanifold
has $r$ contravariant (upper) and $s$ covariant (lower)
indices, $T^{a}{}^{b}{}_{c}{}^{d}{}_{\ldots}$.
On a supermanifold, 
components of tensor fields can be either
commuting or anti-commuting supernumbers.

\item
In order to define their mutual statistics,
the grade $\epsilon (\mathcal{O})=0,1$ is assigned to each object
$\mathcal{O}$ on a supermanifold, 
where $\mathcal{O}$ can be a supernumber, a supertensor field,   
a component of a supertensor field, etc.
When the ordering of two objects $\mathcal{O}_1$ and $\mathcal{O}_2$
is exchanged, a factor 
$(-1)^{\epsilon(\mathcal{O}_1) \epsilon (\mathcal{O}_2)}$ 
arises,
$
\mathcal{O}_1 \mathcal{O}_2 = 
(-1)^{\epsilon(\mathcal{O}_1) \epsilon (\mathcal{O}_2)}
\mathcal{O}_2 \mathcal{O}_1.
$
In order to define the grade 
for the component of a tensor field of arbitrary rank,
we first assign a grade to each of its index.
If index $a$ corresponds to a commuting component of
the coordinates of a point on the supermanifold
$\epsilon(a)=0$, while $\epsilon(a)=1$ otherwise.
For simplicity, a statistical factor
$(-1)^{\epsilon(a) \epsilon (b)}$
is often abbreviated by
$(-1)^{ a b }$, e.g.,
$
X^a Y^b = (-1)^{ab}Y^b X^a
$.
The grade of the component 
$T^{a}{}^{b}{}_{c}{}^{d}{}_{\ldots}$
of the tensor field $T$ is then
$
\epsilon (T^{a}{}^{b}{}_{c}{}^{d}{}_{\ldots})
=
\epsilon(a)+\epsilon(b)+\epsilon(c)+\epsilon(d)+\cdots
$.%
~\cite{Rem: grade of the components of tensor fields}

\item
It is convenient to introduce a shifting rule for 
the left most index of tensors,
\begin{subequations}
\begin{eqnarray}
{}^{a}T^{b\ldots}{}_{c\ldots}
&:=&
T^{ab\ldots}{}_{c\ldots},
\\
{}_{a}T^{b\ldots}{}_{c\ldots}
&:=&
(-1)^{a}
T_{a}{}^{b\ldots}{}_{c\ldots}.
\label{eq: shifting rule for the left most indices}
\end{eqnarray}
\end{subequations}

\item
As with all other objects on a supermanifold,
derivatives also carry a grade.
Thus, derivatives from the left $\overrightarrow{\partial}_{a}$
and the right $\overleftarrow{\partial}_{a}$
need to be distinguished. They are related by
\begin{eqnarray}
\overrightarrow{\partial}_{a}
f
=(-1)^{a\epsilon(f)+a}f
\overleftarrow{\partial}_{a},
\label{eq: left <-> right derivatives}
\end{eqnarray}
where $f$ is a function over the manifold.
In this appendix we will mainly use
derivatives $\overleftarrow{\partial}_{a}$
and 
covariant derivatives $\overleftarrow{\nabla}_{a}$
that act from the right.

\item
Super transposition of matrices, denoted by ${}^{\sim}$,
is defined by
\begin{subequations}
\begin{eqnarray}
{}^{i}K^{\sim}{}_{j}^{\ }
&=&
(-1)^{j(i+j)}\,
{}_{j}^{\ }K^{i},
\\
{}_{j}^{\ }M^{\sim}{}_{i}^{\ }
&=&
(-1)^{i+j+ij}\,
{}_{i}^{\ }M{}_{j}^{\ },
\\
{}^{j} N^{\sim}{}^{j}
&=&
(-1)^{ij}\,
{}^{i}N^{j}.
\end{eqnarray}
\end{subequations}

\end{itemize}

For illustration, we shall apply these rules to
the components $g^{\ }_{ab}$ of the metric tensor field,
and components $g^{ab}$ of the inverse of the metric tensor field.
From the shifting rule
\begin{subequations}
\begin{eqnarray}
{}^{a}g^{b}
=
g^{ab},
&&
\qquad
{}^{\ }_{a}g_{b}^{\ }
=
(-1)^{a}\,
g_{ab}^{\ },
\\
{}^{\ }_{a}g^{\ }_{c}\,
{}^{c}g^{b}
=
{}^{\ }_{a}\delta^{b},
&&
\qquad
{}^{a}g^{c}\,
{}^{\ }_{c}g^{\ }_{b}
=
{}^{a}\delta^{\ }_{b}.
\end{eqnarray}
\end{subequations}
The supersymmetry of the metric tensor field implies
\begin{subequations}
\label{eq: supersymmetry of metric}
\begin{eqnarray}
{}^{\ }_{a}g^{\ }_{b}&=&
(-1)^{a}\,
g_{ab}^{\ }
\nonumber\\
&=&
(-1)^{a+ab}\,
g_{ba}^{\ }
\nonumber\\
&=&
(-1)^{a+b+ab}\,{}^{\ }_{b}g^{\ }_{a},
\label{eq: supersymmetry of metric a}
\\
{}^{a}g^{b}&=&
g^{ab}
\nonumber\\
&=&
(-1)^{ab}g^{ba}
\nonumber\\
&=&
(-1)^{ab}\,
{}^{b}g^{a}.
\label{eq: supersymmetry of metric b}
\end{eqnarray}
\end{subequations}
An index of a component $X^a$ of a supervector field
can be raised, lowered, and shifted according to the rules
\begin{subequations}
\begin{eqnarray}
{}^{a}X=X^{a},
&&
\qquad
{}^{\ }_{a}X
=
(-1)^{a}\,
X^{\ }_{a},
\\
{}_{a}^{\ }X
=
{}^{\ }_{a}g_{b}^{\ }\, {}^{b}X,
&&
\qquad
X_{a}=
X^{b}\,
{}^{\ }_{b}g_{a}^{\ },
\\
{}^{a}_{\ }X
=
{}_{\ }^{a}g^{b}_{\ }\, {}^{\ }_{b}X,
&&
\qquad
X^{a}=
X^{\ }_{b}\,
{}_{\ }^{b}g^{a}_{\ }.
\end{eqnarray}
\end{subequations}

\paragraph{%
The NL$\sigma$M on a Riemannian supermanifold%
          }

The NL$\sigma$M on a Riemannian supermanifold
is defined by the partition function
\begin{subequations}
\begin{eqnarray}
Z&:=&
\int
\mathcal{D}[\phi]\,
e^{-S[\phi]},
\\
S[\phi]&:=&
\frac{1}{4\pi t}
\int_r
\partial^{\ }_{\mu}\phi^{a}\,
{}_{a}^{\ }g^{\ }_{b}(\phi)\,\,
\partial^{\ }_{\mu} {}^{b}\phi,
\end{eqnarray}
\end{subequations}
where 
$
\big(\phi^{a}(r)\big)\in
\mathbb{R}_{\mathrm{c}}^{M}
\times 
\mathbb{R}^{N}_{\mathrm{a}}
$
represents the $M+N$ components of the coordinates of point $r$
on the target manifold, 
with $M$ refering to the number of commuting coordinates
whereas $N$ that of the number of anti-commuting ones,
\cite{DeWitt92}
and $t$ is the NL$\sigma$M
coupling constant.

\paragraph{
The covariant background field method
and the super Riemann normal coordinate expansion
          }

In the background field method,
the covariant vector field $\phi^{a}(r)$
is separated into two parts,
\begin{eqnarray}
\phi^{a}(r)&=&
\psi^{a}(r)
+
\pi^{a}(r),
\end{eqnarray}
whereby $\psi^{a}(r)$ is assumed to be a 
slowly varying solution to the classical
equations of motion
which transforms like a contravariant vector
and $\pi^{a}(r)$ 
represents fluctuations around 
the slow degrees of freedom $\psi^{a}(r)$.

Two steps are needed
to perform the one-loop RG program 
in a covariant fashion.
First, we need to trade the expansion 
in terms of $\pi^{a}$, which is not covariant,
for that of fields $\xi^{a}$ 
that transform like contravariant vectors.
Second, we need to rotate the RNC by a vielbein
to evaluate the functional integral over the fast modes.

The RNC is based on the notion of geodesics on Riemannian manifolds.
Since almost all the notions on a Riemannian manifold
such as forms, connection, the Riemann tensor, geodesics, etc.,
have their counterpart on a Riemannian supermanifold
\cite{DeWitt92},
it is straightforward to develop the super extension of the
RNC expansion.
The expansion of a covariant tensor field $T$ in terms of the
super RNC $\xi^{a}$ is given by
\begin{widetext}
\begin{eqnarray}
T_{b^{\ }_{1}\cdots b^{\ }_{l}}^{\ }(\psi+\pi)
=
T_{b^{\ }_{1}\cdots b^{\ }_{l}}^{\ }(\psi)
\,+\,
T^{\ }_{b^{\ }_{1}\cdots b^{\ }_{l}}
\overleftarrow{\nabla}_{c}(\psi)\,
\xi^{c}
+
\frac{1}{2!}
\Bigg[
T_{b^{\ }_{1}\cdots b^{\ }_{l}}
\overleftarrow{\nabla}_{c^{\ }_{1}}
\overleftarrow{\nabla}_{c^{\ }_{2}}
\hphantom{AAAAAAAAAAAAAAAAAAAAAAAA}
&&
\nonumber\\
-
\frac{1}{3}
\sum_{i=1}^{l}
T^{\ }_{
b^{\ }_{1}\cdots b^{\ }_{i-1}bb^{\ }_{i+1}\cdots b^{\ }_{l}
                  }
(-1)^{ (b_i+b)(b_{i+1}+\cdots + b_{l})}
\left(
R^{b}_{\ b^{\ }_{i}c^{\ }_{1}c^{\ }_{2}}
+
(-1)^{b_{i}c_{1}}
R^{b}_{\ c^{\ }_{1}b^{\ }_{i}c^{\ }_{2}}
\right)
\Bigg]
\xi^{c_2}\xi^{c_1}
+
\cdots.
\hphantom{AAA}
&&
\label{eq: super RNC for T}
\end{eqnarray}
The RNC expansion for $\partial_{\mu} (\psi^{a}+\pi^{a})$ is given by
\begin{eqnarray}
\partial_{\mu} (\psi^{a}+\pi^{a}) 
&=&
\partial_{\mu}\psi^{a}
+
\xi^{a} \overleftarrow{D}_{\mu}
+
\frac{1}{3}
R^{a}{}_{c_1 c_2 c_3}^{\ }
\partial_{\mu}\psi^{c_3}
\xi^{c_2}
\xi^{c_1}
+
\cdots,
\label{eq: super RNC for partial phi}
\end{eqnarray}
\end{widetext}
where $\overleftarrow{D}_{\mu}$ is the right covariant derivative
for $\xi^{a}$
\begin{eqnarray}
\xi^{a}
\overleftarrow{D}_{\mu}
&=&
\partial^{\ }_{\mu}
\xi^{a}
+
\Gamma^{a}_{\ bc}\,
\partial^{\ }_{\mu}\psi^{c}\,
\xi^{b},
\end{eqnarray}
and $\Gamma^{a}_{\ bc}$ is a component of the connection. 
For a Riemannian supermanifold the connection can be
derived from the metric.
With the covariant expansion
Eqs.\ (\ref{eq: super RNC for T})
and   (\ref{eq: super RNC for partial phi})
the action is expanded in terms of $\xi^a$ according to
\begin{eqnarray}
&&
S[\psi,\pi]
-
S[\psi]=
\nonumber\\
&&
\frac{1}{4\pi t}
\int_r
\Big[
\xi^{a}\,
\overleftarrow{D}^{\ }_{\mu}\,\,
{}^{\ }_{a}g^{\ }_{b}(\psi)\,
{}^{b}\xi
\overleftarrow{D}^{\ }_{\mu}\,
+
R^{\ }_{abcd}\,
\partial^{\ }_{\mu}\psi^{d}\,
\xi^{c}\,
\xi^{b}\,
\partial^{\ }_{\mu}\psi^{a}
\Big]
\nonumber\\
&&
+
\cdots.
\label{eq: action expanded in terms of super RNC xi}
\end{eqnarray}

\paragraph{%
Vielbeins on a Riemannian supermanifold%
          }

To integrate over the fast modes in the expansion%
~(\ref{eq: action expanded in terms of super RNC xi}),
we rotate $\xi^a$ to $\zeta^a$
with the linear transformation
\begin{subequations}
\begin{eqnarray}
\zeta^{\hat a}:=
\xi^{a}\,
{}^{\ }_{a}
\hat e^{\hat a}(\psi),
&&\qquad
\xi^{a}=
\zeta^{\hat a}\,
{}^{\ }_{\hat a}
\hat e^{\ a}(\psi),
\\
{}^{\hat a}
\zeta:=
{}^{\hat a}
\hat e_{\ a}(\psi)\,
^{a}\xi,
&&\qquad
{}^{b}\xi=
{}^{ b}\hat e^{\ }_{\hat b}
(\psi)\,
{}^{\hat b}
\zeta,
\end{eqnarray}
\end{subequations}
where
\begin{subequations}
\begin{eqnarray}
{}^{\ }_{a}
\hat e^{\hat a}\,
{}^{\ }_{\hat a}
\hat e^{b}=
{}^{\ }_{a}\delta^{b}_{\ },
&&\qquad
{}^{\ }_{\hat a}
\hat e^{a}\,
{}^{\ }_{a}
\hat e^{\hat b}=
{}^{\ }_{\hat a}
\delta^{\hat b},
\\
{}_{\ }^{a}
\hat e^{\ }_{\hat a}\,
{}_{\ }^{\hat a}
\hat e^{\ }_{b}=
{}{\ }^{a}\delta_{b}^{\ },
&&\qquad
{}_{\ }^{\hat a}
\hat e^{\ }_{a}\,
{}_{\ }^{a}
\hat e^{\ }_{\hat b}=
{}_{\ }^{\hat a}
\delta^{\ }_{\hat b},
\\
(-1)^{\hat a(\hat a + a)}
{}^{\ }_{\hat a}\hat e^{a}
=
{}^{a}\hat e_{\hat a},
&&\qquad
(-1)^{a( a + \hat a)}
{}^{\ }_{a}\hat e^{\hat a}
=
{}^{\hat a}\hat e_{ a}.
\nonumber\\
&&
\end{eqnarray} 
\end{subequations}
Here, the vielbeins (and their inverse) $\hat e$ are defined by
\begin{subequations}
\begin{eqnarray}
{}^{\ }_{\hat a} \hat e^{a}\, 
{}^{\ }_{a}g^{\ }_{b}\, 
{}^{b}\hat e_{\hat b} =
{}^{\ }_{\hat a}\eta^{\ }_{\hat b},
\qquad
{}^{\ }_{a}
\hat e^{\hat a}\,
{}^{\ }_{\hat a}
\eta^{\ }_{\hat b}\, 
{}^{\hat b} \hat e_{b}^{\ }=
{}^{\ }_{a}g^{\ }_{b},
\end{eqnarray}
with
\begin{eqnarray}
&&
{}^{\ }_{\hat a}\eta^{\ }_{\hat b}
=
\left(
\begin{array}{cc}
D^{\ }_{M}
&
0
\\
0
&
D^{\ }_{N}
\end{array}
\right),
\\
&&
D^{\ }_{M}:=
\mathrm{diag}
\Bigg(
\underbrace{
-1, 
\cdots,
-1,
+1,
\cdots,
+1
}_{M}
\Bigg),
\\
&&
D^{\ }_{N}:=
\Bigg(
\underbrace{
\left(
\begin{array}{cc}
0 & +\mathrm{i} \\
-\mathrm{i} & 0
\end{array}
\right),
\cdots,
\left(
\begin{array}{cc}
0 & +\mathrm{i} \\
-\mathrm{i} & 0
\end{array}
\right)
}_{N/2}
\Bigg).
\end{eqnarray}
\end{subequations}
Observe that we are allowing the bosonic part of $\eta$ to 
be pseudo Riemannian, i.e., $\eta$ is allowed to have $-1$ as well as
$+1$ as its diagonal elements.
From now on, latin letters with a hat refer to the coordinates
of the Riemannian manifold in the vielbein basis.

The covariant expansion of the action in terms of $\zeta^{\hat a}$ is given by
\begin{subequations}
\begin{eqnarray}
&&
S[\psi,\pi]
-
S[\psi]
=
\nonumber\\
&&
\frac{1}{4\pi t}
\int_r
\Big[
\zeta^{\hat a}
\hat D^{\ }_{\mu}
{}^{\ }_{\hat a}\eta_{\hat b}^{\ }
{}^{\hat b}\zeta
\hat D^{\ }_{\mu}
+
R^{\ }_{i l_1 l_2 j}
\partial^{\ }_{\mu}\psi^{j}
\zeta^{\hat a}
{}^{\ }_{\hat a}\hat e^{l_2}
\zeta^{\hat b}
{}^{\ }_{\hat b}\hat e^{l_1}
\partial^{\ }_{\mu}\psi^{i}
\Big]
\nonumber\\
&&
+
\cdots,
\end{eqnarray}
where yet another covariant derivative
from the right
\begin{eqnarray}
{}^{\hat a}\zeta
\hat D^{\ }_{\mu}\,
&:=&
\partial^{\ }_{\mu}\,
{}^{\hat a}\zeta
+
{}^{\hat a}
\big(
A^{\ }_{\mu}(\psi)
\big){}^{\ }_{ \hat b}\,
{}^{\hat b}\zeta,
\\
\zeta^{\hat a}
\hat D^{\ }_{\mu}\,
&:=&
\partial^{\ }_{\mu}\,
\zeta^{\hat a}
-
\zeta^{\hat b}\,
{}_{\hat b}^{\ }
\big(
A^{\ }_{\mu}(\psi)
\big)^{ \hat a},
\end{eqnarray}
has been introduced 
together with the $\mathrm{OSp}(M|N)$ spin connection
[if the bosonic part of $\eta$ is pseudo Riemannian,
$\mathrm{OSp}(M|N)$ should be replaced by 
$\mathrm{OSp}(M_1, M_2|N)$ 
where $M_{1/2}$ is the number of $\pm1$ in the diagonal part of
$\eta$
with $M_1 + M_2 = M$],
\begin{eqnarray}
{}^{\hat a}(\omega_{c})^{\ }_{\hat b}
&:=&
{}^{\hat a}e_{a}^{\ }
\Big[
(-1)^{c\hat b}\,\,
\big(
{}^{a}_{\ }
\hat e^{\ }_{\hat b}\,
\overleftarrow{\partial}^{\ }_{c}
\big)\,
+
\Gamma^{a}_{\ cb}\,
{}^{b}_{\ } \hat e^{\ }_{\hat b}\,
\Big],
\nonumber\\
&&
\end{eqnarray}
the $\mathrm{OSp}(M|N)$ gauge field 
\begin{eqnarray}
 {}^{\hat a}_{\ }(A_{\mu}){}^{\ }_{\hat b}
&:=&
(-1)^{c\hat b}\,\,
{}^{\hat a}(\omega_{c})^{\ }_{\hat b}\,
\partial^{\ }_{\mu}\psi^{c},
\\
{}^{\hat a}(A_{\mu})_{\hat b}^{\ }
&=&
-
(-1)^{\hat b(\hat a+\hat b)}\,
{}_{\hat b}^{\ }(A_{\mu})^{\hat a},
\end{eqnarray}
and the $\mathrm{OSp}(M|N)$ field strength tensor
\begin{eqnarray}
{}^{\hat a}(F_{\mu\nu}){}^{\ }_{\hat b}
&=&
\partial_{\mu}\,
{}^{\hat a}(A_{\nu})^{\ }_{\hat b}
-
\partial_{\nu}\,
{}^{\hat a}(A_{\mu})^{\ }_{\hat b}
\nonumber\\
&&
+
{}^{\hat a}(A_{\mu})^{\ }_{\hat c}
{}^{\hat c}(A_{\nu})^{\ }_{\hat b}
-
{}^{\hat a}(A_{\nu})^{\ }_{\hat c}
{}^{\hat c}(A_{\mu})^{\ }_{\hat b}
\nonumber \\
&=&
-\,
{}^{\hat a}\hat e^{\ }_{a}\,\,
R^{a}{}^{\ }_{bcd}\,\,
\partial_{\mu}\psi^{d}\,\,
\partial_{\nu}\psi^{c}\,\,
{}^{b}\hat e^{\ }_{\hat b}.
\end{eqnarray}
\end{subequations}
The components of the Riemann tensor are here
$R^{a}{}^{\ }_{bcd}$.

\paragraph{%
One-loop beta function%
          }

The integration over the fast modes $\zeta$
is performed with the help of the cumulant expansion.
The effective action for the slow modes $\psi$ 
is, up to one-loop,
\begin{subequations}
\begin{eqnarray}
&&
Z:=
\int
\mathcal{D}[\psi]\,
e^{-S[\psi]+\delta S[\psi]},
\\
&&
\delta S[\psi]=
\frac{1}{2}G^{\ }_{0}(0)
 \int_r
\partial^{\ }_{\mu}\psi^{i}\,
(-1)^{i}R^{\ }_{ij}\,
\partial^{\ }_{\mu}\, {}^{j}\psi,
\end{eqnarray}
where 
\begin{eqnarray}
R^{\ }_{ij}&:=&
(-1)^{k(i+1)+l}g^{kl}R_{likj}
\end{eqnarray}
\end{subequations}
is the Ricci tensor.

\paragraph{%
RG flows of high-gradient operators%
          }

The covariant expansion in terms of the super RNC 
applies to the high-gradient operators
\begin{eqnarray}
T_{ijk\ldots }(\phi)\,
\partial_{\mu} \phi^{i}
\partial_{\nu} \phi^{j}
\partial_{\rho} \phi^{k}
\cdots
\end{eqnarray}
as well 
and can be used to compute their anomalous scaling dimensions.
We shall assume that all the covariant derivatives of the tensor field
$T^{\ }_{ijk\ldots }(\phi)$ vanish.
If so we infer the RG transformation law
\begin{widetext}
\begin{subequations}
\label{eq: master formulae if superriemann}
\begin{eqnarray}
\left\langle 
\left[
T^{\ }_{ij\ldots}(\phi)
\partial^{\ }_{\mu}\phi^{i}
\partial^{\ }_{\nu}\phi^{j}
\cdots
\right]^{\ }_{\zeta^2}
\right\rangle
&=&
\left\langle
\left[
T^{\ }_{ij\ldots}
\right]^{\ }_{\zeta^2}
\right\rangle
\partial^{\ }_{\mu}\psi^{i}
\partial^{\ }_{\nu}\psi^{j}
\cdots
\nonumber\\
&&
+
T^{\ }_{ij\ldots}(\psi)
\left\langle
\left[\partial^{\ }_{\mu}\phi^{i}\right]^{\ }_{\zeta^2}
\right\rangle
\partial^{\ }_{\nu}\psi^{j}
\cdots
+
T^{\ }_{ij\ldots}(\psi)
\partial^{\ }_{\mu}\psi^{i}
\left\langle
\left[\partial^{\ }_{\nu}\phi^{j}\right]^{\ }_{\zeta^2}
\right\rangle
\cdots
\nonumber\\
&&
+
T^{\ }_{ij\ldots}(\psi)
\left\langle
\left[\partial^{\ }_{\mu}\phi^{i}\right]^{\ }_{\zeta^1}
\left[\partial^{\ }_{\nu}\phi^{j}\right]^{\ }_{\zeta^1}
\right\rangle
\cdots
+
\cdots.
\end{eqnarray}
Thus, all we need are
\begin{eqnarray}
\left\langle
\left[\partial^{\ }_{\mu}\phi^{i}\right]^{\ }_{\zeta^1}
\left[\partial^{\ }_{\nu}\phi^{j}\right]^{\ }_{\zeta^1}
\right\rangle
&= &
-
\frac{t{d}l}{2}
R^{i}{}^{\ }_{bcd}
\partial^{\ }_{\mu}\psi^{d}
\partial^{\ }_{\nu}\psi^{c}\,
{}^{b}g^{j}
-
\frac{t{d}l}{2} 
\delta^{\ }_{\mu,-\nu}\,
(-1)^{j(k+l)}
R^{i}{}_{l k l^{\ }_{1}}^{\ }\,
{}^{l^{\ }_1}g^{j}\,
\partial^{\ }_{+\rho}\psi^{k}\,
\partial^{\ }_{-\rho}\psi^{l}\, ,
\\
\left\langle 
\left[
\partial_{\mu}\phi^{i}
\right]_{\zeta^2}
\right\rangle
&=&
\frac{t{d}l}{3}
R^{i}{}_{b c d}
\partial_{\mu}\psi^{d}\,
{}^{c}g^{b},
\end{eqnarray} 
on the one hand and
\begin{eqnarray}
\left\langle
\left[
T^{\ }_{b^{\ }_{1}\cdots b^{\ }_{l}}(\phi)
\right]^{\ }_{\zeta^2} 
\right\rangle
=
-
\frac{t{d}l}{2!}
\frac{1}{3}
\sum_{i=1}^{l}
T^{\ }_{
b^{\ }_{1}\cdots b^{\ }_{i-1}bb^{\ }_{i+1}\cdots b^{\ }_{l}
       }
(-1)^{(b^{\ }_{i}+b)(b^{\ }_{i+1}+\cdots+b^{\ }_{l})}
\left(
R^{b}_{\ b^{\ }_{i}c^{\ }_{1}c^{\ }_{2}}
+
(-1)^{b^{\ }_{i}c^{\ }_{1}}
R^{b}_{\ c^{\ }_{1}b^{\ }_{i}c^{\ }_{2}}
\right)
{}^{c^{\ }_2}g^{c^{\ }_1},
\end{eqnarray}
\end{subequations}
\end{widetext}
on the other hand.
Here, we are using the conformal indices defined in
Eq.\ (\ref{eq: def conformal indices}).

\subsection{
NL$\sigma$M on K\"ahler supermanifolds
        }

\paragraph{%
Notations and conventions for Hermitian supermanifolds%
          }

Some additional notation is needed to deal with Hermitian
and K\"ahlerian manifolds.
\begin{subequations}
\begin{itemize}
\item
With the convention that complex conjugation 
of a composite objects reverts the ordering,
$(\mathcal{O}_1\mathcal{O}_2\mathcal{O}_3\cdots)^*=
\cdots (\mathcal{O}_3)^* (\mathcal{O}_2)^* (\mathcal{O}_1)^*
$,
it is useful to introduce sign factors associated to 
the operation of complex conjugation.
For covariant and contravariant holomorphic and anti-holomorphic vectors,
the action of the complex conjugation is defined by
\begin{eqnarray}
(V^{a})^{*}=V^{* a},
\qquad
(V_{a})^{*}=(-1)^a V^{*}_{a}.
\end{eqnarray}
For tensor fields of interest in this paper,
the complex conjugation of a tensor field is 
given by
\begin{eqnarray}
&&
\left[
T^{A_{1}\ldots}{}^{\ }_{\ldots A_{r+s}^{\ }}
\right]^*
=
\label{eq: complex conjugate, tensors }
\\
&&
(-1)^{
\delta^{\ }_{r+s}(A)
\,
+
\,
(A_{r+1}^{\ }+\cdots + A_{r+s}^{\ })
}
\times
T^{A^{*}_{1}\ldots}{}^{\ }_{\ldots A^{*}_{r+s} },
\nonumber
\end{eqnarray}
with
$
\delta^{\ }_{r+s}(A)=
\sum_{t,u=1,t<u}^{q} A_{t}A_{u}
$.%
~\cite{Rem: complex conjugation of tensor fields}

\item 
The grade for an antiholomorphic index 
is equal to that of its holomorphic counterpart,
$\epsilon(a^*)=\epsilon(a)$.

\end{itemize}
\end{subequations}

For illustration, we shall apply these rules to
the components ${}^{\ }_{a}g^{\ }_{b^{*}}$ of the metric tensor field.
Combining
the rule%
~(\ref{eq: complex conjugate, tensors })
with the supersymmetry of the metric%
~(\ref{eq: supersymmetry of metric}),
\begin{eqnarray}
&&
{}^{\ }_{a}g^{\ }_{b^*}
=
(-1)^{a+b+ab}
({}^{\ }_{a^*}g^{\ }_{b})^*,
\quad
{}^{\ }_{a^*}g^{\ }_{b}
=(-1)^{a+b+ab}\,{}^{\ }_{b}g^{\ }_{a^*},
\nonumber\\
&&
\end{eqnarray}
we infer that the metric tensor field is superhermitian
\begin{eqnarray}
g_{b^* a}^{\ }=
(g_{a^* b})^*.
\end{eqnarray}

\paragraph{%
The NL$\sigma$M on a Hermitian supermanifold%
          }

The NL$\sigma$M on a Hermitian supermanifold
is defined by the partition function
\begin{subequations}
\begin{eqnarray}
Z&:=&
\int
\mathcal{D}[z^*,z]\,
e^{-S[z^*,z]},
\\
S[z^*,z]&:=&
\frac{1}{2\pi t}\int_r
\partial^{\ }_{\mu}z^{* a}\,
{}_{a^*}^{\ }g^{\ }_{b}(z^*,z)\,\,
\partial^{\ }_{\mu} {}^{b}z,
\nonumber\\
&&
\end{eqnarray}
\end{subequations}
where $(z^{*a},z^{a})$ represents coordinates
on the complex supermanifold.
For a K\"ahlerian supermanifold,
the metric is derived from the K\"ahler potential $K(z^*,z)$,
\begin{eqnarray}
g_{a b^*}^{\ }&=&
K(z^*,z)
\overleftarrow{\partial}_{a}^{\ }
\overleftarrow{\partial}_{b^*}^{\ }.
\end{eqnarray}
[Here, observe that 
the shifting rule for indices 
(\ref{eq: shifting rule for the left most indices})
and the relation between the left- and right-derivatives
(\ref{eq: left <-> right derivatives})
are compatible.]

\paragraph{%
The covariant background field method
and the super K\"ahler normal coordinate expansion%
          }

In the background field method,
the covariant vector field $(z^{*a},z^{a})(r)$
is separated into two parts,
\begin{eqnarray}
z^{* a}=
\psi^{* a}+\pi^{* a},
\qquad
z^{a}=
\psi^{a}+\pi^{a},
\end{eqnarray}
whereby $(\psi^{*a},\psi^{a})(r)$ is assumed to be a 
slowly varying solution to the classical
equations of motion
which transforms like a contravariant vector
and $(\pi^{*a},\pi^{a})(r)$ 
represents fluctuations around 
the slow degrees of freedom $(\psi^{*a},\psi^{a})(r)$.

The expansion in terms of
$(\pi^*,\pi)$  is traded off for one in terms of
normal coordinates $(w^*,w)$,
the super version of the KNC.
The super KNC expansion for 
a tensor field of rank $(0,r+s)$
with $r$ holomorphic  and $s$ antiholomorphic indices
$T^{\ }_{b_{1}\cdots b_{r}b^{*}_{1}\cdots b^{*}_{s}}(z^*,z)$
is
\begin{widetext}
\begin{eqnarray}
T^{\ }_{b_{1}^{\ }\cdots b_{r}^{\ }c^{*}_{1}\cdots c^{*}_{s}}
(z^*,z)&=&
T^{\ }_{b_{1}^{\ }\cdots b_{r}^{\ }c^{*}_{1}\cdots c^{*}_{s}}
(\psi^*,\psi)
+
T^{\ }_{b_{1}^{\ }\cdots b_{r}^{\ }c^{*}_{1}\cdots c^{*}_{s}}
\overleftarrow{\nabla}_{k_{1}^{\ }}
w^{k_{1}}
+
T^{\ }_{b_{1}^{\ }\cdots b_{r}^{\ }c^{*}_{1}\cdots c^{*}_{s}}
\overleftarrow{\nabla}_{l_{1}^{*}}
w^{*l_{1}}
\nonumber\\
&+&
\frac{1}{2}
T^{\ }_{b_{1}^{\ }\cdots b_{r}^{\ }c^{*}_{1}\cdots c^{*}_{s}}
\overleftarrow{\nabla}_{k_{1}}^{\ }
\overleftarrow{\nabla}_{k_{2}}^{\ }
w^{k_{2}}
w^{k_{1}}
+
\frac{1}{2}
T^{\ }_{b_{1}^{\ }\cdots b_{r}^{\ }c^{*}_{1}\cdots c^{*}_{s}}
\overleftarrow{\nabla}_{l_{1}^{*}}
\overleftarrow{\nabla}_{l_{2}^{*}}
w^{*l_{2}}
w^{*l_{1}}
+
T^{\ }_{b_{1}^{\ }\cdots b_{r}^{\ }c^{*}_{1}\cdots c^{*}_{s}}
\overleftarrow{\nabla}_{k_{1}}
\overleftarrow{\nabla}_{l^{*}_{1}} 
w^{*l_{1}}
w^{k_{1}}
\nonumber\\
&-&
\sum_{i=1}^{r}
(-1)^{(a+b_i)(b_{i+1}+\cdots+c_{s})}
T^{\ }_{b_{1}^{\ }\cdots b_{i-1}^{\ }ab_{i+1}^{\ } \cdots b_{r}^{\ }
c^{*}_{1}\cdots c^{*}_{s}}
R^{a}{}_{b_{i}k_{1}l_{1}^{*}}^{\ }
w^{*l_{1}}
w^{k_{1}}
+\cdots
\nonumber\\
&=&
T^{\ }_{b_{1}^{\ }\cdots b_{r}^{\ }c^{*}_{1}\cdots c^{*}_{s}}
(\psi^*,\psi)
+
T^{\ }_{b_{1}^{\ }\cdots b_{r}^{\ }c^{*}_{1}\cdots c^{*}_{s}}
\overleftarrow{\nabla}_{k_{1}^{\ }}
w^{k_{1}}
+
T^{\ }_{b_{1}^{\ }\cdots b_{r}^{\ }c^{*}_{1}\cdots c^{*}_{s}}
\overleftarrow{\nabla}_{l_{1}^{*}}
w^{*l_{1}}
\nonumber\\
&+&
\frac{1}{2}
T^{\ }_{b_{1}^{\ }\cdots b_{r}^{\ }c^{*}_{1}\cdots c^{*}_{s}}
\overleftarrow{\nabla}_{k_{1}}^{\ }
\overleftarrow{\nabla}_{k_{2}}^{\ }
w^{k_{2}}
w^{k_{1}}
+
\frac{1}{2}
T^{\ }_{b_{1}^{\ }\cdots b_{r}^{\ }c^{*}_{1}\cdots c^{*}_{s}}
\overleftarrow{\nabla}_{l_{1}^{*}}
\overleftarrow{\nabla}_{l_{2}^{*}}
w^{*l_{2}}
w^{*l_{1}}
+
T^{\ }_{b_{1}^{\ }\cdots b_{r}^{\ }c^{*}_{1}\cdots c^{*}_{s}}
\overleftarrow{\nabla}_{l^{*}_{1}}
\overleftarrow{\nabla}_{k_{1}}
w^{k_{1}}
w^{*l_{1}}
\nonumber\\
&-&
\sum_{i=1}^{s}
(-1)^{(a+c_i)(c_{i+1}+\cdots+c_s)}
T^{\ }_{b_{1}^{\ }\cdots  b_{r}^{\ }c^{*}_{1}\cdots c^{*}_{i-1}a^{*}c^{*}_{i+1} \cdots c^{*}_{s}}
R^{a^{*}}{}_{c_{i}^{*} l_{1}^{*} k_{1}^{\ }}
w^{k_{1}}
w^{*l_{1}}
+\cdots.
\label{eq: super Hermi RNC for T}
\end{eqnarray}
The super KNC expansion for
$\partial_{\mu} (\psi^{A}+w^A)$ is given by
\begin{eqnarray}
\partial^{\ }_{\mu}
\left(
\psi^{a}
+
\pi^{a}
\right)
&=&
\partial^{\ }_{\mu}\psi^{a}
+
w^{a}
\overleftarrow{D}^{\ }_{\mu}
+
\frac{1}{2}
R^{a}_{\ c^{\ }_1c^{\ }_2c^{*}_{3}}
(\psi,\psi^*)
\partial^{\ }_{\mu}\psi^{*c^{\ }_{3}}
w^{c^{\ }_2}w^{c^{\ }_1}
+\cdots
\nonumber\\
\partial^{\ }_{\nu}
({}^{a}\psi^*+ {}^{a}\pi^*)
&=&
\partial^{\ }_{\nu}{}^{j}\psi^{*}
+
{}^{a}w^{*}
\overleftarrow{D}^{*}_{\nu}
+
\frac{1}{2}
{}^{a^*}
R_{\ c^{*}_{1}c^{*}_{2} c_3}(\psi,\psi^*)
\partial^{\ }_{\nu}\psi^{c_3} 
w^{* c^{\ }_{2}}
w^{* c^{\ }_{1}}
+
\cdots,
\label{eq: super Hermi RNC for partial phi}
\end{eqnarray}
\end{widetext}
where 
$\overleftarrow{D}^{\ }_{\mu}$
is the right covariant derivative for $(w^{*j},w^{j})$
\begin{subequations}
\begin{eqnarray}
w^{j}\overleftarrow{D}^{\ }_{\mu}
&=&
w^{j}
\nabla_{a}
\partial_{\mu}
\psi^{a}
\nonumber\\
&=&
\partial_{\mu}w^{j}
+
\Gamma^{j}{}_{ab}^{\ }
w^{b}
\partial_{\mu}
\psi^{a},
\\
{}^{j}w^* \overleftarrow{D}^*_{\mu}
&=&
{}^{j}w^* \nabla_{a^*}\partial_{\mu}\psi^{*a}
\nonumber\\
&=&
\partial_{\mu}{}^{j}w^{*}
+
\Gamma^{j^*}{}_{a^*b^*}^{\ }
w^{*b}
\partial_{\mu}
\psi^{* a}
\nonumber\\
&=&
(w^{j}\overleftarrow{D}^{\ }_{\mu})^*,
\end{eqnarray}
\end{subequations}
with $\Gamma^{A}_{\ BC}$ the coefficient of the connection
that satisfies
$
(\Gamma^{a}{}_{bc}^{\ })^{*}
=
(-1)^{c+b+a(b+c)+bc}\,\,
\Gamma^{a^*}{}_{b^*c^*}
$
[see~(\ref{eq: complex conjugate, tensors })].
With the covariant expansion
Eqs.\ (\ref{eq: super Hermi RNC for T})
and   (\ref{eq: super Hermi RNC for partial phi})
the action is expanded in terms of $(w^{*a},w^{a})$
according to
\begin{eqnarray}
&&
S[\psi^*+\pi^*,\psi+\pi]
-
S[\psi^*,\psi]=
\nonumber\\
&&
+
\frac{1}{2\pi t}
\int_r
(\overrightarrow{D}^{*}_{\mu} w^{*a})\,
{}^{\ }_{a^*}g^{\ }_{b}\,
({}^{b}w \overleftarrow{D}^{\ }_{\mu})
\nonumber\\
&&
-
\frac{1}{2\pi t}
\int_r
R_{i^*j k l^*}
\Big(
w^{*l}w^{k} \partial_{\mu}\psi^{j}\partial_{\mu}\psi^{*i}
\nonumber\\
&&
-
\frac{1}{2}
w^{*l}\partial_{\mu}\psi^{k}\partial_{\mu}\psi^{j}w^{*i}
-
\frac{1}{2}
\partial_{\mu}\psi^{*l}
w^{k}
w^{j}
\partial_{\mu}\psi^{*i}
\Big)
\nonumber\\
&&
+\cdots.
\end{eqnarray}

\paragraph{%
Vielbeins on a Hermitian supermanifold%
          }

Introduce the fast degrees of freedom 
$(\zeta^{*},\zeta)$
by the linear transformations
\begin{subequations}
\begin{eqnarray}
&&
\zeta^{\hat a}:=
w^{a}\,
{}^{\ }_{a}
\hat e^{\hat a},
\qquad
w^{a}=
\zeta^{\hat a}\,
{}^{\ }_{\hat a}
\hat e^{\ a},
\\
&&
{}^{\hat a}
\zeta:=
{}^{\hat a}
\hat e_{\ a}\,
{}^{a}w,
\qquad
{}^{b}w=
{}^{ b}\hat e^{\ }_{\hat b}\,
{}^{\hat b}
\zeta,
\end{eqnarray}
and
\begin{eqnarray}
&&
\zeta^{* \hat a}:=
w^{* a}\,
{}^{\ }_{a^*}
\hat e^{\hat a^*},
\qquad
w^{* a}=
\zeta^{* \hat a}\,
{}^{\ }_{\hat a^*}
\hat e^{\ a^*},
\\
&&
{}^{\hat a}
\zeta^*:=
{}^{\hat a^*}
\hat e_{\ a^*}\,
{}^{a}w^*,
\qquad
{}^{b}w^*=
{}^{ b^*}\hat e^{\ }_{\hat b^*}\,
{}^{\hat b}
\zeta^*
,
\end{eqnarray}
where
\begin{eqnarray}
{}^{\ }_{a}
\hat e^{\hat a}\,
{}^{\ }_{\hat a}
\hat e^{b}=
{}^{\ }_{a}\delta^{b}_{\ },
&&
{}^{\ }_{\hat a}
\hat e^{a}\,
{}^{\ }_{a}
\hat e^{\hat b}=
{}^{\ }_{\hat a}
\delta^{\hat b},
\\
{}_{\ }^{a}
\hat e^{\ }_{\hat a}\,
{}_{\ }^{\hat a}
\hat e^{\ }_{b}=
{}{\ }^{a}\delta^{\ }_{b},
&&
{}_{\ }^{\hat a}
\hat e^{\ }_{a}\,
{}_{\ }^{a}
\hat e^{\ }_{\hat b}=
{}_{\ }^{\hat a}
\delta^{\ }_{\hat b},
\\
(-1)^{\hat a(\hat a + a)}
{}^{\ }_{\hat a}\hat e^{a}
=
{}^{a}\hat e_{\hat a},
&&
(-1)^{a( a + \hat a)}
{}^{\ }_{a}\hat e^{\hat a}
=
{}^{\hat a}\hat e_{ a},
\nonumber\\
&&
\\
\left(
{}^{\ }_{a}\hat{e}{}^{\hat a}
\right)^{*}
=
(-1)^{(a+\hat a)a}
{}_{a^*}\hat e ^{\hat a^*},
&&
\left(
{}^{\ }_{\hat a}\hat{e}{}^{a}
\right)^{*}
=
(-1)^{(a+\hat a) \hat a}
{}_{\hat a^*}\hat e ^{a^*}.
\nonumber\\
\end{eqnarray} 
These definitions of the vielbeins 
are equivalent to the non-supersymmetic cases
except the last two lines.
The vielbeins (and their inverse) $\hat e$ are defined such that
\begin{eqnarray}
w^{a}\,{}^{\ }_{a}g^{\ }_{b^*}\,{}^{b}w^*
=
\zeta^{\hat a}\, 
{}^{\ }_{\hat a}\eta^{\ }_{\hat b^*}\, 
{}^{\hat b}\zeta^*
\end{eqnarray}
where the canonical form of the metric $\eta$ is given by
\begin{eqnarray}
{}^{\ }_{\hat a}\eta^{\ }_{\hat b^*}
&=&
\mathrm{diag}
\big(
\mathbb{I}_{M_1},
-\mathbb{I}_{M_2},
\mathbb{I}_{N}
\big).
\end{eqnarray}
\end{subequations}
Note that we are allowing the bosonic part of $\eta$ to be 
pseudo Riemannian, i.e., $\eta$ is allowed to have $-1$ as well as
$+1$ as its diagonal elements.
From now on, latin letters with a hat refer to the coordinates
of the Hermitian manifold in the vielbein basis.

The covariant expansion of the action in terms of 
$(\zeta^{*\hat a},\zeta^{\hat a})$ is given by
\begin{eqnarray}
&&
S[\psi^*+\pi^*,\psi+\pi]
-
S[\psi^*,\psi]=
\nonumber\\
&&
+
\frac{1}{2\pi t}
\int_{r}\,
(\hat D^{* }_{\mu} \zeta^{* \hat a }\,)\,
{}^{\ }_{\hat a^*}\eta_{\hat b}^{\ }\,
({}^{\hat b}\,\zeta\,\hat D^{\ }_{\mu})\,
\nonumber\\
&&
-
\frac{1}{2\pi t}
\int_{r}
R_{i^*j k l^*}
\Big[
(\zeta^{* \hat l}\, {}_{\hat l^*}^{\ }\hat e^{l^*})
(\zeta^{\hat k}\, {}_{\hat k}^{\ }\hat e^{k})
\partial_{\mu}\psi^{j}
\partial_{\mu}\psi^{*i}
\nonumber\\
&&
-
\frac{1}{2}
(\zeta^{* \hat l}\,{}_{\hat l^*}^{\ } \hat e^{l^*})
\partial_{\mu}\psi^{k}
\partial_{\mu}\psi^{j}
(\zeta^{* \hat i}\,{}_{\hat i^*}^{\ } \hat e^{i^*})
\nonumber\\
&&
-
\frac{1}{2}
\partial_{\mu}\psi^{*l}
(\zeta^{ \hat k}\,{}_{\hat k}^{\ }\hat e^{k})
(\zeta^{ \hat j}\,{}_{\hat j}^{\ }\hat e^{j})
\partial_{\mu}\psi^{*i}
\Big]
+\cdots,
\end{eqnarray}
where yet another covariant derivative from the right
\begin{subequations}
\begin{eqnarray}
{}^{\hat a}\zeta \hat{D}_{\mu}
&=&
\partial_{\mu}{}^{\hat a}\zeta
+
{}^{\hat a}(A_{\mu})_{\hat b}^{\ }\,
{}^{\hat b}\zeta,
\\
\zeta^{\hat a} \hat{D}_{\mu}
&=&
\partial_{\mu}\zeta^{\hat a}
-
\zeta^{\hat b}\,
{}_{\hat b}^{\ }(A_{\mu})^{\hat a},
\\
 \hat{D}^*_{\mu}
{}^{\hat a}\zeta^{*}
&=&
\partial_{\mu}{}^{\hat a}\zeta^{*}
+
{}^{\hat a^*}(A_{\mu})_{\hat b^*}
{}^{\hat b}\zeta^{*},
\\
\hat{D}^*_{\mu}
\zeta^{* \hat a } 
&=&
\partial_{\mu}\zeta^{* \hat a}
-
\zeta^{* \hat b}
{}_{\hat b^*}(A_{\mu})^{\hat a^*}
\end{eqnarray}
\end{subequations}
has been introduced 
together with the $\mathrm{U}(M_1,M_2|N)$ 
gauge field $A_{\mu}$.
The gauge field with different index-structures are related by
\begin{subequations}
\begin{eqnarray}
{}^{\hat a}\zeta \hat{D}_{\mu}
=
\zeta^{\hat a} \hat{D}_{\mu}
&\Rightarrow &
 {}^{\hat a}(A_{\mu})_{\hat b}^{\ }
=
-
(-1)^{\hat b (\hat a + \hat b)}\,
{}_{\hat b}^{\ }(A_{\mu})^{\hat a},
\nonumber\\
&&
\\
({}^{\hat a}\zeta\hat{D}_{\mu})^*
=
\hat{D}^*_{\mu}
\zeta^{* \hat a}
&\Rightarrow &
-
{}_{\hat b^*}^{\ }(A_{\mu})^{\hat a^*}
=
\big[
{}^{\hat a}(A_{\mu})_{\hat b}
\big]^{*}.
\nonumber\\
&&
\end{eqnarray}
\end{subequations}
The gauge field is skew superhermitian,
\begin{eqnarray}
{}_{\hat b}^{\ }(A_{\mu})_{\hat c^*}^{\ }
&=&
-
\big({}_{\hat c}^{\ }(A_{\mu})_{\hat b^*}\big)^*
\end{eqnarray}
where indices of $A_{\mu}$ are raised and lowered by $\eta$.
The explicit form of the gauge field in terms of 
the vielbeins is given by
the $\mathrm{U}(M_1,M_2|N)$ spin connection
\begin{subequations}
\begin{eqnarray}
{}_{\hat a}^{\ }\omega^{\hat b}{}_{c^*}^{\ }
&:=&
{}_{\hat a}^{\ }\hat e^{a}\,
\big(
{}_{a}\hat e^{\hat b}
\overleftarrow{\partial}_{c^*}^{\ }
\big),
\\
{}_{\hat a}^{\ }\omega^{\hat b}{}_{c}^{\ }
&:=&
{}_{\hat a}^{\ }\hat e_{p^*}^{\ }\,
\big(
{}^{p^*}\hat e^{\hat b}
\overleftarrow{\partial}_{c}^{\ }
\big)
\quad
\nonumber\\
&=&
{}_{\hat a}^{\ }\hat e^{a} \Big[
({}_{a}^{\ }\hat e^{\hat b}\overleftarrow{\partial}_{c})
-
(-1)^{a+ar+(r+\hat b)c}\,
\Gamma^{r}{}_{ac}^{\ }\,
{}_{r}^{\ }\hat e^{\hat b}
\Big],
\nonumber\\
&&
\\
{}_{\hat a}^{\ }\omega_{\hat b^* c^*}^{\ }
&=&
(-1)^{c(\hat b + \hat a)}
\big(
{}_{\hat b}^{\ }\omega_{\hat a^* c}^{\ }
\big)^{*},
\end{eqnarray}
\end{subequations}
whereby
\begin{eqnarray}
{}_{\hat a}^{\ }
\left(
A_{\mu}
\right)^{\hat b}
&=:&
{}_{\hat a}^{\ }\omega^{\hat b}{}_{c^*}^{\ }
\partial_{\mu}\psi^{*c}
+
{}_{\hat a}^{\ }\omega^{\hat b}{}_{c}^{\ }
\partial_{\mu}\psi^{c}.
\end{eqnarray}
The $\mathrm{U}(M_1,M_2|N)$ field strength tensor is given by
\begin{eqnarray}
{}_{\hat a}^{\ }(F_{\mu\nu}){}^{\hat b}
=
\partial_{\mu}\,
{}_{\hat a}^{\ }(A_{\nu})^{\hat b}
-
\partial_{\nu}\,
{}_{\hat a}^{\ }(A_{\mu})^{\hat b}
\qquad\qquad\qquad\,\,\,
&&
\nonumber\\
+
{}_{\hat a}^{\ }(A_{\mu})^{\hat c}
{}_{\hat c}^{\ }(A_{\nu})^{\hat b}
-
{}_{\hat a}^{\ }(A_{\nu})^{\hat c}
{}_{\hat c}^{\ }(A_{\mu})^{\hat b}
&&
\nonumber\\
=
-
(-1)^{a+ar}\,
{}_{\hat a}^{\ }\hat e^{a}\,
R^{r}{}_{ad c^*}^{\ }\,
\partial_{\nu}\psi^{*c}\,
\partial_{\mu}\psi^{d}\,
{}_{r}^{\ }\hat e^{\hat b}
&&
\nonumber\\
+
(-1)^{a+ar}\,
{}_{\hat a}^{\ }\hat e^{a}\,
R^{r}{}_{ad c^*}^{\ }\,
\partial_{\mu}\psi^{*c}\,
\partial_{\nu}\psi^{d}\,
{}_{r}^{\ }\hat e^{\hat b}.
&&
\end{eqnarray}

\paragraph{%
One-loop beta function%
          }

The integration over the fast modes $(\zeta^*,\zeta)$
is performed with the help of the cumulant expansion.
The effective action for the slow modes $(\psi^*,\psi)$ 
is, up to one-loop,
\begin{subequations}
\begin{eqnarray}
&&
Z:=
\int
\mathcal{D}[\psi^{*},\psi]\,
e^{-S[\psi^{*},\psi]+\delta S[\psi^{*},\psi]},
\\
&&
\delta S[\psi^{*},\psi]=
-
G_{0}(0)
\int_{r}
R^{\ }_{a^{*} b}
\partial^{\ }_{\mu}\psi^{b}\,
\partial^{\ }_{\mu}\psi^{*a},
\end{eqnarray}
where
\begin{eqnarray}
R_{i^* j}^{\ }:=
(-1)^{k+l+il}\,
{}^{l^*}g^{k}\,
R_{ k  i^* l^* j}^{\ }
\end{eqnarray}
\end{subequations}
is the Ricci tensor.

\paragraph{%
RG flows of high-gradient operators%
          }

The covariant expansion in terms of the super KNC 
applies to the high-gradient operators
\begin{eqnarray}
T^{\ }_{IJK\ldots}(z^{*},z)
\partial z_{\mu}^{I}
\partial z_{\nu}^{J}
\partial z_{\rho}^{K}
\cdots
\label{eq: high_grad_op, app}
\end{eqnarray}
as well 
and can be used to compute their anomalous scaling dimensions.
We shall assume that all the covariant derivatives
of the tensor field $T^{\ }_{IJK\ldots}(z^{*},z)$ vanish.
If so we infer the RG transformation law
\begin{widetext}
\begin{subequations}
\label{eq: master formulae if superkahler}
\begin{eqnarray}
\left\langle 
\left[
T^{\ }_{IJ\ldots} (z^*,z)
\partial^{\ }_{\mu}z^{I}
\partial^{\ }_{\nu}z^{J}
\cdots
\right]^{\ }_{\zeta^2}
\right\rangle
&=&
\left\langle
\left[
T^{\ }_{IJ\ldots}(z^*,z)
\right]^{\ }_{\zeta^2}
\right\rangle
\partial^{\ }_{\mu}z^{I}
\partial^{\ }_{\nu}z^{J}
\cdots
+
T^{\ }_{IJ\ldots}(z^*,z)
\left\langle
\left[
\partial^{\ }_{\mu}z^{I}
\right]^{\ }_{\zeta^1}
\left[
\partial^{\ }_{\nu}z^{J}
\right]^{\ }_{\zeta^1}
\right\rangle
\cdots
+
\cdots.
\nonumber\\
&&
\end{eqnarray}
Thus, all we need are 
\begin{eqnarray}
\left\langle
\left[\partial_{\mu}z^{*i}\right]^{\ }_{\zeta^1}
\left[\partial_{\nu}z^{*j}\right]^{\ }_{\zeta^1}
\right\rangle
&=&
-
t{d}l\delta^{\ }_{\mu,-\nu}\,
(-1)^{j(k+l)}
R^{i^*}{}_{l^* k^* l^{\ }_{1}}^{\ }\,
{}^{l^{\ }_1}g^{j^*}\,
\partial^{\ }_{\mu}\psi^{* k}\,
\partial^{\ }_{\nu}\psi^{* l},
\\
\left\langle
\left[\partial^{\ }_{\mu}z^{i  }\right]^{\ }_{\zeta^1}
\left[\partial^{\ }_{\nu}z^{* j}\right]^{\ }_{\zeta^1}
\right\rangle
&=&
- 
t{d}l\delta^{\ }_{\mu,-\nu}
R^{i}{}^{\ }_{bc^*d}
\partial^{\ }_{\mu}\psi^{d}
\partial^{\ }_{\nu}\psi^{*c}\,
{}^{b}g^{j^*},
\\
\left\langle
\left[\partial_{\mu}z^{*i}\right]^{\ }_{\zeta^1}
\left[\partial_{\nu}z^{j }\right]^{\ }_{\zeta^1}
\right\rangle
&=&
-
t{d}l\delta^{\ }_{\mu,-\nu}
R^{i^*}{}^{\ }_{b^* c d^*}
\partial^{\ }_{\mu}\psi^{*d}
\partial^{\ }_{\nu}\psi^{c}\,
{}^{b^*}g^{j},
\\
\left\langle
\left[\partial^{\ }_{\mu}z^{i}\right]^{\ }_{\zeta^1}
\left[\partial^{\ }_{\nu}z^{j}\right]^{\ }_{\zeta^1}
\right\rangle
&=&
-
t{d}l\delta^{\ }_{\mu,-\nu}\,
(-1)^{j(k+l)}
R^{i}{}_{l k l^{* }_{1}}^{\ }\,
{}^{l^{* }_1}g^{j}\,
\partial^{\ }_{\mu}\psi^{k}\,
\partial^{\ }_{\nu}\psi^{l},
\end{eqnarray}
on the one hand and
\begin{eqnarray}
\left\langle 
\left[
T^{\ }_{
b^{\ }_{1}
\cdots 
b^{\ }_{r}
c^{* }_{1}
\cdots 
c^{* }_{s}
       }
(z^*,z)
\right]^{\ }_{\zeta^2}
\right\rangle
&=&
-
t{d}l
\sum_{i=1}^{r}
(-1)^{(a+b^{\ }_{i})(b^{\ }_{i+1}+\cdots+c^{\ }_{s})}
T^{\ }_{
b^{\ }_{1}
\cdots 
b^{\ }_{i-1}
a
b^{\ }_{i+1} 
\cdots 
b^{\ }_{r}
c^{* }_{1}
\cdots 
c^{* }_{s}
       }
R^{a}{}_{b^{\ }_{i}k^{\ }_{1}l^{* }_{1}}^{\ }
{}^{*l^{\ }_{1}}g^{k^{\ }_{1}}
\nonumber\\
&=&
-
t{d}l
\sum_{i=1}^{s}
(-1)^{(a+c^{\ }_i)(c^{\ }_{i+1}+\cdots+c^{\ }_{s})}
T^{\ }_{
b^{\ }_{1}
\cdots  
b^{\ }_{r}
c^{* }_{1}
\cdots 
c^{* }_{i-1}
a^{* }
c^{* }_{i+1} 
\cdots 
c^{* }_{s}
       }
R^{a^{*}}{}_{c^{* }_{i}l^{* }_{1}k^{\ }_{1}}
{}^{k^{\ }_{1}}g^{*l_{1}},
\end{eqnarray}
\end{subequations}
\end{widetext}
on the other hand. 
Here, we are using the conformal indices defined in
Eq.\ (\ref{eq: def conformal indices}).

Since a K\"ahlerian supermanifold is a Riemannian supermanifold,
one can make use of either the RNC or KNC expansion.
Indeed, all the results in this appendix  derived
from  the KNC expansion can actually be obtained
from the RNC expansion as well.
A subtle thing here is that 
the RNC and KNC expansion
for a tensor field $T_{IJK\ldots}(z^*,z)$
and
$\partial_{\mu}z^A$
are different.
\cite{Higashijima00,Higashijima02a}
However,
the RNC 
and KNC expansions
are identical when applied to 
the high-gradient operators (\ref{eq: high_grad_op, app})
and thus give us the same results.

\newpage


\begin{thebibliography}{99}
\bibitem{footnoteDIIIandCI}
The two remaining BdG symmetry classes
are called DIII and CI by AZ. 
They correspond to cases for which time-reversal symmetry
is respected in all realizations of the disorder
whereas spin-rotation symmetry is broken in 
symmetry class DIII
but preserved in symmetry class CI.
\bibitem{footnoteNotationCumulant}
In general we denote the $s$th cumulant of a random
variable by $\langle\langle x^{s}\rangle\rangle$.
\bibitem{antisymmetry of riemann tensor}
The antisymmetry of the Riemann tensor
$R^{A}{ }^{\ }_{BC^*D}$
has to be used.
\bibitem{CommentDomainWalls}
As we will not include domain walls in symmetry class D, 
we replace $\mathrm{O}(N)$ by its connected counterpart
$\mathrm{SO}(N)$ in the following.
\bibitem{Rem: grade of the components of tensor fields}
Here,
we are considering $c$-type (commuting) tensor fields only.
If the tensor field itself is $a$-type (anticommuting),
components of the tensor field pick up an extra
statistical factor.\cite{DeWitt92}
\bibitem{Rem: complex conjugation of tensor fields}
In this paper,
we are always dealing with
real $c$-type vector and tensor fields
and their components 
in the so-called standard basis.
\cite{DeWitt92}
\end{thebibliography}
\end{document}